\documentclass[reprint,twocolumn,longbibliography,superscriptaddress,
showpacs,preprintnumbers,
nofootinbib,
notitlepage,umath,amssymb,
prx,
]{revtex4-1}

\usepackage{graphicx}
\usepackage{dcolumn}
\usepackage{bm}
\usepackage{bbm}
\usepackage{amsthm} 
\usepackage{diagbox}
\usepackage{braket}

\usepackage{graphicx}
\usepackage{enumerate}
\usepackage{subfigure}
\usepackage{epstopdf}
\usepackage[colorlinks,
linkcolor=red,
anchorcolor=black,
citecolor=blue]{hyperref}
\usepackage{amssymb}
\usepackage{amsmath}
\usepackage{framed}
\usepackage{tabularx}
\usepackage[normalem]{ulem}
\usepackage{booktabs}
\usepackage{float}
\usepackage[table]{xcolor}
\usepackage{array}
\usepackage{soul}
\usepackage{tikz}
\hypersetup{colorlinks=true}
\usetikzlibrary{positioning}
\usetikzlibrary{decorations,arrows}
\usetikzlibrary{decorations.pathreplacing}
\usetikzlibrary{chains, scopes, positioning, backgrounds, shapes, fit, shadows, calc, arrows.meta}
\tikzset{>=stealth}
\usepackage{color}
\usepackage{wasysym}
\usepackage{mathtools}
\usepackage{comment}
\usetikzlibrary{through,hobby}
\usetikzlibrary{decorations.markings}
\usetikzlibrary{fadings,shadings}
\usetikzlibrary{plotmarks}
\definecolor{darkblue}{rgb}{0.,0.,0.4}
\definecolor{darkred}{rgb}{0.5,0.,0.}
\definecolor{BlueViolet}{RGB}{138,43,226}
\definecolor{SkyBlue}{RGB}{30,144,255}
\definecolor{DarkGreen}{RGB}{0,150,0}
\definecolor{DarkYellow}{RGB}{0,1,1}
\usetikzlibrary{intersections}
\definecolor{iro100}{cmyk}{1,0,0,0}
\definecolor{iro90}{cmyk}{.9,0,0,0}
\definecolor{iro80}{cmyk}{.8,0,0,0}
\definecolor{iro60}{cmyk}{0,.6,0,0}
\definecolor{iro10}{cmyk}{0,.1,0,0}

\newtheorem{theorem}{Theorem}[section]

\theoremstyle{definition}
\newtheorem{definition}[theorem]{Definition}

\newcommand{\bs}{\boldsymbol}

\newcommand{\E}{\mathcal{E}}
\newcommand{\1}{\text{\uppercase\expandafter{\romannumeral1}}}
\newcommand{\2}{\text{\uppercase\expandafter{\romannumeral2}}}
\newcommand{\3}{\text{\uppercase\expandafter{\romannumeral3}}}
\newcommand{\4}{\text{\uppercase\expandafter{\romannumeral4}}}
\newcommand{\5}{\text{\uppercase\expandafter{\romannumeral5}}}
\newcommand{\6}{\text{\uppercase\expandafter{\romannumeral6}}}

\def\U{\mathrm{U}(1)}
\def\H{\mathcal{H}}
\def\Z{\mathbb{Z}}
\newcommand{\op}[1]{\ket{#1}\!\bra{#1}}
\DeclareMathOperator{\Tr}{Tr}

\newcommand{\meng}[1]{\text {[Meng: #1]}}

\newcommand{\Chong}[1]{ { \color{darkblue} \footnotesize (\textsf{CW}) \textsf{\textsl{#1}} }}

\newcommand{\Jianhao}[1]{ { \color{DarkGreen} [JHZ: #1] }}
\begin{document}
\title{Topological Phases with Average Symmetries:\\
the Decohered, the Disordered, and the Intrinsic}
\author{Ruochen Ma}
\thanks{These authors contributed equally.}
\affiliation{Perimeter Institute for Theoretical Physics, Waterloo, Ontario, Canada N2L 2Y5}
\affiliation{Department of Physics and Astronomy, University of Waterloo, Waterloo, ON N2L 3G1, Canada}
\affiliation{Kadanoff Center for Theoretical Physics, University of Chicago, Chicago, Illinois 60637, USA}
\author{Jian-Hao Zhang}
\thanks{These authors contributed equally.}
\affiliation{Department of Physics, The Pennsylvania State University, University Park, Pennsylvania 16802, USA}
\affiliation{
Department of Physics and Center for Theory of Quantum Matter,
University of Colorado, Boulder, Colorado 80309, USA
}
\author{Zhen Bi}
\affiliation{Department of Physics, The Pennsylvania State University, University Park, Pennsylvania 16802, USA}
\author{Meng Cheng}
\affiliation{Department of Physics, Yale University, New Haven, Connecticut  06511-8499, USA}
\author{Chong Wang}
\affiliation{Perimeter Institute for Theoretical Physics, Waterloo, Ontario, Canada N2L 2Y5}

\begin{abstract}
Global symmetries greatly enrich the landscape of topological quantum phases, playing an essential role from topological insulators to fractional quantum Hall effect. Topological phases in mixed quantum states, originating from \textit{decoherence} in open quantum systems or \textit{disorders} in imperfect crystalline solids, have recently garnered significant interest. Unlike pure states, mixed quantum states can exhibit \textit{average symmetries} -- symmetries that keep the total ensemble invariant but not on each individual state. 
In this work, we present a systematic classification and characterization of average symmetry-protected topological (ASPT) phases applicable to generic symmetry groups, encompassing both average and exact symmetries, for bosonic and fermionic systems. Moreover, we formulate the theory of average symmetry-enriched topological (ASET) orders in disordered bosonic systems. 
Our systematic approach helps clarify nuanced issues in previous literature and uncovers compelling new physics.  Notably, we discover that (1) the definition and classification of ASPT phases in decohered and disordered systems exhibit subtle differences; (2) despite these differences, ASPT phases in both settings can be classified and characterized under a unified framework of defect decoration and spectral sequence; (3) this systematic classification uncovers a plethora of ASPT phases that are \textit{intrinsically mixed}, implying they can exclusively manifest in decohered or disordered systems where part of the symmetry is average; (4) similarly for ASET, we find intrinsically disordered phases exhibiting exotic anyon behaviors -- the ground states of such phases necessarily contain localized anyons, with gapless (yet still localized) excitation spectral.

\end{abstract}

\maketitle
\tableofcontents

\section{Introduction}

Topological quantum phases are often enriched by global symmetries~\cite{wenbook}. The most familiar example is the fractional quantum Hall effect (FQHE). Independent of any global symmetry, the FQHE ground states are characterized by intrinsic \textit{topological orders} that support anyon excitations. Global symmetries such as U(1) charge conservation, however, can further enrich the topological orders, for example by assigning fractional charges to the anyon excitations. FQHE enriched by U(1) charge conservation is a simple example of \textit{symmetry-enriched topological orders} (SET) \cite{gauging3}. Even quantum states with trivial topological orders can be enriched by global symmetries. Familiar examples include topological insulators \cite{KaneRMP, ZhangRMP} and the Haldane spin-$1$ chain \cite{haldane1983nonlinear}. These states do not carry exotic excitations such as anyons, yet they cannot be adiabatically deformed to un-entangled product states (like atomic insulators) without breaking relevant symmetries. Such states are known as \textit{symmetry-protected topological} (SPT) states \cite{Chen:2011pg, XieChenScience, Ashvin2013, ChongWang2014, XieChen_2014, Senthil_2015, Bi_2015a}.

Traditionally the concepts of SPT and SET phases are defined for a pure quantum state $|\Psi\rangle$ -- typically the ground state of some local Hamiltonian $H$. Our understanding of SPT and SET phases in pure states has reached a quite mature stage over the past few decades. For mixed quantum states, however, much less is known. The two relevant physical contexts to study topological phases in mixed states are systems with \textit{decoherence} and \textit{disorders}. In decohered systems (such as noisy intermediate-scale quantum (NISQ) simulators \cite{Preskill_2018, Bernien_2017, Ruben_2023, Iqbal_2023b}), mixed states arise naturally from the coupling between the system and the environment. In a disordered system (such as an imperfect crystalline solid), the Hamiltonian is drawn from a statistically random ensemble $\{H_I\}$, where $I$ labels disorder realizations with a certain probability $P_I$. Given the ubiquity of decoherence and disorder in realistic systems, it is both natural and important to try to extend our understanding of topological phases, including SPT and SET, to mixed states. {\color{black}On a practical level, this understanding could assist us in utilizing topological phases for quantum information purposes, such as measurement-based quantum computation with SPT \cite{Briegel_2009, Nautrup2015Symmetry, Else_2012, Raussendorf_2019} and fault-tolerant quantum computation with topological orders \cite{TopoQCRMP}, even in realistic systems affected by disorder or decoherence.}

In contrast to pure states, there are two distinct types of symmetry for mixed states. An \textit{exact symmetry} (or \textit{strong symmetry}) is a symmetry for each individual pure state $|\Psi_I\rangle$ in the mixed state ensemble. In contrast, an \textit{average symmetry} (or \textit{weak symmetry}) is not necessarily a symmetry for each individual pure state, but is a symmetry for the ensemble density matrix $\rho=\sum_IP_I|\Psi_I\rangle\langle\Psi_I|$. Namely, for an average symmetry operator $g$, we have
\begin{equation}
    g\rho g^{-1}=\rho.
\end{equation}
For example, in a solid with quenched disorder, where the electrons experience a random potential, the crystalline symmetries are (at best) average symmetries, while the U(1) electric charge symmetry remains an exact symmetry.

The central task of this work is to systematically examine how exact and average symmetries enrich topological phases in mixed states, in both decohered and disordered systems. Historically, it was found that some free-fermion SPT phases remain nontrivial \cite{Ringel2012, Mong2012, Fulga2012}, in the sense that the boundary states remain delocalized when part of the protecting symmetries become average due to disorders. For bosonic systems, it was found \cite{deGroot2022} that, at least in (1+1)$d$, SPT phases protected by exact (strong) symmetries are stable against decoherence \cite{deGroot2022}. More recently, it was realized \cite{MaWangASPT} that the general notion of SPT phases can be extended to systems with both exact and average symmetries. These ``average SPT'' (ASPT) phases are characterized by nontrivial topological response, or equivalently decorated defects, involving both the exact and average symmetries. Many examples of ASPT were studied in Refs.~\cite{MaWangASPT, LeeYouXu2022, ZhangQiBi2022}. { Closely related notions, such as average symmetry defect and anomaly, have also been studied recently in the field theory context \cite{Antinucci2023}.}

In this work, we systematically classify and characterize topological phases, both SPT and SET, with average and exact symmetries. This allowed us to clarify many subtle issues from previous literature, and discover new physics that are intrinsically associated with mixed states. 

First, in Sec.~\ref{sec:ASPT} we discuss the general framework to classify ASPT, in both decohered and disordered systems. Even though the physical states in both scenarios can be described by density matrices of the form $\rho=\sum_IP_I|\Psi_I\rangle\langle\Psi_I|$, in a disordered system the state is also further endowed with an ensemble of Hamiltonians $\{H_I\}$ with probability $\{P_I\}$. This makes the physics of the two scenarios quite different. In particular, the classification of ASPT phases will be different in the two scenarios. Pleasantly, despite the physical difference between decohered and disordered systems, the ASPT phases in both scenarios can be characterized and classified under the unified mathematical framework of spectral sequence. The physics behind the spectral sequence framework is the decoration of symmetry defects~\cite{XieChen_2014, Gaiotto:2017zba, SpectralSequence}, which is familiar from the classification of pure state SPT phases.

For decohered systems, we carefully examine the conditions under which the classification based on decorated symmetry defects applies. As a necessary condition, in order for the topological invariants to be well-defined, we find that not only the usual spontaneous symmetry breaking, defined by long-range order in two-point correlation function of order parameters, should be forbidden, but a more subtle form of spontaneous symmetry breaking from an exact/strong symmetry down to an average/weak symmetry can also invalidate the classification. To circumvent the problem, we introduce a notion of invertibility for SRE mixed states, generalizing a similar concept for ground states.

As an appealing consequence of our classification, we discover a plethora of ASPT phases that are \textit{intrinsically mixed}. These are SPT phases that can only exist with average symmetries, in a decohered or disordered system, and by definition cannot exist as pure state SPT. In fact, if we try to deform an intrinsically mixed ASPT to a pure state, for example by reducing the disorder strength, the state reduces to the so-called ``intrinsically gapless SPT''. We discuss many examples of intrinsically mixed ASPT, in both bosonic and fermionic systems, in Sec.~\ref{sec:BerryFree}.

In Sec.~\ref{sec:ASET} we move on to nontrivial intrinsic topological orders in (2+1)$d$ in the presence of quenched disorders, and discuss how average (and exact) symmetries enrich the structure of topological orders. In clean systems, global symmetries can (1) permute different anyon types and (2) fractionalize on certain anyons (such as the fractional U(1) charges of FQHE quasi-particles). Certain enrichment patterns come with 't Hooft anomalies -- such anomalous SET states can be viewed as the surface states of certain SPT phases in one higher dimension. We show that average symmetries can also permute anyons. Furthermore an average symmetry and an exact symmetry can jointly fractionalize on anyons. However, the fractionalization of average symmetry alone becomes ill-defined on the anyons -- unless the fractionalization pattern involves certain 't Hooft anomalies. 

Another interesting feature of disordered systems is that certain obstructions of symmetry enrichment, known as $\H^3$ obstruction, are lifted once the symmetries involved become average. Such intrinsically disordered ASET comes with additional features. For instance, the system will have localized anyons that lead to a gapless spectrum, yet the system still hosts short-range correlated ground states. We discuss many examples, featuring various nontrivial properties of ASETs mentioned above, in Sec.~\ref{sec:ASET}.

We end with a summary and outlook in Sec~\ref{sec:summary}. Some technical details are discussed in the Appendices.

\section{Average SPT: generalities}
\label{sec:ASPT}

In this section, we first give an overview of the basic notions of ASPT in two different physical scenarios, with decoherence (Sec.~\ref{sec:decoheredReview}) and disorders (Sec.~\ref{sec:disorderedReview}), in the simplest cases where the total symmetry $\tilde{G}$ is the direct product of the average symmetry $G$ and the exact symmetries $A$. In Sec.~\ref{sec:SSclassification} we show that for more general symmetry structures, the classification of ASPT can be described using the Atiyah-Hirzebruch spectral sequence, for both the decohered and disordered scenarios (with different input data and consistency conditions). The goal of this Section is to not only review, but significantly systematize and clarify earlier discussions in Ref.~\cite{MaWangASPT}.

\subsection{Decohered ASPT}
\label{sec:decoheredReview}

We now define the notion of a decohered ASPT state, which is relevant for open quantum systems. 
{\color{black} 
\subsubsection{Gapped, symmetrically invertible states}
\label{sec:symmetricallyinvertible}
For pure states, a crucial property of SPT states is \textit{invertibility}. Namely, even though a nontrivial SPT state $|\Psi\rangle\in \mathcal{H}$ cannot be deformed to a trivial product state using a finite-depth symmetric unitary circuit, we can find an ``inverse'' state $|\tilde{\Psi}\rangle$ from an auxiliary Hilbert space $\tilde{\mathcal{H}}$, such that $|\Psi\rangle\otimes|\tilde{\Psi}\rangle$ can be deformed to a trivial product state in $\mathcal{H}\otimes\tilde{\mathcal{H}}$, using a symmetric finite-depth unitary circuit. The invertibility condition provides a foundation for classifying SPT phases using invertible topological quantum field theory \cite{invertible2,invertible3}. In fact, it is often more satisfactory to demand symmetric invertibility as it allows us to naturally incorporate phases like the integer quantum Hall effect. 

Another consequence of symmetric invertibility for pure states, which turns out to be crucial for our purpose, is the property of symmetry localization, which leads to the ``decorated domain wall" picture. To be more concrete, let us consider a state $\ket{\Psi}$ and an on-site unitary symmetry transformation $g$. Denote the restriction of $g$ to a subregion $\Gamma$ by $g_{\Gamma}$. For a symmetric (i.e. no SSB) SRE pure state, we expect that \cite{Else_2014}
\begin{equation}
\label{eq:invertiblestring}
    g_\Gamma \ket{\Psi}= V_{\partial \Gamma}\ket{\Psi},
\end{equation}
where $V_{\partial \Gamma}$ is a unitary operator supported on the boundary $\partial \Gamma$ of $\Gamma$.
In 1D, $V_{\partial \Gamma}$ further decomposes into local unitaries supported at the two ends of $\Gamma$ and this is equivalent to the existence of string order parameters. In general, the group-cohomology classification can be derived if one assumes that $V_{\partial \Gamma}$ and the analogous higher-codimensional defect operators are finite-depth unitary circuits \cite{Else_2014}. Eq. \eqref{eq:invertiblestring} directly follows from symmetric invertibility, which will be proved as a special case of the Theorem \ref{thm:mixedsymfrac} below.

Motivated by the study of pure-state invertible phases, we focus on symmetrically invertible states in our investigation of decohered ASPT phases, which can be similarly defined for mixed states. To facilitate the definition, let us first define symmetric finite-depth quantum channel.

Throughout these definitions, the physical system of interest lives in a (tensor product) Hilbert space $\mathcal{H}$. The strong (weak) symmetries form a group $A$ ($G$). All symmetries are assumed to be on-site and unitary.

\begin{definition}
A quantum channel $\E$ is a symmetric finite-depth channel, if it admits a purification to a unitary $\mathbf{U}$ on an enlarged space $\H\otimes\mathcal{A}$, such that
\begin{equation}
    \E(\rho) = \mathrm{tr}_\mathcal{A}(\mathbf{U} \rho\otimes |0_\mathcal{A} \rangle\langle 0_\mathcal{A} | \mathbf{U}^\dagger).
    \label{eq:defsymFDC}
\end{equation}
Specifically, (1) the space $\mathcal{A}$, which transforms trivially under $A$ while acted on by the weak symmetry $G$ via $g_\mathcal{A}$, is obtained by associating an ancilla with each site in $\H$; (2) $|0_\mathcal{A} \rangle \in \mathcal{A}$ is a product state symmetric under $g_\mathcal{A}$; (3) $\mathbf{U}$ is a local unitary circuit, with each gate commuting with $a\otimes \mathbf{1}_\mathcal{A}$ and $g\otimes g_\mathcal{A}$, where $a$ and $g$ are the generators of strong and weak symmetries in $\H$, and $\mathbf{1}_\mathcal{A}$ is the identity in $\mathcal{A}$.
\end{definition}

{\definition
Given a state $\rho$ (in Hilbert space $\mathcal{H}$) and a set of strong ($A$) and weak ($G$) symmetries, we call $\rho$ symmetrically invertible if there is a state $\tilde{\rho}$ from an auxiliary Hilbert space $\tilde{\mathcal{H}}$, such that (1) each element $a \in A$ (and $g \in G$) acts onsite and diagonally in $\H\otimes \tilde{\H}$ as $\mathbf{a}=a\otimes \tilde{a}$ (and $\mathbf{g}=g\otimes \tilde{g}$); (2) $\rho\otimes\tilde{\rho}$ is two-way connected to a pure product state $|0\rangle\in \H\otimes\tilde{H}$ through some symmetric finite-depth channels { $\E_1:\rho\otimes\tilde{\rho}\to|0\rangle\langle0|$ and $\E_2:|0\rangle\langle0|\to\rho\otimes\tilde{\rho}$}. 
}

According to Ref. \cite{2020Piroli}, symmetrically invertible states always exhibit exponential decay of the correlation functions of local operators and obey an area law for the mutual information between a region and its complement.}

Another virtue of focusing on symmetrically invertible states  is that it mitigates the complication due to the non-invertibility of quantum channels. To make this point more manifest, let us assume a slightly stronger form of symmetric invertibility: we demand $\rho\otimes\rho^*$ to be two-way connected to a pure product state through symmetric finite-depth channels. In other words, we demand the complex conjugation $\rho^*$ to be the inverse of $\rho$. This is motivated by the anticipation that nontrivial topological structure from quantum mechanics should manifest as some nontrivial phase factors -- this intuition is true at least for invertible pure states, and may even be true for general symmetrically invertible mixed states (we are not aware of counter examples). The strong form of symmetric invertibility leads to the following appealing result: if two strongly symmetrically invertible states are one-way connected, namely $\exists$ $\mathcal{E}^{12}_{\rm SFD}: \rho_1\to\rho_2$, then the two states are automatically two-way connected, namely $\exists$ $\mathcal{E}^{21}_{\rm SFD}: \rho_2\to\rho_1$. The ``inverse'' channel can be constructed in a few simple steps: $\rho_2 \to \rho_2\otimes \rho_1\otimes\rho_1^*\to \rho_2\otimes \rho_1\otimes\rho_2^* \to \rho_1$, where we have used $(\mathcal{E}^{12}_{\rm SFD})^*: \rho^*_1\to\rho^*_2$ in the second arrow, and strong symmetric invertibility $\rho_2\otimes\rho_2^*\to|0\rangle\langle0|$ in the last arrow.

{\color{black}
As mentioned earlier, symmetric invertibility allows strong symmetries to be localized on mixed states, generalizing a similar result for pure states. More formally, we have the following theorem:

\begin{theorem}
    If $\rho$ is (strongly) symmetrically invertible, then for any strong symmetry $a\in A$ of $\rho$ and a large but finite region $\Gamma$, there exists an \sout{unitary} { operator} $V_{\partial\Gamma}$ supported on the boundary $\partial\Gamma$ such that
    \begin{equation}
        a_\Gamma\rho=V_{\partial \Gamma}\rho,
        \label{eq:mixedelsenayak}
    \end{equation}
    { where $V_{\partial\Gamma}$ has operator norm (the largest singular value) $\|V_{\partial \Gamma}\|=1$.}
    \label{thm:mixedsymfrac}
\end{theorem}
\begin{proof} 
Since $\rho$ is symmetrically invertible, we have 
\begin{equation}
    \mathbf{U}\rho \otimes \tilde{\rho} \otimes |0_\mathcal{A}\rangle \langle 0_\mathcal{A} | \mathbf{U}^\dagger = | 0 \rangle \langle 0| \otimes \rho_\mathcal{A},
\end{equation}
where $|0_\mathcal{A}\rangle$ is a symmetric product state on an ancillary Hilbert space $\mathcal{A}$. 
Note that $a$ is extended to act on-site on the Hilbert space $\tilde{\H}$ as $\tilde{a}$, and we denote $\mathbf{a}=a\otimes \tilde{a}$. $\mathbf{U}$ is a \emph{locally} symmetric unitary circuit acting on $\H \otimes \tilde{\mathcal{H}} \otimes \mathcal{A}$, meaning that each gate in $\mathbf{U}$ commutes with $a\otimes \tilde{a}$. We thus have
\begin{equation}
\begin{split}
    \mathbf{a}_\Gamma \rho \otimes \tilde{\rho} \otimes |0_\mathcal{A} \rangle\langle 0_\mathcal{A} | &=  \mathbf{a}_\Gamma  \mathbf{U}^\dagger | 0 \rangle \langle 0 | \otimes \rho_\mathcal{A} \mathbf{U}\\
    &=\mathbf{a}_\Gamma  \mathbf{U}^\dagger \mathbf{a}_\Gamma^{-1} \mathbf{U}\cdot \mathbf{U}^\dag | 0 \rangle \langle 0 | \otimes \rho_\mathcal{A} \mathbf{U}\\
    &=\mathbf{V}_{\partial \Gamma}\rho\otimes\tilde{\rho}\otimes \ket{0_{\mathcal{A}}}\bra{0_{\mathcal{A}}}.
\end{split}
\label{eq:invertiblestrongsymm}
\end{equation}
where $\mathbf{a}_\Gamma = a_\Gamma \otimes \tilde{a}_\Gamma$ represents the strong symmetry acting on $\H\otimes\tilde{\H}$ within a large but finite region $\Gamma$, and the operator $\mathbf{V}_{\partial\Gamma} = \mathbf{a}_\Gamma \mathbf{U}^\dagger \mathbf{a}_\Gamma^\dagger \mathbf{U}$ is a unitary supported solely on the boundary $\partial \Gamma$ because each gate of $\mathbf{U}$ is strongly symmetric.  

{ Tracing out $\tilde{\H}\otimes\mathcal{A}$ from Eq.~\eqref{eq:invertiblestrongsymm} gives Eq.~\eqref{eq:mixedelsenayak} with $V={\rm Tr}_{\tilde{\H}\otimes\mathcal{A}}\mathbf{V}\tilde{\rho}\otimes|0_{\mathcal{A}}\rangle\langle0_{\mathcal{A}}|$. The norm one property of $V$ can be shown as follows: the definition of $V$ implies that the $\Vert V\Vert\leq 1$. On the other hand, Eq. ~\eqref{eq:mixedelsenayak} implies that $\Vert V\Vert\geq 1$. Together we must have $\Vert V\Vert=1$.}
\end{proof}


{ We note that, although $V_{\partial\Gamma}$ is not guaranteed to be unitary, it acts as a unitary on the image of $\rho_{\partial\Gamma}$ (the reduced density operator on the boundary region). For this reason, it may be physically reasonable to practically treat $V_{\partial\Gamma}$ as a unitary, which is what we shall do in subsequent discussions.}

Also, note that $\mathbf{U}$ and $\mathbf{a}_\Gamma$ commute with the weak symmetry acting simultaneously on $\H\otimes \tilde{\H} \otimes \mathcal{A}$; therefore, $V_{\partial \Gamma}$ must commute with the weak symmetry $G$ acting on $\H$. As a $G$-symmetric locality-preserving unitary, $V_{\partial\Gamma}$ can at most pump a $G$-symmetric invertible phase to the boundary $\partial \Gamma$. Therefore, we conclude that in a symmetrically invertible mixed state, a domain wall of the strong symmetry traps a $G$-symmetric invertible phase in one lower dimension. 
 We will see more concrete examples in Sec.~\ref{sec:clusteredge}.

 In $1d$, the boundary operator decomposes as $V_{\partial \Gamma} = V_L V_R$, where $V_L$ and $V_R$ are supported near the left and right endpoints of $\Gamma$, respectively. Furthermore, for $a, \, b \in A$, the operators satisfy the relation (up to a phase):
\begin{equation}
V_R^a V_R^b \propto V_R^{ab}.
\end{equation}

 \begin{figure}
\begin{tikzpicture}[scale=1]
\draw[draw=DarkGreen, thick, dashed] (0,0)circle (2.5);
\tikzstyle{sergio}=[rectangle,draw=none]
\filldraw[fill=violet!20, draw=black, thick] (0,0)circle (2);
\filldraw[fill=yellow!20, draw=black, thick] (0,0)circle (1);
\filldraw[fill=yellow!20, draw=DarkGreen, thick, dashed] (0,0)circle (0.5);
\path (0,0) node [style=sergio,color=red] {\large$C_1$};
\path (-0.4,0.4) node [style=sergio,color=red] {\large$A^-$};
\path (0.55,-0.55) node [style=sergio,color=red] {$B_1$};
\path (1.57,-1.57) node [style=sergio,color=red] {$B_2$};
\path (2,-2) node [style=sergio,color=red] {$C_2$};
\path (1.5,0.33) node [style=sergio,color=red] {\large$|\partial A|$};
\path (-1.075,1.075) node [style=sergio,color=red] {\Large$\partial A$};
\path (-1.75,1.75) node [style=sergio,color=red] {\Large$\bar{A}^-$};
\draw[thick, <->] (1,0) -- (2,0);
\end{tikzpicture}
\caption{A circular tripartition $A^-$, $\partial A$, and $\bar{A}^-$ of the system. $B_1$ and $B_2$ are buffer region of $\partial A$ with finite width.}
\label{Fig: tripartition}
\end{figure}



Theorem \ref{thm:mixedsymfrac} provides a sufficient condition for a strong symmetry to effectively localize near the boundary when acting on a large but finite region. A natural question then arises: under what conditions on a mixed state does a weak symmetry exhibit similar localization? It turns out that certain constraints must be imposed on the information-theoretic properties of the mixed state, which we refer to as being a \emph{gapped mixed state}{ , or \emph{gapped Markovian state} following Ref.~\cite{lessa2024mixedstate}}:

\begin{definition}
Consider a tripartite separation of the system (see Fig. \ref{Fig: tripartition}), where $A^-$ is a disk regime, $\partial A$ is a buffer regime surrounding $A$ with the width $|\partial A|$, and $\bar{A}^-$ is the rest of the system. A mixed state $\rho$ is defined as ``gapped'' if for arbitrary tripartition of this form, we have (here $S(A^-)$ is the von Neumann entropy of the regime $A^-$)
\begin{enumerate}[a.]
\item The mutual information (MI) between the two disjoint regions decays exponentially:
\begin{align}
I_\rho(A^-:\bar{A}^-)&=S(A^-)+S(\bar{A}^-)-S(A^-\bar{A}^-)\nonumber\\
&=D\left(\rho_{A^-\cup \bar{A}^-}\big\Vert\rho_{A^-}\otimes\rho_{\bar{A}^-}\right)\nonumber\\
&\sim e^{-|\partial A|/\xi_{\mathrm{MI}}},
\label{Eq: MI}
\end{align}
with a finite correlation length $\xi_{\mathrm{MI}}$. Here $D\left(\rho_{A^-\cup \bar{A}^-}\big\Vert\rho_{A^-}\otimes\rho_{\bar{A}^-}\right)$ is the quantum relative entropy between density matrices $\rho_{A^-\cup \bar{A}^-}$ and $\rho_{A^-}\otimes\rho_{\bar{A}^-}$;
\item The conditional mutual information (CMI) also decays exponentially:
\begin{align}
I_\rho &(A^-:\bar{A}^-|\partial A)\nonumber\\
&=S(A^-\cup\partial A)+S(\partial A\cup\bar{A}^-)\nonumber\\
&-S(\partial A)-S(A^-\cup\partial A\cup\bar{A}^-)\nonumber\\
&\sim e^{-|\partial A|/\xi_{\mathrm{CMI}}},
\label{Eq: CMI}
\end{align}
with a finite $\xi_{\mathrm{CMI}}$, referred to as the Markov length in \cite{sang2024stability}.
\end{enumerate}
\label{def:infoSRE}
\end{definition}


Def. \ref{def:infoSRE} captures the physical notion of a mixed state being short-range correlated: The mutual information serves as a measure of the total bipartite correlation between $A^-$ and $\bar{A}^-$, and Eq. \eqref{Eq: MI} implies that all connected correlation functions of the form \begin{equation} 
C^{(2)}(x,y) = \Tr(\rho O_x O_y) - \Tr(\rho O_x) \Tr(\rho O_y)
\end{equation} 
decay exponentially with the distance $|x-y|$ \cite{2008arealaw}. Additionally, the conditional mutual information, which quantifies the global correlations between $A^-$ and $\bar{A}^-$ that cannot be inferred from their individual correlations with the buffer region $\partial A$, also remains small over large distances. We now demonstrate that for a mixed state satisfying Def. 
\ref{def:infoSRE}, a weak symmetry, when applied to a finite but sufficiently large region, effectively localizes near the boundary. 
}
{\color{black}
\begin{theorem}
\label{Thm: weak symmetry localization}
For a gapped mixed state $\rho$, a weak symmetry $g \in G$ can always be effectively localized. Specifically, when a weak symmetry operator $W$ is truncated to a large region $A$, the truncated operator $W_A$ acts as a boundary quantum channel $\E_B$, such that
\begin{align} 
W_A \rho W_A^\dag = \E_B[\rho], \label{eq:weakfrac} 
\end{align} 
where $\E_B$ is supported in a region $B=B_1\cup\partial A\cup B_2$ near the boundary of $A$, as shown in Fig. \ref{Fig: tripartition}. Moreover, for 1$d$ systems, $\E_{B}$ can be decomposed into two local quantum channels $\E_L$ and $\E_R$ that are located at the left and right boundaries, respectively, namely $\E_{B}=\E_L\circ\E_R$. 
\end{theorem}

\begin{proof}
 Consider the mixed state that is obtained from conjugating $W_A$ to $\rho$, namely
\begin{align}
\rho'=W_A \rho W_A^\dag.
\label{def:weakpartial}
\end{align}
 Without loss of generality, we assume $\xi_{\mathrm{MI}} = \xi_{\mathrm{CMI}} = \xi$. We then divide the entire system into three regions: $A^-$, $\partial A$, and $\bar{A}^-$, as illustrated in Fig. \ref{Fig: tripartition}. The region $\partial A$ is chosen such that $|\partial A| \gg \xi$, and $A$ is selected so that $A^- \subset A$, with the distance from the boundary of $A$ to both $A^-$ and $\bar{A}^-$ being much larger than $\xi$. We denote the onsite weak symmetry operator within $A^-$ as $W_{A^-}$, and the operator acting within $A \setminus A^-$ as $W_{A \setminus A^-}$. Clearly, the total symmetry operator in $A$ satisfies $W_A = W_{A^-} W_{A \setminus A^-}$.



Let us first demonstrate Eq. \eqref{eq:weakfrac} in arbitrary spatial dimensions. One has
\begin{equation}
    \begin{split}
        \rho_{A^-\cup\bar{A}^-} & = \Tr_{\partial A} (\rho) \simeq \rho_{A^-} \otimes\rho_{\bar{A}^-} \\
        & = W_{A^-}\rho_{A^-} W_{A^-}^\dagger \otimes\rho_{\bar{A}^-} \\
        & = W_{A^-}\Tr_{\partial A} (W_{A \setminus A^-} \rho W_{A \setminus A^-}^\dagger) W_{A^-}^\dagger \\
        & \simeq \rho'_{A^-\cup\bar{A}^-},
    \end{split}
    \label{eq:reducedsame}
\end{equation}
where $\simeq$ denotes an error that is exponentially small in $|\partial A|$. In Eq. \eqref{eq:reducedsame}, we utilized the following: (1) in the first line, Eq. \eqref{Eq: MI}; and (2) in the second line, the fact that the reduced density matrix of $\rho$ on any arbitrary region is weakly symmetric.

The channel $\E_B$ in Eq. \eqref{eq:weakfrac} can now be constructed as follows. First, note that since $\rho'$ is obtained from $\rho$ via an onsite unitary transformation as defined in Eq.\eqref{def:weakpartial}, the CMI remains unchanged:
\begin{equation}
    I_{\rho'}(\partial A; \bar{ B} | B\setminus \partial A) = I_{\rho}(\partial A; \bar{B} | B\setminus \partial A),
    \label{eq:rho'CMI}
\end{equation}
which decays exponentially with the distance between $\partial A$ and $\bar{B}$ according to Eq.\eqref{Eq: CMI}. Consequently, we have:
\begin{equation}
    \begin{split}
        &\mathcal{R}_B\circ\Tr_{\partial A}(\rho) = \mathcal{R}_B(\rho_{A^-\cup \bar{A}^-}) \\
        & \simeq \mathcal{R}_B\circ\Tr_{\partial A}(\rho') \simeq \rho',
    \end{split}
\end{equation}
where the second line uses the following observations: (1) Given that the CMI of $\rho'$ in Eq.\eqref{eq:rho'CMI} decays exponentially with the distance between $\partial A$ and $\bar{B}$, there exists a \emph{local} channel $\mathcal{R}_B$ acting on region $B$ such that:
\begin{equation}
    \Vert \mathcal{R}_B\circ\Tr_{\partial A}(\rho') - \rho' \Vert_1 \simeq \exp[-(|B|-|\partial A|)/\xi].
\end{equation}
The channel $\mathcal{R}_B$ is chosen as the Petz recovery channel for the partial trace $\Tr_{\partial A}$, as discussed in Ref. \cite{2015Petzrecovery}. (2) The CMI in Eq. \eqref{eq:rho'CMI} indeed decays exponentially with $|B| - |\partial A|$, following from the assumption that $\rho$ is gapped. Specifically, this behavior is guaranteed by Eq. \eqref{Eq: CMI}, applied to the geometry illustrated in Fig. \ref{Fig: tripartition}, and can be derived using the chain rule of CMI:
\begin{equation}
    \begin{split}
        & I_\rho(\partial A;\bar{B}|B\setminus \partial A)\\
        =& I_\rho(\partial A;C_1|B_1\cup B_2)+ I_\rho(\partial A;C_2|B_1\cup B_2\cup C_1)\\
        =& I_\rho(\partial A\cup B_2;C_1|B_1)-I_\rho(B_2;C_1|B_1) \\
        +& I_\rho(\partial A\cup B_1 \cup C_1;C_2|B_2)-I_\rho(B_1\cup C_1;C_2|B_2),
    \end{split}
\end{equation}
where all four terms in the last two lines are exponentially small in $|B| - |\partial A|$. This completes the proof of Eq. \eqref{eq:weakfrac}, where $\E_B = \mathcal{R}_B\circ \Tr_{\partial A}$.

Finally, we show that in $1d$, the channel $\E_B$ factorizes as $\E_B = \E_L \circ \E_R$. This follows straightforwardly: in $1d$, the partial trace channel factorizes into two channels localized near the left and right components of $\partial A$, respectively, i.e., $\Tr_{\partial A} = \Tr_{L} \circ \Tr_{R}$. Given that the mixed state is gapped, each component can be recovered independently near its respective region.
\end{proof}

After examining the conditions under which strong and weak symmetries localize on a mixed state, we now use them to define topological invariants for gapped, symmetrically invertible mixed states.

}

{\color{black}
\subsubsection{SPT invariants in 1$d$}

{Let us examine 1$d$ systems with a strong symmetry $A$ and a weak symmetry $G$. Let $g_1, g_2 \in G$ be two elements of the weak symmetry, with corresponding truncated symmetry operators $W_S(g_1)$ and $W_S(g_2)$ defined on a finite region $S$. By Theorem \ref{Thm: weak symmetry localization}, it follows that for a gapped mixed state $\rho$, the application of $W_S(g_i)$ ($i=1,2$) effectively localizes as a quantum channel near $\partial S$, specifically:
\begin{align}
W_S(g_i) \rho W_S^\dag(g_i) = \E_{\partial S}^{g_i}[\rho].
\label{eq:weakfra}
\end{align}
We sequentially conjugate $\rho$ with $W_S(g_1)$ and $W_S(g_2)$, obtaining:
\begin{equation} 
\begin{split} W_S(g_1)W_S(g_2) \rho W_S^\dag(g_2) W_S^\dag(g_1) = \E_{\partial S}^{g_1} \circ \E_{\partial S}^{g_2}[\rho] \\ = W_S(g_1 g_2) \rho W_S^\dag(g_1 g_2) = \E_{\partial S}^{g_1 g_2}(\rho). 
\end{split} 
\end{equation}
This implies that the channels $\{ \E_{\partial S}^g , g \in G \}$ form a linear representation of $G$. Furthermore, in 1$d$, each channel $\E_{\partial S}^{g_i}$ factorizes into two channels, $\E_L^{g_i}$ and $\E_R^{g_i}$, supported at the left and right boundaries, respectively (by Theorem \ref{Thm: weak symmetry localization}). Consequently, the channels $\{ \E_L^g \}$ also form a representation of $G$:
\begin{align} 
\E_L^{g_1} \circ \E_L^{g_2} = \E_L^{g_1 g_2}, 
\label{eq:channellinearrep}
\end{align}
which must be linear (with a trivial projective phase), as quantum channels are by definition completely positive. The same conclusion applies to $\{ \E_R^g \}$. Consequently, there are no nontrivial SPT phases that are protected solely by a weak symmetry.

Now we consider a strong symmetry $a$, and for simplicity in this section, we assume the total symmetry is $A \times G$, i.e., $A$ and $G$ commute. As a result, the density operator in Eq. \eqref{eq:weakfra} retains the same charge under $a$ as $\rho$. In one dimension, this implies that each Kraus operator in $\E_L^g$ and $\E_R^g$ also carries a well-defined charge under $a$. One can define
\begin{equation}
    (\E^{g}_L)^\dagger (a)=\chi_{g}(a) a,
    \label{eq:defweaklocinv}
\end{equation}
where $g\in G$ and $\chi_{g}(a)$ is a $\U$ phase. It is clear that
\begin{equation} 
\chi_g(a_1 a_2) = \chi_g(a_1) \chi_g(a_2), \end{equation} 
indicating that for any $g \in G$, $\chi_g(\cdot)$ defines a character (charge) of $A$, valued in $H^1(A, \U)$. Furthermore, from Eq.\eqref{eq:channellinearrep}, we obtain
\begin{equation} 
\chi_{g_1}(a) \chi_{g_2}(a) = \chi_{g_1 g_2}(a), \, \forall a \in A. \end{equation} 
Thus, $\chi_g$ defines a group homomorphism from $G$ to the group of $A$ characters, valued in $H^1(G, H^1(A, \U))$.  

Finally, by Theorem \ref{thm:mixedsymfrac}, a strong symmetry fractionalizes on a symmetrically invertible mixed state: \begin{equation} 
a_\Gamma \rho = V_L^a V_R^a \rho, \end{equation} 
where $a_\Gamma$ is the strong symmetry generator restricted to an interval $\Gamma$, and $V_L^a$, $V_R^a$ are unitaries supported near the left and right endpoints of $\Gamma$, respectively. Although the symmetry generators $\{ a_\Gamma \}$ for $a \in A$ form a linear representation of $A$, the endpoint operators $\{ V_L^a \}$ may form a projective representation, labeled by an element $\omega \in H^2(A,\U)$. 

In addition, recall that $V_{\partial \Gamma}=V_L^a V_R^a$ commute with $G$. As a result, $V_L^a$ carries a well-defined $G$-charge, given by
\begin{equation}
    g V_L^a g^{-1}=\chi'_a(g) V_L^a.
    \label{eq:def1dmixedinvariant}
\end{equation}
It is straightforward to show that $\chi'_a(g)$ defines an element  in $\H^1(A, H^1(G, \U))$.

Now we show that the two invariants $\chi \in H^1(G, H^1(A, \U))$, as defined in Eq.\eqref{eq:defweaklocinv}, and $\chi' \in H^1(A, H^1(G, \U))$ defined in Eq.\eqref{eq:def1dmixedinvariant}  are equal. This equivalence can be demonstrated by considering two large overlapping intervals, $\Gamma_1 = (x, y)$ and $\Gamma_2 = (\frac{x+y}{2}, \frac{3y-x}{2})$. Denoting the strong symmetry restricted to interval $\Gamma_1$ as $a_{\Gamma_1}$ and the weak symmetry similarly, we have:
\begin{equation}
\begin{split}
& a_{\Gamma_1} g_{\Gamma_2} \rho g_{\Gamma_2}^\dagger = \chi_g(a) V_L V_R \E_L \circ \E_R(\rho) \\
& = g_{\Gamma_2} a_{\Gamma_1} \rho g_{\Gamma_2}^\dagger = \chi'_a(g) V_L V_R \E_L \circ \E_R(\rho),
\end{split}
\end{equation}
which implies that for any $a \in A$ and $g \in G$, $\chi_g(a)= \chi'_a(g)$.

Lastly, we show that two gapped, symmetrically invertible mixed states $\rho_1$ and $\rho_2$ share the same invariants $\omega$ and $\chi$ if they are two-way connected by symmetric finite-depth channels. This establishes that $\omega$ and $\chi$ (or equivalently, the $H^1(G, H^1(A, \U))$ class) are genuine topological invariants for mixed-state SPT phases.

By virtue of strong symmetry localization $a_\Gamma \rho_1 = V_L V_R \rho_1$ (Theorem \ref{thm:mixedsymfrac}), the state $\rho_1$ exhibits a string order parameter: \begin{equation} 
\Tr(\rho_1 V_L^\dagger a_\Gamma V_R^\dagger) = 1, \end{equation}
where $\Gamma$ is a finite but large interval, and $V_L$ and $V_R$ are supported near the left and right endpoints of $\Gamma$, respectively. Since $\rho_1$ can be prepared from $\rho_2$ by a symmetric finite-depth local channel (see Eq.\eqref{eq:defsymFDC}), we have
\begin{equation}
\begin{split}
    1=&\mathrm{tr}_{\H,\mathcal{A}}(\mathbf{U} \rho_2 \otimes |0_\mathcal{A} \rangle \langle 0_\mathcal{A} |\mathbf{U}^\dagger V_L^\dagger a_\Gamma V_R^\dagger) \\
    =& \mathrm{tr}_{\H,\mathcal{A}}(\rho_2 \otimes |0_\mathcal{A} \rangle \langle 0_\mathcal{A} | \tilde{V}_L^\dagger a_\Gamma \tilde{V}_R^\dagger) \\
    =& \mathrm{tr}_{\H,\mathcal{A}}(\rho_2 \otimes |0_\mathcal{A} \rangle \langle 0_\mathcal{A} | W_L\tilde{V}_L^\dagger W_R\tilde{V}_R^\dagger)
    \label{eq:invarianceof stringorder}
\end{split}
\end{equation}
where $\H$ denotes the physical Hilbert space. In Eq.\eqref{eq:invarianceof stringorder} we used the following observations: (1) Since $\mathbf{U}$ is a symmetric finite-depth unitary acting on $\H \otimes \mathcal{A}$, we have $\mathbf{U}^\dagger V_L^\dagger a_\Gamma V_R^\dagger \mathbf{U} = \tilde{V}_L^\dagger a_\Gamma \tilde{V}_R^\dagger$, where $\tilde{V}_{L/R}^\dagger$ are supported near the left and right endpoints of $\Gamma$, respectively. In particular, $\tilde{V}_{L/R}$ carries the same representation under $A$ and $G$ as $V_{L/R}$. (2) In the third line, we used the localization of the strong symmetry on $\rho_2$, i.e., $\rho_2 a_\Gamma = \rho_2 W_L W_R$. As a result, $W_L$ and $\tilde{V}_L$, and consequently $V_L$, must carry the same representation under both $A$ and $G$. This implies that $\rho_1$ and $\rho_2$ share the same string order parameter, and hence the same SPT invariant, valued in $H^2(A, \U) \times H^1(A, H^1(G, \U))$. 
}

\subsubsection{Classification in 1$d$}
\label{1d classification}

We summarize the results in the following theorem:
\begin{theorem} In one dimension,
\begin{enumerate}[1.]
\item a gapped, symmetrically invertible mixed state with $A \times G$ symmetry is associated with an element in 
\begin{equation} H^1(G,H^1(A,\U)) \oplus H^2(A,\U); \label{1dclassificationAtimesG} \end{equation} 
\item two gapped, symmetrically invertible states correspond to the same element in Eq. \eqref{1dclassificationAtimesG} if they can be connected in both directions by symmetric finite-depth channels. 
\end{enumerate}
\label{Thm:1dASPTclassification}
\end{theorem}

We conjecture that the classification here is \emph{complete}, in the sense that two (symmetrically invertible and gapped) mixed states with the same group cohomology invariants are indeed connected by symmetric quantum channels.

In addition, we also need to check that all elements in Eq. \eqref{1dclassificationAtimesG} can be physically realized. When the symmetry group takes the form $A\times G$, one can start from a $A\times G$ pure state SPT with the given invariant, and apply a quantum channel to make $G$ weak. For example, in the ``fixed-point" model, one can simply apply a dephasing channel with a non-trivial $G$ character. We will provide a general construction in Sec.~\ref{sec:fixedpointmodel}.

Compared to the ground state classification $\H^2(G\times A, \U)$, the classification in Eq. \eqref{1dclassificationAtimesG} misses a factor $\H^2(G, \U)$, i.e. those that are entirely protected by the weak symmetry. Hence the decohered SPT phases are a subset of the ground state phases.

We also remark that the SPT invariants are well-defined as long as $A$ and $G$ commute. For example, the total symmetry group (including both strong and weak) can form a nontrivial central extension of $G$ by $A$ and the topological invariants can be defined in the same way. In this case, however, the classification may contain elements that do not occur in the ground state classification. This is because certain choices of the invariants are not compatible with the group structure and the SRE nature of gapped ground states. We will elaborate on these cases in Sec.\ref{sec:BerryFree}. 

\subsubsection{Example: cluster chain and edge state}
\label{sec:clusteredge}
Within the context of symmetrically invertible mixed states, let us consider a one-dimensional qubit chain with $\Z_2\times\Z_2^{\rm ave}$ symmetry as an illustrative example, where the exact $\Z_2$ acts on even-sites as $\prod_{i=2n}X_i$ ($i$ labeling the lattice sites), and the average $\Z_2^{\rm ave}$ acts on odd-sites as $\prod_{i=2n+1}X_i$. ASPT phases with this symmetry are classified by $\H^1(\Z_2,\H^1(\Z_2,\U))=\Z_2$, with one nontrivial phase. A representative density matrix of the nontrivial phase, on a closed chain with $2N$ sites, is
\begin{align}
\begin{gathered}
\rho_{\rm{cluster}}=\frac{1}{2^N}\sum_{z_{2n+1}=\pm1} |\Psi_{\{z_{2n+1}\}}\rangle\langle \Psi_{\{z_{2n+1}\}}|\\
|\Psi_{\{z_{2n+1}\}}\rangle=\bigotimes\limits_{i=2n+1}|Z_i=z_i\rangle\bigotimes\limits_{j=2n}|X_j=z_{j-1}z_{j+1}\rangle
\label{Eq:cluster}
\end{gathered}.
\end{align}
Essentially, we have a classical ensemble of $\Z_2^{\rm ave}$ domain-wall configurations, and at each domain wall, a nontrivial exact $\Z_2$ charge is decorated. It is not difficult to check that $\rho_{\rm{cluster}}$ is symmetrically invertible, by explicitly constructing symmetric finite-depth channels connecting $\rho_{\rm{cluster}}\otimes\rho_{\rm{cluster}}$ to a pure product state. 

The state $\rho_{\rm cluster}$ can be more compactly written as
\begin{equation}
    \rho_{\rm cluster}=\frac{1}{2^N}\prod_{n}\frac{1+Z_{2n-1}X_{2n}Z_{2n+1}}{2},
\end{equation}
or $X_{2n}\rho_{\rm cluster}=Z_{2n-1}Z_{2n+1}\rho_{\rm cluster}$. The strong $\Z_2$ symmetry can thus be localized:
\begin{equation}
    \prod_{i=n}^m X_{2i}\rho_{\rm cluster}=Z_{2n-1}Z_{2m+1}\rho_{\rm cluster}.
\end{equation}
The defect operator $Z_{2n-1}$ is charged under the weak $\Z_2$ symmetry. Similarly, the localization of the weak symmetry can be directly verified:
\begin{equation}
    \prod_{i=n}^{m-1} X_{2i+1} \rho_{\rm cluster}\prod_{i=n}^{m-1} X_{2i+1} = Z_{2n}Z_{2m}\rho_{\rm cluster}Z_{2m}Z_{2n}.
\end{equation}

We can thus characterize the ASPT phase using the following string order parameter: 
\begin{equation}
\langle Z_{2n-1}X_{2n}X_{2n+2}...X_{2m}Z_{2m+1} \rangle \xrightarrow{|m-n|\gg1} O(1).
\label{eq:string}
\end{equation}
In contrast, it was shown in Ref.~\cite{MaWangASPT} that states prepared from a pure product state with symmetric finite-depth channels cannot have the above string order parameter being $O(1)$.  Furthermore, if we deform the average cluster state using a finite-depth symmetric channel, as long as we keep the state symmetrically invertible (see Sec.~\ref{sec:symmetricallyinvertible}), the $O(1)$ string order parameter will survive -- the only possible change is that the end point operator may evolve from $Z_i$ to some other local operator charged under the week $\mathbb{Z}_2$ symmetry.  Therefore, the density matrix in Eq. \ref{Eq:cluster} represents a nontrivial ASPT. A similarly robust order parameter is strange correlators defined in Refs.~\cite{ZhangQiBi2022, LeeYouXu2022}.}

We now show that the string order parameter Eq.~\eqref{eq:string} implies nontrivial edge correlations, similar to the clean SPT. Consider an open chain, say from $i=1$ to $i=L$. Far away from the two boundaries the system should be indistinguishable from the closed cluster chain. This means that the string order parameter Eq.~\eqref{eq:string} should be $\sim O(1)$ as long as the two ends are not too close to the boundary. But if the system has the exact (strong) $\Z_2$ symmetry, $\prod_{i=2k}X_i\rho=\pm\rho$. So the string order parameter can be equivalently expressed as
\begin{equation}
    \langle X_2...X_{2n-2}Z_{2n-1}\cdot Z_{2m+1}X_{2m+2}...X_{2\lfloor\frac{L}{2} \rfloor}\rangle \xrightarrow{|m-n|\gg1} O(1).
\end{equation}
Furthermore, average symmetry requires that
\begin{equation}
    \langle X_2...X_{2n-2}Z_{2n-1}\rangle=\langle Z_{2m+1}X_{2m+2}...X_{2\lfloor\frac{L}{2} \rfloor}\rangle=0.
\end{equation}
Therefore a nontrivial edge correlation is enforced by the $\Z_2\times\Z_2^{\rm ave}$ and the bulk topology.\footnote{In Ref.~\cite{MaWangASPT} the same state was analyzed, and it was concluded that there was no nontrivial edge correlation, and the only feature from the open boundary was that the exact $\Z_2$ charge would fluctuate within the ensemble. What was not appreciated in Ref.~\cite{MaWangASPT} was that the fluctuating $\Z_2$ charge breaks the $\Z_2$ from exact to average symmetry, and since $\Z_2^{\rm ave}\times\Z_2^{\rm ave}$ does not have a nontrivial topological phase, the edge correlation disappears once the symmetry is lowered.}

A well-known application of the cluster chain is to serve as a resource state for measurement-based quantum computation and teleportation \cite{Briegel_2009, Nautrup2015Symmetry, Else_2012, Raussendorf_2019}. The standard protocol, however, suffers from instability from arbitrarily weak decoherence errors. Recently Ref.~\cite{zhang2024quantum} showed that the 2$d$ average cluster state (with average $1$-form $\Z_2^{(1)}$ symmetry) can be a resource state for the faithful teleportation of a 1$d$ repetition code, and the threshold is mapped to the 2$d$ random-bond Ising model on the Nishimori line. The result can be further generalized to the faithful quantum teleportation of a 2$d$ surface code through a 3$d$ average cluster state protected by a pair of 1-form $\Z_2^{(1)}$ symmetries \cite{Xu2024teleportation}. In particular, there is a finite threshold of faithful quantum teleportation which belongs to the universality class of a 3$d$ random-plaquette gauge model \cite{Takeda_2004}.

\subsubsection{Strong-to-weak spontaneous symmetry breaking}
\label{sec:swSSB}

In the previous section we derive a topological classification of mixed state phases, using two key assumptions: symmetric invertibility, and gapped Markovian. It is instructive to consider mixed states that do not obey these conditions. We will restrict to states which exhibit short-range correlations for local observables, i.e. they have a finite correlation length in the usual sense. In fact, we may even impose a stronger condition of ``SRE mixed states"  introduced in \cite{MaWangASPT}. Briefly, a mixed state is SRE if it can be prepared from a trivial product state by a finite-depth local quantum channel, namely
\begin{align}
\rho=\E[\ket{0}\bra{0}].
\end{align}
It follows from this definition that the $\rho$ has short-range correlations.

One can immediately see that a symmetrically invertible state is SRE.
Nevertheless,  an interesting feature of mixed states is that it is possible to have a state $\rho$ that is SRE (in the above sense) but not symmetrically invertible, and is not Markovian.  
In fact, a simple example of such a state is the strong-to-weak SSB (SW-SSB) state 
\begin{equation}
  \rho_X\propto \mathbbm{1}+\prod_iX_i,
\end{equation}
where the normalization is omitted, in a 1D qubit chain with a strong $\Z_2$ symmetry generated by $X=\prod_i X_i$. Clearly, $\rho_X$ is invariant under the strong symmetry as $X\rho_X=\rho_X$. To see that $\rho$ is SRE, we can start from the pure product state $|X_i=+1\rangle$, we apply the following depth-1 strongly-symmetric quantum channel:
\begin{equation}
    \E_i^{ZZ}(\rho)=\frac12(\rho+ Z_iZ_{i+1}\rho Z_iZ_{i+1}),
\end{equation}
and $\E^{ZZ}=\prod_i \E_i$. When applied to the product state $\ket{X_i=1}$, we obtain the $\mathbbm{1}+X$ state.
\footnote{We thank Yijian Zou for pointing out this construction.} As a matter of fact, if we choose $X$-basis, the density matrix $\rho_X\propto\mathbbm{1}+X$ is the maximally mixed state within the $\Z_2$ charge even sector, namely
\begin{align}
\rho_X\propto\mathbbm{1}+X=\sum_{\prod_ix_i=1}\ket{\{x_i\}}\bra{\{x_i\}},
\end{align}
where $\ket{\{x_i\}}$ is a product state with even $\Z_2$ charge. 

However, we will now show that $\rho_X$ is not symmetrically invertible. Suppose on the contrary that it is. Then for an interval $I=(x,y)$, the strong symmetry $X$ can be localized (Theorem \ref{thm:mixedsymfrac}): $X_I\rho=O_xO_y\rho$, where $O_x$ ($O_y$) are unitaries localized around $x$ ($y$). The localization implies that $O_x^\dag X_I O_y^\dag\rho=\rho$, 
\begin{equation}
    \Tr O_x^\dag X_I O_y^\dag\rho = 1.
\end{equation}
However, one can directly verify that $\langle O_x^\dag O_y^\dag X_I\rangle=0$ for the initial state $\rho_X$, for any choice of local $O_{x,y}$, which leads to a contradiction. Therefore, the density matrix $\rho_X$ is not symmetrically invertible. 

One can also show that $\rho_X$ is not Markovian. In fact, take $A, B, C$ to be three neighboring intervals that covers the entire system. Then it is straightforward to find that $I(A:C|B)=\ln 2$. Intuitively, $\rho_X$ reduced to any subsystem becomes the maximally mixed state, but on the entire system there is one global constraint of $X=1$ from the strong symmetry. 

In fact, as shown by recent work, states like $\rho_X$ exhibit a new type of collective phenomenon intrinsic to mixed states, where strong symmetry is spontaneously broken to weak symmetry, dubbed strong-to-weak spontaneous symmetry breaking (SWSSB). A more general and systematic account of SWSSB, including its formal definitions and fundamental properties such as stability, local indistinguishability, spontaneity, and (non-)invertibility, can be found in Ref. \cite{Lessa:2024gcw}. Below we will discuss a few characteristic features of $\rho_X$ to illustrate the physics of SWSSB.

One reason that $\rho_X$ represents spontaneous strong-to-weak symmetry breaking is because under any symmetry-breaking local channel, no matter how weak it is, the state is ``collapsed" to one with only weak symmetry, and the original strongly-symmetric state can not be recovered by any local channel. For example, consider the following symmetry-breaking ``measurement" channel:
\begin{equation}
    \mathcal{E}_Z(\rho)=(1-p)\rho+pZ_i \rho Z_i.
\end{equation}
Note that the channel is applied just on one site $i$. Then we find
\begin{equation}
    \mathcal{E}_Z(\rho_X)\propto (1-p)(\mathbbm{1}+X) + p (\mathbbm{1}-X) = \mathbbm{1}+(1-2p)X.
\end{equation}
When $p=1$ one finds the state $\mathbbm{1}-X$, which is also strongly symmetric but with total charge $X=-1$. However, for any $0<p<1$, $\mathcal{E}_Z(\rho_X)$ is only weakly symmetric under $X$. On the other hand, locally $\mathcal{E}_Z(\rho_X)$ and $\rho_X$ are indistinguishable for any $p$. In fact, on any subsystem, $\mathcal{E}_Z(\rho_X)$ reduces to the maximally mixed state. Thus we expect that there is no way to locally recover $\rho_X$ from $\mathcal{E}_Z(\rho_X)$. This is also reminiscent to SSB in pure state, where a cat state $|+\rangle\equiv|\!\uparrow\uparrow\uparrow...\rangle+|\!\downarrow\downarrow\downarrow...\rangle$ is locally indistinguishable from $|-\rangle\equiv|\uparrow\uparrow\uparrow...\rangle-|\downarrow\downarrow\downarrow...\rangle$, and any superposition of the two cat states will explicitly break the symmetry. 

It is also natural to ask whether the SWSSB order can be characterized by certain long-range correlations.
Recall that a weak symmetry is \textit{explicitly} broken if the one-point function ${\rm tr}(\rho O(x))\neq0$ for some charged operator $O(x)$: {\color{black}
If $\rho$ is weakly symmetric under $G$, then by definition, we have ${\rm tr}(\rho O(x)) = {\rm tr}(G^{-1} \rho G O(x)) = e^{i\theta} {\rm tr}(\rho O(x))$, where $e^{i\theta} \neq 1$ represents the $G$ charge of $O(x)$. This implies ${\rm tr}(\rho O(x)) = 0$. Consequently, the \emph{spontaneous} breaking of a weak $G$ symmetry is characterized by the two-point function ${\rm tr}(\rho O(x) O(y)^\dagger)$. 
Similarly, a strong symmetry implies that $O(x)\rho O^{\dagger}(x)$, where $O(x)$ carries a nontrivial charge, is orthogonal to $\rho$, as they contain states in different charge sectors. Consequently, the fidelity $F(\rho, O(x)\rho O^{\dagger}(x))$ is constrained by the strong symmetry to be zero. To characterize the \emph{spontaneous} breaking of strong symmetry, we examine two-point functions of charged operators. Specifically, we define a state to exhibit strong-to-weak SSB if the ordinary two-point correlator vanishes exponentially (indicating that the weak symmetry remains unbroken), while the ``fidelity" correlator is nonzero:
\begin{equation}
\label{eq:swSSB}
    F\left(\rho, O(x)O^\dag(y)\rho O^{\dagger}(x)O(y) \right)\sim O(1),
\end{equation}
for some charged operator $O(x)$, $O(y)$ with $|x-y|\to\infty$. As an immediate sanity check, we notice that such strong-to-weak SSB is only possible for mixed state since for pure state the fidelity correlator is simply the square of the ordinary correlator. We note that a similar notion of strong-to-weak SSB (defined using R\'enyi-2 correlator) has also been discussed recently in Ref.~\cite{LeeJianXu}.}

For the example $\rho_X$, consider the $\Z_2$ charged operator $O(x)=Z_x$. One can readily verify that Eq.~\eqref{eq:swSSB} indeed holds. In fact, if we write $\rho$ in $Z$-basis, it is a mixture (convex sum) of cat (GHZ) states:
\begin{equation}
    \rho \propto \sum_{s} (|s\rangle+X|s\rangle)(\langle s|+\langle s|X),
\end{equation}
where $s$ is a bit string in the $Z$-basis. 
{ In this basis, the fidelity $Z$-correlator becomes exactly the Edward-Anderson correlator used for spin-glass orders $\overline{|\langle Z_xZ_y\rangle_s|}$, where the overline means averaging over the bit strings. This analogy provides an intuitive picture for the spontaneous breaking of strong (exact) to weak (average) symmetry.}

As proved in \cite{Lessa:2024gcw}, in general a SWSSB state defined by Eq. \eqref{eq:swSSB} is not symmetrically invertible, and can not be gapped Markovian. Thus they are excluded when considering mixed-state SPTs,
just as usual SSB is avoided when considering pure-state SPT.

{\color{black}
\subsubsection{Higher dimensions}

We have shown in Sec. \ref{1d classification} that bosonic symmetrically invertible and gapped Markovian mixed states with $A\times G$ symmetry in 1D can be classified according to topological invariants valued in
\begin{equation} H^1(A,H^1(G,\U)) \oplus H^2(A,\U). 
\end{equation}  
Here the $H^1(A, H^1(G, \U))$ describes the $G$ charges (i.e. 0d $G$ SPTs) carried by $A$ defects.

We now discuss generalizations to higher dimensions. From now on in this section we only consider mixed states that are symmetrically invertible and gapped Markovian. We claim that SPT phases in $D<4$ space dimensions with symmetry $\tilde{G}=A\times G$ ($A$ being exact and $G$ being average) are classified \cite{MaWangASPT} by
\begin{equation}
\bigoplus_{p=1}^{D+1}\H^p\left(A, \H^{D+1-p}\left(G,\U\right)\right).
\label{eq:decoheredclassificationdp}
\end{equation}
This classification can be justified as follows. According to Theorem \ref{thm:mixedsymfrac}, in a symmetrically invertible mixed state, a domain wall of the strong symmetry traps a $G$-symmetric invertible phase in one lower dimension. In fact, by considering multi-domain wall junctions \cite{SpectralSequence}, one can see that the same conclusion holds for defects of $A$ in each co-dimension. Therefore, a symmetric invertible state is classified by the $G$-symmetric invertible phases on the defects of $A$, whose composition must satisfy the fusion rule of $A$ defects, or the Berry phase acquired when we compose the $A$ defect operators $V_{\partial \Gamma}$ in different orders \cite{Else_2014}. 

Furthermore, in Appendix \ref{App:ASPTclassification}, we demonstrate that a pure-state SPT protected solely by $G$ (or invertible phases that do not require a protecting symmetry) can be trivialized by a symmetric finite-depth channel. However, a pure-state SPT protected by $A\times G$ (or just $A$ alone) can not be connected to a trivial state via symmetric finite-depth channels. These arguments give further support to our result, that the classification of decohered ASPT with $A\times G$ symmetry is therefore given by Eq. \eqref{eq:decoheredclassificationdp}.

On the other hand, instead of thinking about an $A$ defect decorated by $G$-symmetric invertible states, the Künneth theorem allows us to equivalently consider a $G$ defect decorated by an $A$-symmetric invertible state, which is more intuitive and easier to generalize to more complex group structures. Specifically, we can reformulate the classification in Eq. \eqref{eq:decoheredclassificationdp} as
\begin{equation}
\begin{split}
    &\bigoplus_{p=1}^{D+1}\H^p(A, \H^{D+1-p}(G, \U)) \\
    = &\H^{D+1}(A\times G, \U)/\H^{D+1}( G,\U) \\
     = & \bigoplus_{p=0}^{D}\H^p(G, \H^{D+1-p}(A, \U)).
\end{split}
\label{eq:weakdefect}
\end{equation}
Compared with the group-cohomology classification of a pure state SPT with $A \times G$ symmetry, the term $\H^{D+1}(G, \U)$ is absent in the mixed state classification, while all others remain nontrivial. The physical interpretation of Eq. \eqref{eq:weakdefect} is as follows: a pure $A \times G$ SPT state is a superposition of $G$ domain wall configurations, and each codimension-$p$ defect can be decorated with an $A$-SPT in $D-p$ space dimensions. All possible decoration patterns are labeled by elements in $\H^p\left(G, \H^{D+1-p}\left[A, \U\right]\right)$ \cite{XieChen_2014}.
To be more concrete, a representative wavefunction of a clean SPT phase has the following form
\begin{align}
|\Psi\rangle=\sum\limits_{\mathcal{D}}\sqrt{p_{\mathcal{D}}}e^{i\theta_{\mathcal{D}}}|\Psi_{\mathcal{D}}\rangle|a_{\mathcal{D}}\rangle
\end{align}
where $|a_{\mathcal{D}}\rangle$ in the ``ancilla'' space describing the quantum state of defect network of $G$, $|\Psi_{\mathcal{D}}\rangle$ is the decorated $A$-symmetric invertible phase, and $e^{i\theta_{\mathcal{D}}}$ is a superposition phase factor that encodes the information in the $\H^{D+1}(G,\U)$ term. 
Once the $G$ degrees of freedom are decohered such that $G$ becomes a weak symmetry, the density matrix becomes a mixed ensemble describing the classical convex sum of $G$ defect configurations,
\begin{align}
\rho=\sum\limits_{\mathcal{D}}p_{\mathcal{D}}|\Psi_{\mathcal{D}}\rangle\langle\Psi_{\mathcal{D}}| \otimes |a_\mathcal{D} \rangle \langle a_\mathcal{D}|.
\label{eq:decohereddensitymatrx}
\end{align}
One can see from Eq. \eqref{eq:decohereddensitymatrx} that the relative phases of different $G$ defect configurations are no longer well-defined (equivalently, the phase factors from the bra and ket in $\rho$ cancel out). In contrast, the $A$ symmetric invertible state $| \Psi_\mathcal{D}\rangle$ remains intact as long as $A$ is a strong symmetry. From the perspective of symmetry defects, the effective boundary channel of a truncated weak symmetry, as described in Eq. \eqref{eq:weakfrac}, can be interpreted as pumping an $A$-symmetric invertible phase to the boundary.

In Sec \ref{sec:SSclassification}, when generalizing decohered ASPT phases to cases with non-direct product group structures, we adopt the physical picture described above: a classical convex sum of weak symmetry defects is decorated by invertible states protected by the strong symmetry. In addition to providing clarity in the physical picture, this construction has an additional advantage — it connects to the established construction of pure state SPT phases using the spectral sequence \cite{SpectralSequence,Gaiotto:2017zba}, where the decoration is protected by a normal subgroup of the entire symmetry. (One can verify that in a mixed state, the strong symmetry is always a normal subgroup of the full symmetry group.) Notably, this construction provides the full classification, as established by Theorem \ref{thm:mixedsymfrac} and the discussion in this section, for $A \times G$ symmetry, though it serves only as a constructive method for more general group structures. Whether this method offers a complete classification remains an open question for future study.
}

\subsection{Disordered ASPT}
\label{sec:disorderedReview}

We now review the notion of disordered ASPT, which is relevant for zero-temperature systems with disordered Hamiltonians.

We consider an ensemble of disordered Hamiltonians. For concreteness, the Hamiltonian takes the form
\begin{equation}
\label{eq:Ham}
H_I=H_0+\sum_i(v^{I}_i\mathcal{O}_i+h.c.),
\end{equation}
where $v^{I}_i$ is a quenched disorder potential drawn from a classical probability distribution $P[v^I]$ ($I$ labeling a particular realization and $i$ labeling a lattice site), $\mathcal{O}$ is a local operator, and $H_0$ is the non-random part of the Hamiltonian. We require the disorder to be at most short-range correlated, namely $\overline{v^*_iv_j}$ (averaged over the classical probability $P[v^I]$) should decay exponentially with $|i-j|$.

We now consider the ensemble of ground states $\{|\Omega_I\rangle\}$ of the Hamiltonians $\{H_I\}$. We call the ensemble short-range entangled (SRE) if each $|\Omega_I\rangle$ is short-range entangled with a finite correlation length $\xi_I$ that is upper-bounded in the entire ensemble.\footnote{\color{black} The correlation length $\xi$ is defined as follows: there exists another state $\ket{\tilde{\Psi}}$ in an ancillary Hilbert space and a local unitary of depth $\xi$ acting on the combined system, such that $\mathbf{U} \ket{\Psi}\otimes \ket{\tilde{\Psi}}=\ket{0}$ results in a product state. This definition allows us to include all invertible states (e.g., beyond cohomology states) in our discussion.} {\color{black}In particular, we exclude the possibility of large rare regions in topologically distinct phases, with long-range entangled boundaries of such rare regions. Practically this requirement can be satisfied by assuming the disorder distribution to be strictly bounded. It may be possible to impose only a soft bound on $\xi_I$ to allow for rare-region effects, which is an interesting direction for future investigation}. Superficially, the ensemble gives a density matrix $\rho=\sum_IP_I|\Psi_I\rangle\langle\Psi_I|$ and the situation appears similar to the decohered open system. However, a crucial difference for the disordered Hamiltonian system is that the states $\{|\Psi_I\rangle\}$ form a preferred basis for the ensemble. For example, two states $|\Psi_I\rangle$ and $|\Psi_{I'}\rangle$ may have equal probability of realization, which means $|\Psi_I\rangle+|\Psi_{I'}\rangle$ is an equally good eigenstate of the density matrix $\rho$. But in the disordered setting the latter state has no physical meaning, {\color{black}i.e., it is not the ground state of any Hamiltonian in the ensemble.} This makes the disordered systems physically quite different from the decohered systems. However, as we will see later in Sec.~\ref{sec:SSclassification}, there is a unified mathematical framework for the classification of SPT phases for the two different settings.

Following the decohered case, we can now define exact and average symmetries. A symmetry $A$ is exact if it commutes with $H_I$ for any disorder realization $I$. An important difference with the decohered systems is that in disordered systems, time-reversal symmetry can be exact. A symmetry $G$ is average if any element $g\in G$ takes a realization $H_I$ to a different realization $H_{I'}=gH_Ig^{-1}$ with $P[v^{I'}]=P[v^I]$. In other words, the disorder potential $v$ may transform nontrivially under $G$, but the probability $P[v]$ is symmetric under $G$ transforms. We call the ensemble of states $\{|\Psi_I\rangle\}$ symmetric if both the exact and average symmetries are not spontaneously broken. To align with the discussion for decohered systems, we also demand that the entire ensemble of states $\{|\Psi_I\rangle\}$ to transform identically under the exact symmetry $A$. This can be viewed as a ``canonical ensemble'' for a disordered system -- the condition is imposed for convenience and is not strictly required \footnote{This condition of fixed total charge under $A$ is necessary for decohered mixed states, otherwise there is no difference between exact and average symmetries since one can always simultaneously diagonalize the density matrix and the average symmetry operator. For disordered systems, however, the fixed-charge condition is optional, since there is the preferred basis given by the Hamiltonians and we are not allowed to freely re-diagonalize the density matrix.}.

We define two SRE ensembles (call them $\{H_I,|\Omega_I\rangle\}$ and $\{H'_I,|\Omega'\rangle\}$) to be in the same ASPT phase if $\{H_I\}$ can be continuously deformed\footnote{\color{black}Here by ``continuous deformation'' we consider varying parameters in the Hamiltonian and disorder distribution, while maintaining the system at ground state. In gapped clean system this is equivalent to an adiabatic time-evolution from $H$ to $H'$. However, in disordered systems the two notions may not be equivalent due to rare local resonances during adiabatic evolution\cite{KhemaniNandkishoreSondhi}.} to $\{H'_I\}$ while keeping all the conditions listed above throughout the deformation: (a) the disorder potentials remain short-range correlated; (b) the symmetries (both exact and average) are not broken explicitly or spontaneously; and (c) the ground states remain short-range entangled. We note that the conditions imposed here are slightly simpler than those originally discussed in Ref.~\cite{MaWangASPT}, and in Appendix~\ref{App:DefASPT} we show that they are largely equivalent.

The disordered bosonic ASPT, as defined above,  with symmetry $\tilde{G}=A\times G$ ($A$ being exact and $G$ being average) are classified (see Ref.~\cite{MaWangASPT} and Appendix \ref{App: spectral sequence}) by
\begin{equation}
    \bigoplus_{p=0}^{D-1}\H^p(G, h_I^{D+1-p}(A)),
    \label{eq:disorderedclassificationdp}
\end{equation}
where $h_I^q(A)$ is the classification of invertible phases in $q$ spacetime dimension with symmetry $A$ (for bosonic systems at $q<3$ it is simply the group-cohomology $\H^q(A,\U)$).

{\color{black} We now justify the classification in Eq. \eqref{eq:disorderedclassificationdp}. From the definition of SRE ensembles, two observations can be made: (1) all ground states in the ensemble belong to the same $D$-dimensional $A$-symmetric invertible phase; (2) a decorated domain wall picture also applies to SRE ensembles. To see (1), consider two disorder realizations, $H_\1$ and $H_{\2}$, with ground states $\ket{\Psi_\1}$ and $\ket{\Psi_{\2}}$, respectively. The spatial independence of the disorder potential guarantees the existence of a third disorder realization, $H_{\1'}$, within the ensemble, where $H_{\1'} = H_{\1}$ in a large region $R$ with a diameter much larger than the correlation length $\xi$ of the ensemble, and $H_{\1'} = H_{\2}$ in the complementary region $\bar{R}$. The ground state of $H_{\1'}$ should coincide with $\ket{\Psi_\1}$ deep within $R$ and with $\ket{\Psi_{\2}}$ deep within $\bar{R}$. Therefore, the SRE nature of the ensemble (i.e., the upper bound on $\xi$) implies that $\ket{\Psi_{\1}}$ and $\ket{\Psi_{\2}}$ belong to the same invertible phase.}

{\color{black} Observation (2) can be understood through a similar argument. Consider two disorder potentials, $v(x)$ and $v'(x)$, where $x$ denotes the spatial coordinate. Suppose $v'(x) = g v(x) g^{-1}$ within a large region $R$ with $g \in G$, while $v'(x) = v(x)$ outside $R$. Since the probability distribution is symmetric under $G$, the two disorder realizations occur with the same probability. Their respective ground states, $\ket{\Psi_v}$ and $\ket{\Psi_{v'}}$, should coincide deep outside $R$ while differing only by an average symmetry deep within $R$. As a result, we have
\begin{equation}
    g_R^{-1}\ket{\Psi_{v'}} = V_{\partial R} \ket{\Psi_{v}},
    \label{eq:DWdecorationdisorder}
\end{equation}
where the non-trivial effect of $V_{\partial R}$ on $\ket{\Psi_{v}}$ is localized solely to the boundary $\partial R$. Here, $g_R$ represents an average symmetry action within $R$. From observation (1), we know that $\ket{\Psi_{v}}$ and $\ket{\Psi_{v'}}$ are in the same invertible phase, meaning they can be connected by an $A$-symmetric local unitary circuit $U$, so that $\ket{\Psi_{v'}} = U \ket{\Psi_{v}}$. This implies that on the state $\ket{\Psi_{v}}$, the effect of the boundary unitary $V_{\partial R}$ can be realized by a $D$-dimensional $A$-symmetric local unitary circuit $g_R^{-1}U$, which thus only results in the pumping of a (potentially trivial) $(D-1)$-dimensional $A$-symmetric invertible phase on $\partial R$. Analogous to the case of pure states [cf. Eq. \eqref{eq:invertiblestring}], Eq. \eqref{eq:DWdecorationdisorder} establishes the decorated domain wall picture for SRE ensembles -- although defining a domain wall for the average symmetry requires comparing two distinct realizations within the ensemble.

Carrying out the same analysis for (multi-)junctions of domain walls \cite{SpectralSequence}, we obtain the classification of disordered ASPT phases given in Eq. \eqref{eq:disorderedclassificationdp}, whose physical interpretation becomes evident: on each $G$-domain wall of codimension $p$, we can decorate with an $A$-symmetric invertible state in $D-p$ dimensions.} Compared to the clean case (with exact $\tilde{G}$), the $p=D$ and $p=D+1$ terms are missing. Similar to the decohered case, the $p=D+1$ term is absent because the $G$-domain walls proliferate classically (probabilistically) without a superposition phase factor. 

The absence of the $p=D$ term in Eq.~\eqref{eq:disorderedclassificationdp}, which describes decorating a zero-dimensional defect with an $A$ charge, is more interesting. In the definition of SRE ensembles, we only demanded each ground state wavefunction $|\Omega_I\rangle$ to be short-range entangled, and made no requirement on the energy spectrum -- we do not demand $H_I$ to be gapped. This is appropriate for disordered systems -- for example, even a fully localized Anderson insulator, with un-entangled product-state wavefunction, can be gapless. Once we forgo the requirement on the energy gap, zero-dimensional states with different symmetry charges can now be deformed to each other by continuously tuning the Hamiltonian (a zero-dimensional state is always SRE by definition). This means that decorating zero-dimensional defects will not produce nontrivial phases. Nevertheless, certain patterns of zero-dimensional decoration will come with nontrivial consequences, with an intriguing connection to the physics of localization -- we discuss this aspect in detail in Sec.~\ref{sec:localization}.

{\color{black} Finally, we remark on the connection between the theoretical model, i.e., disordered ensembles, and a single instance relevant to experiments. Physically, the SRE ensemble describes a family of disordered systems whose ground states, although typically gapless, still exhibit only short-range (invertible) entanglement. Moreover, two randomly prepared samples are generally adiabatically connected to each other\footnote{\color{black} Despite being gapless, Hastings proved that the ground states of two disordered Hamiltonians can still be adiabatically connected if there exists a connecting path with localization, i.e., a finite mobility gap \cite{Hastingsdisorderedadiabatic}.}. An example of an SRE ensemble may be the integer quantum Hall state in the presence of disorder. Similar to an SPT state in gapped clean systems, the bulk of a disordered ASPT state (in a single instance) is often trivial: it is insulating, with exponentially decaying correlation functions. One might ask whether any non-trivial features can be observed at the boundary of a disordered ASPT state, analogous to the usual diagnostics for a clean SPT.

Indeed, it has been demonstrated in Ref. \cite{MaWangASPT} that the concept of 't Hooft anomaly can also be extended to the boundary of disordered ASPT states. Specifically, if the 't Hooft anomaly (labeled by the bulk ASPT phase) is nontrivial, then for a single instance (which can be considered as a particular realization within the ensemble), a symmetric boundary must be long-range entangled with probability 1 in the thermodynamic limit. More precisely, for any finite correlation length $\xi_0 > 0$, the probability of a boundary state having a correlation length $\xi \leq \xi_0$ vanishes as the system size $L \to \infty$. We emphasize that this statement does not rely on ensemble averaging.

One classic example is the random singlet phase (RSP) in a disordered spin-1/2 chain, which can be viewed as the boundary of an ASPT protected by $SO(3)$ spin rotation and average lattice translation symmetry. It is known that the RSP exhibits singlet pairs at arbitrarily large distances, resulting in a vanishing spin gap and a power-law decay in the average correlation function \cite{fisher1994random}. Another interesting observation is that, since the nontriviality of the RSP is imposed by a topological constraint (i.e., an ``average" 't Hooft anomaly), there is no reason to expect self-averaging in various observables. For instance, the spin-spin correlation is dominated by rare, long-distance singlet pairs rather than by typical spin pairs. As a result, the spin correlation function may fluctuate significantly between different disorder realizations. This can be contrasted with a model lacking such a nontrivial topological constraint, such as a disordered chain with an even number of spin-1/2 particles per unit cell, where a featureless SRE state with exponentially decaying correlation functions is allowed.
}

{ From the bulk point of view, the nontriviality of an ASPT is manifested on the decorated states on defects (such as domain walls) associated with the average symmetry. Such defects can be physically defined even for a single disorder realization -- for example, for an average $\Z_2$ symmetry the domain walls can be defined as $D-1$ surfaces across which the random $\Z_2$-breaking field changes sign. What matters is that the defects (decorated with the nontrivial lower-dimensional states) appear on arbitrarily large length scale -- in other words, the defects percolate throughout the system. The percolation of the defects is what makes the notions of average symmetry and ASPT well defined even for a single sample, albeit in a more heuristic manner. }

\subsubsection{Interplay with localization physics}
\label{sec:localization}



We now return to zero-dimensional decorations in the context of disordered ASPT. As discussed in Sec.~\ref{sec:disorderedReview}, decorating with (0+1)$d$ SPT (namely charges of the exact symmetry) does not lead to nontrivial phases. This is because a (0+1)$d$ SPT is nontrivial only if we demand an energy gap, but in disordered systems, we do not require each individual disorder realization to be gapped -- we only demand the ground state to be short-range entangled (SRE). In disordered systems, Anderson localization provides a natural mechanism to have a gapless but SRE ground state.

Let us illustrate with a familiar example. Consider a lattice-free fermion system with $\U$ charge conservation and lattice translation $\Z^d$ symmetries, with the simplest tight-binding Hamiltonian
\begin{equation}
H=-t\sum_{\langle i,j\rangle} (c_j^\dag c_{i}+\text{h.c.})+\mu\sum_i c_i^{\dagger}c_i.
\end{equation}

For $\mu>2|t|$ or $\mu<-2|t|$, the system is an atomic insulator with $\U$ charge per site $q=0$ or $q=1$, respectively. The system is metallic for intermediate $\mu$, with a long-range entangled ground state. In fact, any symmetric state interpolating between the two different atomic insulators must be long-range entangled, as guaranteed by the Lieb-Schultz-Mattis (LSM) theorem \cite{lieb1961two,Oshikawa2000,Hastings2004}. 

Now we add disorders that break the exact translation symmetry to an average symmetry:
\begin{equation}
H_\text{disorder}=-\sum_j \epsilon_j c_j^\dag c_j.
\end{equation}
As is well known, for sufficiently strong disorder the metallic states (with an intermediate chemical potential $\mu$) become Anderson localized insulators. The ground states of the Anderson insulators are SRE -- they are essentially product states of fermions sitting at random locations. The probability $P_i$ for a site $i$ to be occupied can smoothly change from $0$ to $1$. This gives a smooth interpolation between the two atomic insulators with charge filling $q=0$ and $q=1$.

Another feature of the above localized intermediate state is that it is generally gapless, with localized excitations at arbitrarily small excitation energy in the thermodynamic limit. In order to be more precise, let us fix the total $\U$ charge of the system for the entire disordered ensemble -- for example, at density $\nu$ we can fix $Q=\lfloor\nu L\rfloor$ where $L$ is the system size and $\lfloor...\rfloor$ is the integer part. Then the system must be gapless as long as the Hamiltonians are bounded, i.e. the distribution of $\epsilon_j$ is bounded (possibly with some rare tails). This is because for a large enough system, we can always find an excitation, e.g. moving a particle from an occupied site to an unoccupied site, that costs arbitrarily small energy. This forced gaplessness can be viewed as the remnant of the LSM constraint for fractional charge filling. 

{\color{black}Note that the above filling-enforced gaplessness in localized system may seem trivial if we consider Gaussian-like disorder distributions, in which case even Anderson insulators at integer filling will in general be gapless. However, at integer filling (say one electron per site), the state can become gapped if we choose an appropriate form of disorder distribution -- for example we can choose the disorder strength to be strictly bounded $|\epsilon_j|\leq \delta$ for $\delta$ smaller than the energy gap at the clean limit. Such a choice would be impossible at fractional filling if we fix the total charge (say to be $Q=\lfloor\nu L\rfloor$).

We expect that the above idealized picture, where each ground state is a trivial product state, can be extended to the situation where each ground state is short-range entangled (SRE). Essentially, each SRE ground state can be written as $|\Psi_I\rangle=U^{FD}_{I}|\Psi_0\rangle$, where $|\Psi_0\rangle$ is a product state, with on-site charge eigenvalues $q_j:=\langle c_j^{\dagger}c_j\rangle_{\Psi_0}\in\Z$ chosen to minimize the depth of the circuit $U^{FD}_{I}$. We can then interpret $\{q_j\}$, which is a set of random integers summing up to $Q=\lfloor\nu L\rfloor$, as effectively the on-site charge occupation numbers of $|\Psi_I\rangle$.  We can then generalize earlier arguments on the filling-enforced gaplessness and conclude that short-range entangled ground states should be gapless at fractional average charge filling. Since long-range entangled (more precisely non-invertible) ground states automatically have vanishing energy gap (between the absolute ground state and the lowest-lying excited state), we conclude that
\begin{framed}
    \textbf{(Disordered LSM constraint)}: for a disordered system with bounded disorder strength,  exact $\U$ and average lattice translation symmetry, and fractional charge filling $\nu$, the energy gap above each ground state must be vanishing in the thermodynamic limit.
\end{framed}
We emphasize that what we presented here is only a physically plausible argument. A rigorous proof is more involved and will be left for future works.
}

In the above example, the lattice sites should be viewed as defects of the translation symmetries \cite{correspondence}. To generalize the above observations to general ASPT with (0+1)$d$ decorations, all we have to do is to replace the lattice sites with the average $G$-defects and to replace $\U$ charge with general Abelian representations of the exact symmetries. The only subtlety is that we need Anderson localization for generic interacting systems -- in other words, we need many-body localization (MBL) \cite{MBLReview}. Crucially, we only need MBL for low-energy states, with vanishing energy density. Although not rigorously proven for the most general setting, it seems reasonable to assume that such low-energy MBL can be achieved in any dimension without fine-tuning \cite{ElgartKlein2022}. With the localization assumption in mind, we conclude that
\begin{framed}
Two disordered ASPT states, with different exact symmetry charge decorations on (0+1)$d$ average symmetry defects, can be smoothly deformed to each other. If the disorder strength is bounded, the intermediate states must have localized excitations with excitation energy vanishing in the thermodynamic limit.
\end{framed}

We emphasize that the story of localized states is only relevant for disordered systems, and does not affect the discussion on decohered topological phases. The reason is that even though each individual state in the ensemble of an Anderson insulator is a trivial product state, the disordered ensemble viewed as a density matrix $\rho=\sum_IP_I|\Psi_I\rangle\langle\Psi_I|$ spontaneously breaks the strong symmetry down to a weak symmetry, making it not \textit{symmetrically invertible}, as defined in Sec.~\ref{sec:symmetricallyinvertible}. Indeed, the example of $\rho_X\propto \mathbbm{I}+X$ discussed in Sec.~\ref{sec:symmetricallyinvertible} is exactly a $\Z_2$ version of the Anderson insulator ensemble. At a concrete level, for disordered systems, an SSB from strong to weak symmetry is measured by the Edward-Anderson correlator $\overline{\langle Z_i Z_j\rangle^2}$. {\color{black}Importantly, for disordered systems, the calculation should be performed in the basis of Hamiltonian ground states ${|\Psi_I\rangle}$. Any other basis that involves superpositions of different $|\Psi_I\rangle$'s does not correspond to the physical ground state of any random Hamiltonian in the disorder ensemble.} For decohered systems, however, any basis decomposition (purification) of the density matrix is equally legitimate. As explained in Sec.~\ref{sec:swSSB}, we can choose a convenient basis to interpret the density matrix $\rho_X\propto \mathbbm{I}+X$ as an ensemble of spin-glass orders, indicating an nonzero Edward-Anderson correlator.

\subsection{Mathematical framework: Spectral sequence}
\label{sec:SSclassification}

In general, for given average symmetry $G$ and exact symmetry $A$, the total symmetry $\tilde{G}$ does not have to be the direct product $A\times G$. Instead, we only require $A$ to be a normal subgroup of the full symmetry group. In the fermionic case, $A$ contains the fermion parity $\Z_2^f$ as a subgroup. $G$ and $A$ fit into the following short exact sequence:
\begin{equation}
1\rightarrow A\rightarrow \tilde{G}\rightarrow G\rightarrow 1.
\label{group extension}
\end{equation}

For a $D+1$ dimensional SPT, the classification via generalized cohomology theory can be understood by decorations on $G$ domain walls/defects. Mathematically the consistency conditions for domain wall decorations are organized into an (Atiyah-Hirzebruch) spectral sequence \cite{Gaiotto:2017zba, SpectralSequence} (see Appendix~\ref{App: spectral sequence} for a brief review), whose $E_2$ page is given by 
\begin{equation}
\bigoplus_{p+q=D+1}E_2^{p,q}=\bigoplus_{p+q=D+1}\H^p(G, h^q(A)).
\label{LHS}
\end{equation}
Here $h^q(A)$ is the classification of invertible phases in $q$ spacetime dimension with symmetry $A$ that can be decorated on $G$-defects. Physically, the term $\H^p(G, h^q(A))$ means decorating invertible states on codimension-$p$ $G$ defects. The decoration pattern must be such that on any $G$ defect there is no 't Hooft anomaly, so it is possible to decorate an invertible state. For example, the condition that any codimension-$(p+1)$ defect is anomaly-free leads to the cocycle condition for the decoration on codimension-$p$ defects, thus the group cohomology. 

The exact form of $h^q(A)$ will depend on the physical context. In particular:
\begin{itemize}
\item For the standard (pure state, clean) SPT, $h^q(A)$ is the classification of all invertible phases in $q$ spacetime dimension with symmetry $A$. Note $h^q(A)$ itself can be computed from the classification of invertible phases (without any symmetry) using the same kind of spectral sequence.
\item For decohered ASPT, $h^q(A)$ is almost the classification of invertible phases in $q$ spacetime dimension with symmetry $A$, except (1) $h^0(A)=0$ for reasons explained in Sec.~\ref{sec:decoheredReview}, and (2) $h^q(A)$ does not contain invertible phases that do not require the $A$ symmetry at all (for example the chiral $E_8$ state in (2+1)$d$) -- this is because such invertible states can be easily trivialized or prepared by a finite-depth quantum channel \cite{MaWangASPT}. Hence the classification of decohered ASPT phases in bosonic systems will be reduced to the Lyndon-Hochschild-Serre (LHS) spectral sequence that replaces $h^q(A)$ by $\H^q[A, \U]$ for $q\geq1$.
\item For disordered ASPT, $h^q(A)$ is almost the classification of invertible phases in $q$ spacetime dimension with symmetry $A$, except $h^0(A)=0$ and $h^1(A)=0$ for reasons explained in Sec.~\ref{sec:disorderedReview}.
\end{itemize}

This is not the end of the story. Importantly, not every decoration pattern given by the $E_2$ page can actually be realized as an ASPT state. A legitimate decoration pattern must satisfy certain consistency conditions. Mathematically, the obstructions to the consistency is given by the ``differentials":
\begin{equation}
d_r: E_2^{p,q}\rightarrow E_2^{p+r, q-r+1}.
\label{obstruction}
\end{equation}
$d_r$ maps to a decorated domain wall configuration in $(D+2)$ dimension. For bosonic systems with $\tilde{G}=A\times G$ these differentials automatically vanish, so we obtain Eq.~\eqref{eq:decoheredclassificationdp} and \eqref{eq:disorderedclassificationdp} as the classifications. For fermion systems or boson systems with nontrivial group extension, the differentials may not vanish and represent obstructions for certain decoration patterns. Note that explicit expressions for the differentials are available for both bosonic and fermionic systems up to three spatial dimensions~\cite{SpectralSequence, general2}.

The physical meanings of these differentials correspond to the three consistency conditions of constructing an SPT, as
\begin{enumerate}[1.]
\item $r\leq D-p$: The decorated $G$-defects can be gapped without breaking $A$-symmetry.
\item $r=D-p+1$: $A$-charge is preserved during a continuous deformation of the $G$-defect network. Since $h^1(A)=0$ for disordered ASPT phases, this obstruction automatically vanishes for disordered ASPT. 

Physically, for clean SPTs, a nontrivial $d_{D-p+1}$ obstruction implies that $A$ charge will change (i.e. not be conserved) when we change the $G$-defect configurations through some local operation. In the disordered setting, however, we change the $G$-defect configurations by drawing a different disorder realization from the ensemble. Then the change of $A$-charge no longer requires actual charge non-conservation -- all we need is a localized mode, or more precisely a local conserved charge operator $Q^{\rm{local}}_A$ that changes its ground state eigenvalue as the disorder realization (which determines the $G$-defect configurations) changes locally. The sample-to-sample fluctuation of the local charge makes the situation similar to the localized states that interpolate between different (0+1)$d$ decorations discussed earlier in Sec.~\ref{sec:localization}. Following the logic in Sec.~\ref{sec:localization}, we also conclude that the localization-enabled ASPT with (0+1)$d$ decorations has to be gapless in the thermodynamic limit, as long as the disorder strength is bounded. Given the finite density of localized states at zero energy, such localization-enabled states will also be dubbed ``compressible''.
\item $r=D-p+2$: There is no Berry phase accumulated after a closed path of continuous $F$-move deformations due to the single-valued property of the SPT wavefunction. Since $h^0(A)=0$ for both decohered and disordered ASPT, this obstruction automatically vanishes in these two contexts. Physically, the $G$-defects only proliferate probabilistically to form ASPT phases, so there is no need to assign consistent Berry phases.
\end{enumerate}

At the level of $E_2$ page, it might appear that there are fewer nontrivial ASPT phases than standard clean SPT phases (absence of $E_2^{p,1}$ for disordered ASPT and absence of $E_2^{p,0}$ for both disordered and decohered ASPT). However, this also means that there are fewer potential obstructions for ASPT phases since each obstruction corresponds to some nontrivial topological phase in one dimension higher. This opens the possibility of ASPT phases that are \textit{intrinsically} disordered or decohered, in the sense that they cannot be viewed as a clean SPT perturbed by disorder or decoherence. Such ``intrinsic ASPT'' will be one of the main focuses of this work. We dub the disordered ASPT phases enabled by vanishing $d_{D-p+1}$ obstructions \textit{localization-enabled} ASPT or \textit{compressible} ASPT, and ASPT phases (both disordered and decohered) enabled by vanishing $d_{D-p+2}$ obstructions \textit{Berry-free} ASPT.

The differential (\ref{obstruction}) is also called the \textit{trivialization} map for SPT phases labeled by elements in $E_2^{p+r,q-r+1}$ with decorated domain wall configurations in one higher dimension. The images of the $d_r$ map give trivial SPT phases. The physical meaning of the trivialization map is that the images of $d_r$ are the states with anomalous SPT states \cite{anomalousSPT} on the boundary which is SRE, and the corresponding bulk states should be topologically trivial. We emphasize that the decoherence/disorder does not affect the trivialization of the ASPT phases: the anomalous SPT states on the boundary in the clean systems might be trivial product states in the presence of decoherence or disorder, which are also SRE and manifest the topologically trivial bulk states.

\section{Intrinsic ASPT: examples}
\label{sec:BerryFree}

In this Section, we discuss the ``Berry-free" and ``compressible'' intrinsic ASPT phases in more detail. The Berry-free intrinsic ASPT phases are enabled by the vanishing of the Berry phase obstruction $d_{D-p+2}$ and the compressible intrinsic ASPT phases are enabled by the vanishing of the charge-decoration obstruction $d_{D-p+1}$, as we discussed in Sec.~\ref{sec:SSclassification}. The Berry-free ASPT can appear in both decohered and disordered settings -- the only difference is that if a phase comes from a (0+1)$d$ decoration it will be trivial in the disordered setting. For Berry-free ASPT states we will mostly not distinguish the two settings in this Section. The compressible ASPT can only appear in disordered settings.

\subsection{Fixed-point model for bosonic ASPT}
\label{sec:fixedpointmodel}
We will describe a class of ``fixed-point" lattice models for Berry-free ASPT phases. The model is a generalization of the group-cohomology model for bosonic SPT phases~\cite{Chen:2011pg}. 

Again we denote by $G$ the average symmetry, $A$ the exact symmetry, and the group extension  $\tilde{G}$ as defined in Eq. (\ref{group extension}). We denote elements of $\tilde{G}$ by $x=(g,a)$, where $g\in G$ and $a\in A$. In ($D$+1)-dimension, the input to the model is a (homogeneous) ($D$+1)-cochain $\nu(x_0,\dots, x_{D+1})$, that satisfies the obstructed cocycle condition:
\begin{equation}
(d\nu)(x_0,\cdots, x_{D+2})=O_{D+2}(g_0, \cdots, g_{D+2}).
\label{eq:obstruction}
\end{equation}
Here $d$ is the coboundary operator, $g_i$ is the $G$-grading of $x_i$. $O_{D+2}$ is a $(D+2)$-cocycle in $\H^{D+2}[G, \U]$. To construct a clean SPT state, we will need $O_{D+2}=1$. For the ASPT construction, this is not necessary. 

We will illustrate the construction in (2+1)$d$ in the following, but the same construction works in any dimension. We will work with a triangular lattice. The system consists of a $\tilde{G}$ spin on each site,
with an orthonormal basis labeled by group elements, i.e. $\{\ket{x}\}_{x\in \tilde{G}}$.  A natural $\tilde{G}$ symmetry action is given by the left multiplication:
\begin{equation}
U_y\ket{x}=\ket{yx}, y\in \tilde{G}.
\end{equation}
In addition, the lattice has to be equipped with a branching structure, which is essentially an ordering of all sites. For each triangle face $\Delta$ of the lattice, denote by $i,j,k$ the three vertices whose ordering satisfies $i<j<k$. We also denote by $s(\Delta)=\pm 1$ the orientation, i.e. whether $i,j,k$ is clockwise or counter-clockwise.
 
Firstly, we review the standard group cohomology construction when $O_{4}=1$. Define the $\tilde{G}$-invariant state on the site $j$:
\begin{equation}
\ket{0^{\tilde{G}}_j}=\frac{1}{\sqrt{|\tilde{G}|}}\sum_{x\in \tilde{G}} \ket{x_j},
\label{eq:singlesite}
\end{equation}
and then the trivial paramagnetic state for the whole system:
\begin{equation}
\ket{\Psi_0^{\tilde{G}}}=\bigotimes_j\ket{0^{\tilde{G}}_j}.
\end{equation}
$\ket{\Psi_0^{\tilde{G}}}$ is the ground state of the following local Hamiltonian:
\begin{equation}
H_\text{trivial}=-\sum_j\ket{0^{\tilde{G}}_j}\bra{0^{\tilde{G}}_j}.
\end{equation}

Let us now define the following finite-depth local unitary circuit 
\begin{equation}
V=\sum_{\{x\}}\prod_{\Delta_{ijk}}\nu^{s(\Delta_{ijk})}(1,x_i,x_j,x_k)\ket{\{x\}}\bra{\{x\}}.
\end{equation}
$V$ can be viewed as a composition of unitary gates each acting on a triangle. Since all the gates are diagonal in the $\ket{\{x\}}$ basis, they commute with each other and as a circuit $V$ has depth 1.  When $O_4=1$, one can show that the local gates do not preserve $\tilde{G}$ individually, but the unitary $V$ as a whole does.

The SPT state is given by 
\begin{equation}
\ket{\Psi_{\rm SPT}}=V\ket{\Psi_0^{\tilde{G}}}.
\end{equation}
The commuting-projector parent Hamiltonian for this state is
\begin{equation}
\begin{gathered}
H=VH_\text{trivial}V^\dag=-\sum_j B_j,\\
B_j=V\ket{0^{\tilde{G}}_j}\bra{0^{\tilde{G}}_j}V^\dag.
\end{gathered}
\end{equation}
Here $B_j$ is an operator that acts on the hexagon centered at $j$.

Now we consider what goes wrong if the cocycle $\nu$ is obstructed by a nontrivial $[O_4]$. We find that the state $\ket{\Psi_{\rm SPT}}$ is no longer symmetric under $\tilde{G}$.  More precisely, it is no longer invariant under the symmetry transformations $U_x$ when $x\in \tilde{G}$ has a nontrivial $G$-grading: $U_x\ket{\Psi_{\rm SPT}}\neq e^{i\varphi}\ket{\Psi_{\rm SPT}}$. However, for $a\in A$, the state is still invariant: $U_a\ket{\Psi_{\rm SPT}}=\ket{\Psi_{\rm SPT}}$. 

Based on this observation, we now show how to construct the intrinsic ASPT phase. It is now more convenient to think of the $\tilde{G}$ spin as a $G$ spin and an $A$ spin, and treat the $G$ spins as quenched disorder configurations. The configuration of $G$ spins will be collectively denoted as $\{g\}$.  
In addition, for $h\in G$ under $U_h$ the classical $G$ spin configuration $\{g\}$ transformsd d to $\{hg\}$.

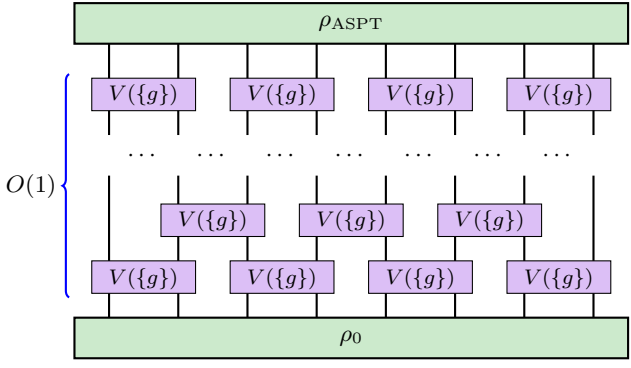
\begin{figure}
\begin{tikzpicture}[xscale=0.46,yscale=0.54]
\tikzstyle{sergio}=[rectangle,draw=none]
\draw[thick] (0,-0.75) -- (0,2.75);
\draw[thick] (2,-0.75) -- (2,2.75);
\draw[thick] (4,-0.75) -- (4,2.75);
\draw[thick] (6,-0.75) -- (6,2.75);
\draw[thick] (8,-0.75) -- (8,2.75);
\draw[thick] (10,-0.75) -- (10,2.75);
\draw[thick] (12,-0.75) -- (12,2.75);
\draw[thick] (14,-0.75) -- (14,2.75);
\draw[thick] (0,6) -- (0,3.75);
\draw[thick] (2,6) -- (2,3.75);
\draw[thick] (4,6) -- (4,3.75);
\draw[thick] (6,6) -- (6,3.75);
\draw[thick] (8,6) -- (8,3.75);
\draw[thick] (10,6) -- (10,3.75);
\draw[thick] (12,6) -- (12,3.75);
\draw[thick] (14,6) -- (14,3.75);
\filldraw[thick, fill=DarkGreen!20] (-1,-0.75) -- (15,-0.75) -- (15,-1.75) -- (-1,-1.75) -- cycle;
\path (7,-1.25) node [style=sergio]{$\rho_0$};
\filldraw[fill=BlueViolet!30, draw=black] (-0.5,-0.15)--(2.5,-0.15)--(2.5,0.65)--(-0.5,0.65)--cycle;
\path (1,0.2) node [style=sergio]{\footnotesize $V(\{g\})$};
\filldraw[fill=BlueViolet!30, draw=black] (-0.5+4,-0.15)--(2.5+4,-0.15)--(2.5+4,0.65)--(-0.5+4,0.65)--cycle;
\path (1+4,0.2) node [style=sergio]{\footnotesize $V(\{g\})$};
\filldraw[fill=BlueViolet!30, draw=black] (-0.5+8,-0.15)--(2.5+8,-0.15)--(2.5+8,0.65)--(-0.5+8,0.65)--cycle;
\path (1+8,0.2) node [style=sergio]{\footnotesize $V(\{g\})$};
\filldraw[fill=BlueViolet!30, draw=black] (-0.5+4+8,-0.15)--(2.5+4+8,-0.15)--(2.5+4+8,0.65)--(-0.5+4+8,0.65)--cycle;
\path (1+4+8,0.2) node [style=sergio]{\footnotesize $V(\{g\})$};
\filldraw[fill=BlueViolet!30, draw=black] (1.5,1.25)--(4.5,1.25)--(4.5,2.05)--(1.5,2.05)--cycle;
\path (3,1.65) node [style=sergio]{\footnotesize $V(\{g\})$};
\filldraw[fill=BlueViolet!30, draw=black] (1.5+4,1.25)--(4.5+4,1.25)--(4.5+4,2.05)--(1.5+4,2.05)--cycle;
\path (3+4,1.65) node [style=sergio]{\footnotesize $V(\{g\})$};
\filldraw[fill=BlueViolet!30, draw=black] (1.5+4+4,1.25)--(4.5+4+4,1.25)--(4.5+4+4,2.05)--(1.5+4+4,2.05)--cycle;
\path (3+4+4,1.65) node [style=sergio]{\footnotesize $V(\{g\})$};
\path (3,3.25) node [style=sergio]{$\cdots$};
\path (5,3.25) node [style=sergio]{$\cdots$};
\path (7,3.25) node [style=sergio]{$\cdots$};
\path (9,3.25) node [style=sergio]{$\cdots$};
\path (11,3.25) node [style=sergio]{$\cdots$};
\path (13,3.25) node [style=sergio]{$\cdots$};
\path (1,3.25) node [style=sergio]{$\cdots$};
\filldraw[thick, fill=DarkGreen!20] (-1,6) -- (15,6) -- (15,7) -- (-1,7) -- cycle;
\path (7,6.5) node [style=sergio]{$\rho_{\rm ASPT}$};
\draw[decorate, decoration={brace,raise=2pt}, blue, thick] (-1,-0.25) -- (-1,5.25);
\path (-2.25,2.5) node [style=sergio]{$O(1)$};
\filldraw[fill=BlueViolet!30, draw=black] (-0.5,5.15)--(2.5,5.15)--(2.5,4.35)--(-0.5,4.35)--cycle;
\path (1,4.75) node [style=sergio]{\footnotesize $V(\{g\})$};
\filldraw[fill=BlueViolet!30, draw=black] (-0.5+4,5.15)--(2.5+4,5.15)--(2.5+4,4.35)--(-0.5+4,4.35)--cycle;
\path (1+4,4.75) node [style=sergio]{\footnotesize $V(\{g\})$};
\filldraw[fill=BlueViolet!30, draw=black] (-0.5+8,5.15)--(2.5+8,5.15)--(2.5+8,4.35)--(-0.5+8,4.35)--cycle;
\path (1+8,4.75) node [style=sergio]{\footnotesize $V(\{g\})$};
\filldraw[fill=BlueViolet!30, draw=black] (-0.5+4+8,5.15)--(2.5+4+8,5.15)--(2.5+4+8,4.35)--(-0.5+4+8,4.35)--cycle;
\path (1+4+8,4.75) node [style=sergio]{\footnotesize $V(\{g\})$};
\end{tikzpicture}
\caption{Quantum circuit as the entangler of the ASPT density matrix $\rho_{\rm ASPT}$, from a trivial density matrix $\rho_0$. $O(1)$ depicts the finite-depth nature of $V(\{g\})$.}
\label{MQC}
\end{figure}

Instead of $\ket{0^{\tilde{G}}}$, we define a trivial paramagnet for the $A$ spins: 
\begin{equation}
\ket{0^A}=\bigotimes_j\Big(\frac{1}{\sqrt{|A|}}\sum_{a\in A}\ket{a_j}\Big),
\end{equation}
and then a trivial $G$ spin ensemble:
\begin{equation}
    \rho_0=\frac{1}{|G|^{N_v}}\sum_{\{g\}} \op{0^A}\otimes\op{\{g\}},
\end{equation}
where $N_v$ is the number of vertices of the lattice. This state can be prepared by measuring $G$ spins in the $\ket{0^{\tilde{G}}}$ state (but no post selection). It is easy to verify that the density matrix $\rho_0$ does not have an exact $G$ symmetry, but still invariant under the average $G$ symmetry. 

The follow-up finite-depth unitary circuit $V(\{g\})$ can still be defined in the same way, but now it is viewed as an operator that acts on the $A$ spins conditioned on the $G$ spins.  More explicitly:
\begin{equation}
V(\{g\})=\sum_{\{a\}}\prod_{\Delta_{ijk}}\nu^{s(\Delta_{ijk})}(1,x_i,x_j,x_k)\ket{\{a\}}\bra{\{a\}}.
\end{equation}
We denote it as $V(\{g\})$ to emphasize the $G$ spin dependence. Importantly, even though $\nu$ is not a 3-cocycle of the group $\tilde{G}$, by definition it is a 3-cocycle of $A$ and therefore $V(\{g\})$ is a (globally) $A$-symmetric finite-depth circuit. On the other hand, under $h\in G$ the unitary transforms as
\begin{equation}
    U_h V(\{g\})U_h^\dag = e^{i\phi(h;\{g\})}V(\{hg\}).
\end{equation}
Fixing $\{g\}$, now we can define an $A$-SPT state:
\begin{equation}
\ket{\Psi_{\rm ASPT}(\{g\})}=V(\{g\}) \ket{0^A}.
\end{equation}
It then follows that 
\begin{equation}
    U_h\ket{\Psi_{\rm ASPT}(\{g\})}=e^{i\phi(h;\{g\})}\ket{\Psi_{\rm ASPT}(\{hg\})}.
\end{equation}
Here $e^{i\phi(h;\{g\})}$ is a phase factor that can be expressed as a product over $O_4$, but the exact expression is not important to us -- the factors from the bra and ket will cancel out in the density matrix.
The parent Hamiltonian of a state $\ket{\Psi(\{g\})}$ in the ensemble is given by
\begin{equation}
H(\{g\})= - V(\{g\})\Big(\sum_j \ket{0^A_j}\bra{0^A_j}\Big)V^\dag(\{g\}).
\end{equation}

The collection of states $\ket{\Psi(\{g\})}$ for all the $G$ spins forms a statistical ensemble. More formally, we can write it as the following density matrix:
\begin{equation}
    \rho_{\rm ASPT}=\sum_{\{g\}}p(\{g\})\ket{\Psi_{\rm ASPT}(\{g\})}\bra{\Psi_{\rm ASPT}(\{g\})}.
\end{equation}
Here $p(\{g\})$ is a probability distribution of $G$ spins. In the simplest case, we can simply set $p$ to be a constant independent of $\{g\}$. The density matrix evidently has $G$ average symmetry, but the $A$ symmetry remains exact. The finite-depth quantum circuit as the entangler of an ASPT density matrix is illustrated in Fig. \ref{MQC}.

Lastly, we discuss two concrete examples. The first example is in (1+1)$d$, with $G=\Z_2, A=\Z_2$ and $\tilde{G}=\Z_4$. We will go through the construction of this example in Sec. \ref{decohered vs gapless}.

The next example is in (2+1)$d$, with $G=\text{SO}(5), A=\Z_2$ and $\tilde{G}=\text{Spin}(5)$. Each state in the ensemble is a Levin-Gu SPT state \cite{LevinGu} protected by the exact $\Z_2$ symmetry. To see why $G$ must be an average symmetry, note that otherwise, we can gauge $A$ to find a double semion topological order with four anyons $\{1,s,s',b\}$. It is enriched by the SO(5) symmetry, where $b$ transforms as the spinor representation of SO(5) (required by the group extension). Because $b=s\times s'$, one of $s$ or $s'$ must also transform as a spinor representation, and we will assume it is $s$. Then $s'$ transforms linearly under SO(5). In other words, we have effectively a semion topological order $\{1,s\}$ with $s$ being a SO(5) spinor. This SET is known to have a nontrivial SO(5) 't Hooft anomaly \cite{wang2017deconfined}. Therefore, the original SPT state can not exist with $G$ being an exact symmetry. However, since the only obstruction is the 't Hooft anomaly of $G$, once $G$ becomes an average symmetry the obstruction no longer matters.

\subsection{Berry-free ASPT and gapless SPT}
\label{decohered vs gapless}
We now discuss the relationship between the ``Berry-free" intrinsic ASPT phases and the recently discussed intrinsically gapless SPT (igSPT) phases \cite{gaplessSPT, Ma2022, wen2022igSPT}. Firstly, let us review the physics of igSPTs in $(1+1)d$~\cite{gaplessSPT}. They can be constructed using a ``slab", where the top boundary is an ``anomalous" gapped $\tilde{G}$ SPT state, obstructed only by a differential mapped into $[\omega]\in\H^{3}(G, \U)$. The bottom boundary instead has a gapless theory , e.g. a conformal field theory (CFT), where the $G$ symmetry acts faithfully in the low-energy theory with a 't Hooft anomaly given by $[\omega^{-1}]$, and the $A$ symmetry does not act. Together the whole slab is free of any anomaly and can be realized with a non-anomalous $\tilde{G}$ symmetry.

Starting from an igSPT state, we consider turning on a random $G$ symmetry-breaking perturbation. Without loss of generality, we assume that the random perturbation is relevant, so it drives the gapless theory into a disordered SRE ensemble. The result is expected to be an intrinsically ASPT state. In the other direction, any symmetry-preserving clean limit of a Berry-free intrinsic ASPT state must be an igSPT state.

Let us consider an example, with $A=\Z_2, G=\Z_2$ and the extension is $\tilde{G}=\Z_4$. For the clean system, there is no nontrivial gapped SPT phase because $\H^2(\Z_4,\U)=\Z_1$. If we look closer, there is a nontrivial $E_2^{1,1}$ term in the LHS spectral sequence, which is however obstructed by a nontrivial $d_2$ differential into $\H^3(\Z_2, \U)$. The same anomaly is realized by a (1+1)$d$ free boson CFT, and together we can construct an igSPT state. We now describe a solvable lattice model for this state. A similar model was studied in \cite{Li:2022jbf}.

The Hilbert space of the model consists of Ising spins $\sigma$ on the sites and $\tau$ on links. The symmetries are defined as 
\begin{equation}
U_g=\prod_j \sigma^x_j e^{i\frac{\pi}{4}\sum_j (1-\tau_{j+1/2}^x)},~U_a=\prod_j \tau_{j+1/2}^x.
\end{equation}
Here $g/a$ is the generator of $G/A$. The unitaries are on-site, and satisfy $U_a^2=1,~U_g^2=U_a$.  So this is a non-anomalous $\Z_4$ symmetry.

We define a projector:
\begin{equation}
P=\prod_j P_j,~P_j= \frac{1+\sigma^z_j \tau_{j+1/2}^x\sigma^z_{j+1}}{2}.
\end{equation}
Physically, in the subspace $P=1$ an Ising domain wall $\sigma_j^z\sigma_{j+1}^z=-1$ is decorated by a charge $\tau_{j+1/2}^x=-1$. So $P=1$ enforces domain wall decoration corresponding to the nontrivial element in $\H^1(G, \H^1(A, \U))$. In this subspace, $U_g$ takes the following form:
\begin{equation}
U_g=\prod_j \sigma^x_j e^{i\frac{\pi}{4}\sum_j (1-\sigma_{j}^z\sigma_{j+1}^z)},
\end{equation}
which takes the form of the anomalous $\Z_2$ symmetry of the Levin-Gu edge model~\cite{CZX, LevinGu}. It is also easy to see that $U_a$ becomes the identity in this low-energy subspace, at least in the bulk of the spin chain. Define 
\begin{equation}
    \tilde{\sigma}_j^x=\tau_{j-1/2}^z\sigma^x_j\tau_{j+1/2}^z,~
    \tilde{\sigma}_j^z=\sigma_j^z.
\end{equation}
$\tilde{\sigma}_j^x$ and $\tilde{\sigma}_j^z$ generate the entire algebra of operators that commute with $P$. They satisfy the usual commutation relations of Pauli operators.

Now we consider the following Hamiltonian:
\begin{align}
H_{\text{LG}}=&-\sum_j  \tilde{\sigma}^x_j(1-\tilde{\sigma}_{j-1}^z\tilde{\sigma}_{j+1}^z)P.
\label{1D Z4 gapless SPT}
\end{align}
Notice that since the Hamiltonian is written in terms of the $\tilde{\sigma}^{x,z}$ operator it commutes with $P$. In the low-energy space, Eq. \eqref{1D Z4 gapless SPT} is identical to the edge Hamiltonian of the Levin-Gu model~\cite{LevinGu}.  In addition, the Hamiltonian conserves the number of Ising domain walls, and thus also conserves the $\Z_4$ charge.  The low-energy effective theory of this model is a $c=1$ free boson, or a Luttinger liquid, with an anomalous $U_g$ symmetry transformation. Thus the model realizes an igSPT phase.

To obtain an ASPT phase, we can proceed in two ways. First,  we add some random Ising disorder $-\sum_jh_j\sigma_j^z$ with $h_j=\pm 1$ to break $G=\mathbb{Z}_2$ symmetry. It can be shown that this disorder is a relevant perturbation to the Luttinger liquid. In the strong disorder limit, we can ignore $H_{\rm LG}$ because it does not commute with the random disorder we have added. The Hamiltonian $H_{\mathcal{D}}$ and ground-state wavefunction $|\Psi_{\mathcal{D}}\rangle$ for a specific disorder realization $\{h_j\}$ is 
\begin{equation}
\begin{gathered}
 H_{\mathcal{D}}=\sum\limits_jP_j+h_j\sigma_j^z\\
|\Psi_{\mathcal{D}}\rangle=\bigotimes_j|\sigma_j^z=h_j\rangle\otimes|\tau_{j+1/2}^x=h_jh_{j+1}\rangle
\end{gathered}. 
\end{equation}
Thus we obtain a disorder ensemble $\{\ket{\Psi_{\cal D}}\}$.

Alternatively, we construct the following mixed state: 
\begin{align}
\rho=\sum\limits_{\mathcal{D}}p_{\mathcal{D}}|\Psi_{\mathcal{D}}\rangle\langle\Psi_{\mathcal{D}}|,
\label{rhoZ4}
\end{align}
where $p_{\mathcal{D}}$ is the classical probability distribution of the disorder realizations.  We now show explicitly that starting from the igSPT (pure) state, one can apply a finite-depth quantum channel to obtain $\rho$. Denote the density matrix of the igSPT state by $\rho_{\rm igSPT}$. First we apply the following quantum channel ${\cal E}^z$:
\begin{equation}
{\cal E}^z={\cal E}^z_1\circ {\cal E}^z_2\circ \cdots, \:{\cal E}^z_j[\rho]=\frac{\rho+\sigma^z_j\rho\sigma^z_j}{2}.
\end{equation}
Notice that this quantum channel only preserves $U_g$ on average, but preserves $U_a$ exactly.
After this step ${\cal E}^z[\rho_{\rm igSPT}]$ already takes the form given in Eq. \eqref{rhoZ4}, but the probability distribution $p_{\cal D}$ is long-ranged. In fact, the correlation function of $\sigma^z$ is the same as that in the pure state.

Then we apply another quantum channel ${\cal E}^x$:
\begin{equation}
\begin{split}
{\cal E}^x &={\cal E}^x_1\circ {\cal E}^x_2\circ \cdots, \\
{\cal E}^x_j[\rho]&=\frac12\rho+\frac14\tilde{\sigma}^x_j\rho \tilde{\sigma}^x_j + \frac14\tilde{\sigma}^x_j\tilde{\sigma}^z_{j-1}\tilde{\sigma}^z_{j+1}\rho \tilde{\sigma}^x_j\tilde{\sigma}^z_{j-1}\tilde{\sigma}^z_{j+1}.
\end{split}
\end{equation}
This channel preserves  $U_a$ exactly and $U_g$ on average. 
It is straightforward to check that $({\cal E}^x\circ {\cal E}^z)[\rho_{\rm igSPT}]$ gives a decohered SPT state with $p_{\cal D}\propto 1$. Notice that this does not imply that the igSPT and the decohered ASPT are in the same phase, as there is no finite-depth local quantum channel that takes $\rho$ (which has only short-ranged correlation functions) to the igSPT (which has power-law correlation functions). 

{\color{black}The above connection between (intrinsic) ASPT and (intrinsic) gapless SPT can also be used to study gapless SPT. In dimension $d>1$ the bulk theory of a gapless SPT is typically rather complicated and may require significant amount of fine-tuning (to avoid trivial IR fate such as spontaneous symmetry-breaking). Our result shows that we can eliminate the long-range correlation (and the associated instability) in the bulk through certain types of decoherence, and the topological aspects (such as boundary states discussed in Sec.~\ref{sec:clusteredge}) will remain. An interesting future direction is to utilize decohered ASPT to better understand various edge states of gapless SPT, especially in $d>1$.}

\subsection{Fermionic intrinsic ASPT}

We now turn to the fermionic case. For simplicity, we will assume that the ``bosonic" symmetry group $G$ becomes average,  while the fermion parity conservation $\Z_2^f$ remains an exact symmetry. In the decohered case, it means that the bath is bosonic, so the coupling between the system and the environment preserves fermion parity of the system. The total symmetry group $\tilde{G}$ is a central extension of $G$ by $\Z_2^f$. We further assume that $G$ is a finite group to simplify the discussion.  

The relevant groups of fermionic invertible phases that can be decorated on $G$-defects, up to (3+1)$d$, are
\begin{equation}
    h^0=\U, h^1=\Z_2, h^2=\Z_2, h^3=\Z, h^4=0,
\end{equation}
where $h^1=\Z_2$ is generated by a complex fermion, $h^2=\Z_2$ is generated by the Majorana chain, and $h^3=\Z$ is generated by the $p+ip$ superconductor.

Following the prescription in Sec.~\ref{sec:SSclassification},  for disordered phases we set $h^0=h^1=0$. Thus for disordered ASPT states, the relevant groups are
\begin{equation}
    h^0=0, h^1=0, h^2=\Z_2, h^3=\Z, h^4=0.
\end{equation}

For decohered ASPT, the groups become
\begin{equation}
    h^0=0, h^1=\Z_2, h^2=\Z_2, h^3=\Z_{16}, h^4=0.
\end{equation}
The change of $h^3$ is because 16 copies of $p+ip$ superconductors are adiabatically equivalent to a $E_8$ state, which becomes trivialized for decohered phases (assuming the bath to be bosonic).

We now study the classifications in more details. Suppose the spatial dimension is $D\leq 3$. Let us consider the following two terms on the $E_2$ page: $\H^{D-1}(G, h^2)$ and $\H^D(G, h^1)$, corresponding to decorations of 1-dimensional $G$ defect junctions by Majorana chains, and $0$-dimensional $G$ junctions by complex fermions.   The potential differentials are 
\begin{align}
    d_2&: \H^{D-1}(G, \Z_2)\rightarrow \H^{D+1}(G, h^1),\\
    d_2&: \H^D(G, \Z_2)\rightarrow \H^{D+2}(G, h^0),
\end{align}
and
\begin{equation}
    d_3: \H^{D-1}(G, \Z_2)\rightarrow \H^{D+2}(G, h^0).
\end{equation}
Explicit expressions for the differentials can be found in \cite{general1, general2}. 

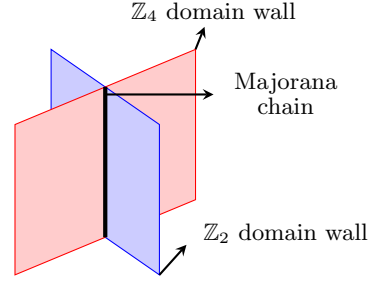
\begin{figure}
\begin{tikzpicture}[xscale=1.2,yscale=1]
\tikzstyle{sergio}=[rectangle,draw=none]
\filldraw[fill=red!20, draw=red] (0,1)--(-1,0.5)--(-1,-1.5)--(0,-1)--cycle;
\filldraw[fill=blue!20, draw=blue] (-1,0.5)--(-0.4,0)--(-0.4,-2)--(-1,-1.5)--cycle;
\filldraw[fill=blue!20, draw=blue] (-1,0.5)--(-1.6,1)--(-1.6,-1)--(-1,-1.5)--cycle;
\filldraw[fill=red!20, draw=red] (-1,0.5)--(-2,0)--(-2,-2)--(-1,-1.5)--cycle;
\draw[ultra thick] (-1,0.5)--(-1,-1.5);
\draw[->,thick] (0,1)--(0.1,1.3);
\path (0.2,1.5) node [style=sergio]{$\Z_4$ domain wall};
\path (1,-1.45) node [style=sergio]{$\Z_2$ domain wall};
\draw[->,thick] (-0.4,-2)--(-0.1,-1.6);
\draw[->,thick] (-1,0.4)--(0.2,0.4);
\path (1,0.55) node [style=sergio]{Majorana};
\path (1,0.25) node [style=sergio]{chain};
\end{tikzpicture}
\caption{(3+1)$d$ intrinsically decohered fermionic ASPT state from decorating a Majorana chain on the junction of 
$\Z_4$ (red) and $\Z_2$ (blue) domain walls.}
\label{Majorana chain decoration}
\end{figure}

First consider disordered ASPT phases, where $h^0=h^1=0$ and $h^2=\Z_2$. For the $\H^{D-1}(G, \Z_2)$ part, both $d_2$ and $d_3$ automatically vanish. Thus we conclude that any element of $\H^{D-1}(G, \Z_2)$ gives a disordered ASPT phase. Let us list a few examples of intrinsically disordered ASPT phases: 
\begin{enumerate}[1.]
\item $D=1, G=\Z_2, \tilde{G}=\Z_4^f$. This example is a Majorana chain with $\Z_4^f$ symmetry (i.e. a charge-$4e$ superconductor in the clean limit), which has a $d_2$ obstruction in the clean case.  Below we will describe a concrete model realization of this state in a 1D Kitaev chain with random pairing.
\item $D=2, G=\Z_2^{\mathsf{T}}, \tilde{G}=\Z_2^{\mathsf{T}}\times \Z_2^f$. Here $\Z_2^{\mathsf{T}}$ is the time-reversal symmetry. The Majorana decoration is classified by $\H^1(\Z_2^{\mathsf{T}},\Z_2)=\Z_2$, and the nontrivial class is obstructed by $d_2$ in the clean case. Note that if $\tilde{G}=\Z_4^{\mathsf{T}f}$, then the $d_2$ obstruction vanishes and the result is the well-known class D\3 topological superconductor in 2D (see Appendix \ref{App: construction} for more details).
\item $D=3, G=\Z_2, \tilde{G}=\Z_2\times\Z_2^f$. The Majorana decoration is classified by $\H^2(\Z_2, \Z_2)=\Z_2$. The nontrivial class is obstructed by $d_2$ in the clean case.
\item $D=3, G=\Z_2\times\Z_4, \tilde{G}=\Z_2\times\Z_4\times\Z_2^f$. The Majorana decoration is classified by $\H^2(\Z_2\times\Z_4, \Z_2)=\Z_2^3$. Interestingly, one of them is only obstructed by $d_3$ in the clean case. More explicitly, denote group elements of $\Z_2\times\Z_4$ by $(a_1,a_2)$, where $a_1=0,1$ and $a_2=0,1,2,3$. The nontrivial cocycle in $\H^2(G, \Z_2)$ that describes Majorana chain decoration on the junction of the $\Z_4$ and $\Z_2$ domain walls (see Fig. \ref{Majorana chain decoration}) is given by 
\begin{equation}
n_2(a,b)=a_1b_2~(\mathrm{mod}~2).
\label{Majorana decoration}
\end{equation}
We explicitly check that the $d_2$ obstruction vanishes and $d_3$ is nontrivial.
    \end{enumerate}
 More examples of obstructed fermionic phases can be found in Ref.~\cite{Manjunath2023}.
 
For decohered ASPTs,  the differentials of the cases with Majorana-chain decoration have been analyzed in the disordered case, and the only difference is that $h^1=\Z_2$ for decohered ASPTs, so the $d_2$ differential needs to vanish to ensure the fermion parity conservation. An example of such intrinsically decohered fermionic ASPT is given by Eq.~\eqref{Majorana decoration} for $G=\Z_4\times\Z_2$. 

For the $\H^D$ part (complex fermion decoration), the $d_2$ differential always vanishes. An example of a ``Berry-free" ASPT phase with complex fermion decoration is $G=\Z_2$, $\tilde{G}=\Z_2\times\Z_2^f$ in $D=3$. The fermionic SPT phase corresponding to the nontrivial element in $\H^3(G, \Z_2)=\Z_2$ is obstructed by $d_2$ in the clean case, and can now be realized an intrinsically decohered ASPT phase. 

Exactly solvable lattice models of fermionic ASPT phases (without $p+ip$ decorations) can be constructed following \cite{general2}. We outline the construction in Appendix \ref{App: spectral sequence} for $D=2$. Below we describe a simple model realization of Majorana chain with an average $\Z_4^f$ symmetry, an example of localization-enabled compressible ASPT phase.

First let us consider a chain of spinless fermions, with the following Hamiltonian:
\begin{equation}
    H=-\sum_j (c_j^\dag c_{j+1}+\text{h.c.})+\sum_j \Delta_j (c_j c_{j+1}+\text{h.c.}).
    \label{kitaev-wire}
\end{equation}
The $\Z_4^f$ symmetry is generated by $g:c_j\rightarrow ic_j$. The pairing $\Delta_j\rightarrow -\Delta_j$ under the $g$ symmetry. When $\Delta_j$ is uniform and nonzero, this is the well-known Hamiltonian of a Kitaev chain, explicitly breaking the $\Z_4^f$ symmetry.
When $\Z_4^f$ is exact, we must have $\Delta=0$ and the ground state is a gapless metal\footnote{There are other terms that can gap out the metal without breaking $\Z_4^f$, such as a translation-breaking potential. However, to keep the system in a nontrivial Kitaev chain, we need the amplitude of such potential $\mu$ to be smaller than $\Delta$. As we recover the exact $\Z_4^f$ symmetry by taking $\Delta\to0$, we need to take $\mu\to0$ first.}. When $\Z_4^f$ is an average symmetry, we can turn on a random pairing term with symmetric probability: $P[\Delta_j]=P[-\Delta_j]$. This random pairing term will localize the metallic ground state, resulting in a random Kitaev chain. 

In the Hamiltonian Eq. \eqref{kitaev-wire}, if the configuration $\Delta_j$ contains one sign-changing domain wall, we find it harbors a localized complex fermion zero mode. Thus when the sign of the pairing potential is disordered, we expect that there are low-energy states filling the superconducting gap, which we confirm numerically.  However, this zero mode is protected by the time reversal symmetry (i.e. the complex conjugation).\footnote{In fact, $\Delta>0$ and $\Delta<0$ belong to distinct topological phases labeled by $\nu=\pm 1$ where $\nu\in \Z_8$ is the topological invariant for topological superconductors with $T^2=1$ time-reversal symmetry (the BDI class), so there must be a protected zero mode at the interface.} In order to obtain a localized state, we lift the local degeneracy from the zero modes by having complex hopping or the pairing terms that break the time-reversal symmetry.

\begin{figure}
\begin{tikzpicture}[scale=1.1]
\tikzstyle{sergio}=[rectangle,draw=none]
\draw[blue, thick] (-3,-0.4) -- (-3.5,-0.4);
\draw[blue, thick] (-3,0.4) -- (-2,-0.4);
\draw[blue, thick] (-2,0.4) -- (-1,0.4);
\draw[blue, thick] (0,0.4) -- (-1,-0.4);
\draw[blue, thick] (0,-0.4) -- (1,0.4);
\draw[blue, thick] (2,-0.4) -- (1,-0.4);
\draw[blue, thick] (2,0.4) -- (3,-0.4);
\draw[blue, thick] (3.5,0.4) -- (3,0.4);
\draw[thick] (-3.5,0)--(3.5,0);
\draw[thick] (-1,0)--(-1,0.15);
\draw[thick] (-2,0)--(-2,0.15);
\draw[thick] (1,0)--(1,0.15);
\draw[thick] (2,0)--(2,0.15);
\draw[thick] (0,0)--(0,0.15);
\draw[thick] (-3,0)--(-3,0.15);
\draw[thick] (3,0)--(3,0.15);
\draw[ball color=blue] (-1,0.4) circle (0.08);
\draw[ball color=blue] (-1,-0.4) circle (0.08);
\draw[ball color=blue] (0,0.4) circle (0.08);
\draw[ball color=blue] (0,-0.4) circle (0.08);
\draw[ball color=blue] (1,0.4) circle (0.08);
\draw[ball color=blue] (1,-0.4) circle (0.08);
\draw[ball color=blue] (-2,0.4) circle (0.08);
\draw[ball color=blue] (-2,-0.4) circle (0.08);
\draw[ball color=blue] (2,0.4) circle (0.08);
\draw[ball color=blue] (2,-0.4) circle (0.08);
\draw[ball color=blue] (3,0.4) circle (0.08);
\draw[ball color=blue] (3,-0.4) circle (0.08);
\draw[ball color=blue] (-3,0.4) circle (0.08);
\draw[ball color=blue] (-3,-0.4) circle (0.08);
\path (-3.3,0.6) node [style=sergio] {$\gamma_{A}$};
\path (-3.3,-0.6) node [style=sergio] {$\gamma_{B}$};
\path (-3.9,0) node [style=sergio] {(a)};
\draw[blue, thick] (-3,-0.4-2) -- (-3.5,-0.1-2);
\draw[blue, thick] (-3,0.4-2) -- (-2,-0.4-2);
\draw[blue, thick] (-2,0.4-2) -- (-1,-0.4-2);
\draw[blue, thick] (0,-0.4-2) -- (-1,0.4-2);
\draw[blue, thick] (0,0.4-2) -- (1,-0.4-2);
\draw[blue, thick] (2,-0.4-2) -- (1,0.4-2);
\draw[blue, thick] (2,0.4-2) -- (3,-0.4-2);
\draw[blue, thick] (3.5,0.1-2) -- (3,0.4-2);
\draw[thick] (-3.5,0-2)--(3.5,0-2);
\draw[thick] (-1,0-2)--(-1,0.15-2);
\draw[thick] (-2,0-2)--(-2,0.15-2);
\draw[thick] (1,0-2)--(1,0.15-2);
\draw[thick] (2,0-2)--(2,0.15-2);
\draw[thick] (0,0-2)--(0,0.15-2);
\draw[thick] (-3,0-2)--(-3,0.15-2);
\draw[thick] (3,0-2)--(3,0.15-2);
\draw[ball color=blue] (-1,0.4-2) circle (0.08);
\draw[ball color=blue] (-1,-0.4-2) circle (0.08);
\draw[ball color=blue] (0,0.4-2) circle (0.08);
\draw[ball color=blue] (0,-0.4-2) circle (0.08);
\draw[ball color=blue] (1,0.4-2) circle (0.08);
\draw[ball color=blue] (1,-0.4-2) circle (0.08);
\draw[ball color=blue] (-2,0.4-2) circle (0.08);
\draw[ball color=blue] (-2,-0.4-2) circle (0.08);
\draw[ball color=blue] (2,0.4-2) circle (0.08);
\draw[ball color=blue] (2,-0.4-2) circle (0.08);
\draw[ball color=blue] (3,0.4-2) circle (0.08);
\draw[ball color=blue] (3,-0.4-2) circle (0.08);
\draw[ball color=blue] (-3,0.4-2) circle (0.08);
\draw[ball color=blue] (-3,-0.4-2) circle (0.08);
\path (-3.3,0.6-2) node [style=sergio] {$\gamma_{A}$};
\path (-3.3,-0.6-2) node [style=sergio] {$\gamma_{B}$};
\path (-3.9,0-2) node [style=sergio] {(b)};
\draw[blue, thick] (-3,-0.4-4) -- (-3.5,-0.1-4);
\draw[blue, thick] (-3,0.4-4) -- (-2,-0.4-4);
\draw[blue, thick] (-2,0.4-4) -- (-1,-0.4-4);
\draw[blue, thick] (0,0.4-4) -- (-1,0.4-4);
\draw[blue, thick] (0,-0.4-4) -- (1,0.4-4);
\draw[blue, thick] (2,-0.4-4) -- (1,-0.4-4);
\draw[blue, thick] (2,0.4-4) -- (3,-0.4-4);
\draw[blue, thick] (3.5,0.1-4) -- (3,0.4-4);
\draw[thick] (-3.5,0-4)--(3.5,0-4);
\draw[thick] (-1,0-4)--(-1,0.15-4);
\draw[thick] (-2,0-4)--(-2,0.15-4);
\draw[thick] (1,0-4)--(1,0.15-4);
\draw[thick] (2,0-4)--(2,0.15-4);
\draw[thick] (0,0-4)--(0,0.15-4);
\draw[thick] (-3,0-4)--(-3,0.15-4);
\draw[thick] (3,0-4)--(3,0.15-4);
\draw[ball color=blue] (-1,0.4-4) circle (0.08);
\draw[ball color=blue] (-1,-0.4-4) circle (0.08);
\draw[ball color=blue] (0,0.4-4) circle (0.08);
\draw[ball color=blue] (0,-0.4-4) circle (0.08);
\draw[ball color=blue] (1,0.4-4) circle (0.08);
\draw[ball color=blue] (1,-0.4-4) circle (0.08);
\draw[ball color=blue] (-2,0.4-4) circle (0.08);
\draw[ball color=blue] (-2,-0.4-4) circle (0.08);
\draw[ball color=blue] (2,0.4-4) circle (0.08);
\draw[ball color=blue] (2,-0.4-4) circle (0.08);
\draw[ball color=blue] (3,0.4-4) circle (0.08);
\draw[ball color=blue] (3,-0.4-4) circle (0.08);
\draw[ball color=blue] (-3,0.4-4) circle (0.08);
\draw[ball color=blue] (-3,-0.4-4) circle (0.08);
\path (-3.3,0.6-4) node [style=sergio] {$\gamma_{A}$};
\path (-3.3,-0.6-4) node [style=sergio] {$\gamma_{B}$};
\path (-3.9,0-4) node [style=sergio] {(c)};
\path (-1,-4.8) node [style=sergio] {$j=1$};
\path (0,-4.8) node [style=sergio] {$j=2$};
\path (1,-4.8) node [style=sergio] {$j=3$};
\end{tikzpicture}
\caption{Illustration of a random Kitaev chain with average $\Z_4^f$ symmetry. (a) A typical Majorana bond configuration. (b) A uniform bond configuration. (c) Two nearby domain walls on top of the uniform configuration. The total fermion parity of (b) and (c) differ by $(-1)$.}
\label{Fig:RandomKitaev}
\end{figure}
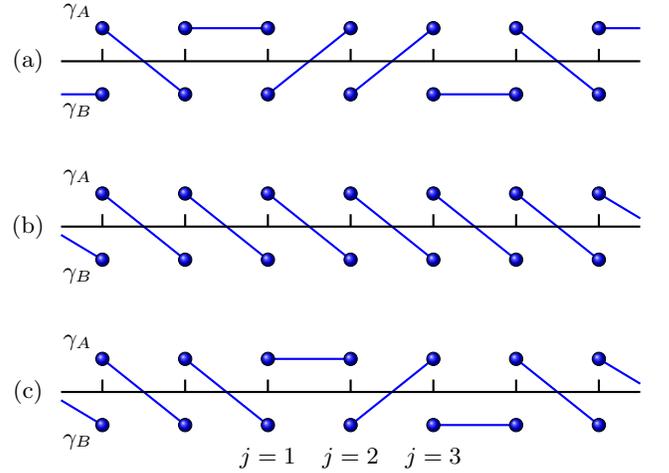

To better appreciate the nature of the random Kitaev chain with average $\Z_4^f$ symmetry, it is more illuminating to consider the following ``fixed point'' model. On each lattice site $j$ we have a complex fermion, which we write as two Majoranas $c_j=\frac{1}{2}(\gamma_{A,j}-i\gamma_{B,j})$. The random Hamiltonian takes the form
\begin{equation}
    H=-\sum_j\sum_{r_j=A,B}ig_j^{r_j,r_{j+1}}\gamma_{r_j,j}\gamma_{r_{j+1},j+1},
\end{equation}
where $g$ is a random coupling constant taking value in $\{0,\pm1\}$ such that every lattice link $(j,j+1)$ is covered by exactly one Majorana bond, which makes each state in the ensemble a nontrivial Kitaev chain. For example, if $g_j^{AB}=1$, then we must have $g_j^{AA}=g_j^{BB}=g_j^{BA}=0$, and $g_{j+1}^{BA}=g_{j+1}^{BB}=g_{j-1}^{AA}=g_{j-1}^{BA}=0$. An example of such a random Majorana bond configuration is shown in Fig.~\ref{Fig:RandomKitaev}~(a). Besides these nearest-neighbor constraints, the random coupling $g$ should be uncorrelated in long distance, generated by a probability functional $P[g_j]$.

Now we examine the condition on $P[g_j]$ imposed by the average $\Z_4^f$ symmetry, which is generated by $U=\exp{(\frac{\pi}{4}\sum_j\gamma_{j,A}\gamma_{j,B})}$. Some simple algebra shows that
\begin{equation}
\begin{gathered}
U(i\gamma_j^A\gamma_{j+1}^B)U^{\dagger}=-i\gamma_j^B\gamma_{j+1}^A\ U(i\gamma_j^A\gamma_{j+1}^A)U^{\dagger}=i\gamma_j^B\gamma_{j+1}^B
\end{gathered}.
\end{equation}
The minus sign above will be important. Now for $P[g]$, we should have
\begin{equation}
    P(g^{AB})=P(-g^{BA}), \hspace{10pt} P(g^{AA})=P(g^{BB}).
    \label{eq:bondProb}
\end{equation}
We can now further simplify the model by having $g^{AB},g^{AA},g^{BB}\in\{0,1\}$ and $g^{BA}\in\{0,-1\}$, generated by a $\Z_4^f$-symmetric probability functional $P[g]$.

Now what do all these mean for the total fermion parity $i^{L}\prod_{j=1}^L\gamma_{j, A}\gamma_{j, B}$? If we have only $AB$ bonds or $BA$ bonds, the total fermion parity on a ring is fixed ($-1$ for periodic boundary conditions). But when we have domain walls between the two bonding patterns, as required by the average $\Z_4^f$ symmetry, the fermion parity starts to change locally. The most illuminating case is when two domain walls are right next to each other, as shown in Fig.~\ref{Fig:RandomKitaev}~(c). In this configuration all links $(j,j+1)$ with $j<1$ and $j>3$ are $AB$ bonds, while the intermediate bonds are given by $g_1^{AA}=1$, $g_2^{BA}=-1$ and $g_3^{BB}=1$. A simple calculation shows that, compared with the configuration with no domain wall at all ($g_j^{AB}=1$ for all $j$, as in Fig.~\ref{Fig:RandomKitaev}~(b)), this configuration has an additional $(-1)$ fermion parity. This means that even though the domain wall behaves like an $\Z_2$ object (the bonding configurations return to the original ones after passing through two domain walls), ``fusing'' two nearby domain walls together will change the fermion parity. As a result, in the clean limit, the domain walls cannot condense to recover the exact $\Z_4^f$ symmetry. This is nothing but the manifestation of the $d_2$ obstruction.  In general, the fermion parity of the state depends on the configurations of the $g$'s and fluctuates randomly within the ensemble. In a disordered system the nontrivial domain-wall fusion does not lead to any obstruction for short-range entanglement, since the domain walls are pinned by the disorders. The above discussion eventually leads to a long-distance picture, which contains localized fermions (that carry nontrivial fermion parity) randomly located at the $\Z_4^f$-domain walls. If we take the absolute ground state of each Hamiltonian realization, different states in the ensemble will have different total fermion parity. If we take a ``canonical ensemble'' and fix the total fermion parity for the entire ensemble, then half of the ensemble will be put in excited states (to match the fermion parity), and the excitation spectral above such states will in general be gapless (assuming a bounded distribution of $g_j$).

\section{Average symmetry-enriched topological orders}
\label{sec:ASET}

\subsection{General structures}
\label{sec: SETGeneral}

In this section we consider disordered topologically ordered phases with average symmetry in (2+1)$d$. We will assume that the system is bosonic. In general, bosonic topological order in the ground state of a gapped Hamiltonian is described by a mathematical structure called unitary modular tensor category (UMTC)~\cite{Kitaev:2005hzj}, or the anyon theory, denoted by $\mathcal{C}$. Physically, $\mathcal{C}$ consists of a set of topological charges (i.e. anyon types), as well as consistency data specifying their fusion and braiding. When the anyon theory is nontrivial, the state is said to be long-range entangled (LRE).

 To analyze LRE phases with disorder, let us first define the notion of LRE ensembles, which is a straightforward extension of SRE ensembles. To formulate the definition it is convenient to choose a ``reference state", which can be the ground state of a gapped Hamiltonian in the clean system, and identify its topological order described by the anyon theory ${\cal C}$. We require that all states in the ensemble are smoothly connected to the reference state, thus described by the same topological order ${\cal C}$. Such an ensemble is said to be LRE with topological order ${\cal C}$. Notice that this definition is meaningful only in disordered systems, where the disorder realizations naturally provide a preferred basis for the density matrix. For a general mixed state which has no preferred basis, the definition of topological order is much more subtle and is beyond the scope of this work.

 Just like SRE ensembles, LRE ensembles can be enriched by exact and average symmetries. To characterize such average SET orders, we begin with a review of the classification of SET phases in clean systems, and then indicate how it should be modified for ensembles. 
 
 Let us consider the ground state of a gapped Hamiltonian with a global symmetry $\tilde{G}$, whose topological order is described by a UMTC $\mathcal{C}$. $\tilde{G}$ can enrich the topological order in three ways~\cite{gauging3, TarantinoSET2016, TeoSET2015}:
\begin{enumerate}[1.]
    \item  There is a group homomorphism from $\tilde{G}$ to the group of auto-equivalence maps $\mathrm{Aut}(\mathcal{C})$ of $\mathcal{C}$,
\begin{equation}
    \rho\, : \, \tilde{G} \to \mathrm{Aut}(\mathcal{C}).
\end{equation}
Here ${\rm Aut}(\cal{C})$ consists of all the permutations of anyon types which keep the fusion and braiding properties invariant~\footnote{Note that more precisely, ${\rm Aut}(\cal{C})$ is the group of braided tensor auto-equivalences of $\cal{C}$ and there can be nontrivial elements which do not permute any anyons. However, such examples are only known to occur for very complicated $\cal{C}$, and for simplicity, we do not consider them.}. Basically, $\rho$ tells us how $\tilde{G}$ permutes anyons.
 \item The anyons may carry fractionalized quantum numbers under $\tilde{G}$. In particular, given $\rho$, there is a possible obstruction to symmetry fractionalization, which is an element in $\H^3_{\rho}(\tilde{G},\mathcal{A})$, where $\mathcal{A}$ is the group of Abelian anyons. When the $\H^3$ class vanishes, distinct symmetry fractionalization classes form a torsor over $\H^2_{\rho}(\tilde{G},\mathcal{A})$.
 \item Once $\rho$ and the symmetry fractionalization of anyons are known, we then need to specify the fusion and braiding properties of $\tilde{G}$ symmetry defects. 
 In particular, given $\rho$ and the symmetry fractionalization of anyons, the global symmetry may have a 't Hooft anomaly valued in $\H^4(\tilde{G}, \U)$. When the $\H^4$ anomaly class vanishes, distinct equivalence classes form a torsor over $\H^3(\tilde{G},\U)$, up to further identifications~\cite{aasen2021torsorial, ChengPRR2020}.  
\end{enumerate}

It is again useful to think of the SET phases in terms of fluctuating symmetry defect lines. Each defect line is associated with an anyon permutation action given by $\rho$. The $\H^3$ obstruction means that defect fusion may fail to be associative: an F move of defect lines may nucleate an extra Abelian anyon, violating the locality requirement. When the $\H^3$ class vanishes, the extra Abelian anyon can be ``absorbed" into decorations of tri-junctions of defects by Abelian anyons. In-equivalent patterns of decorations are classified by a torsor over $\H^2_\rho(\tilde{G}, \cal{A})$. Lastly, once we have well-defined defect fusions, including decorations on tri-junctions, there may be a Berry phase in the space of states with defects, which gives the $\H^4$ anomaly. From this interpretation, it is clear that with a non-trivial $\H^3_\rho(\tilde{G}, \cal{A})$ class the map $\rho$ does not make sense in a pure (2+1)$d$ system\footnote{In field theory, an alternative interpretation of the $\H^3$ obstruction is that the 0-form symmetry $G$ and the 1-form symmetry $\cal{A}$ form a nontrivial 2-group structure.}. The $\H^4$ anomaly means that the $\tilde{G}$ symmetry has a 't Hooft anomaly. Examples include the surface of a (3+1)$d$ bosonic SPT state or (2+1)$d$ lattice models that satisfy Lieb-Schultz-Mattis-type theorems~\cite{ChengPRX2016}.

Let us now consider a LRE ensemble with topological order ${\cal C}$. The symmetry group $\tilde{G}$ fits into the short exact sequence (\ref{group extension}), with exact symmetry $A$ and average symmetry $G$.  We again expect $\tilde{G}$ to be mapped to $\mathrm{Aut}(\mathcal{C})$ through a group homomorphism $\rho$, which is an invariant of the disorder ensemble. The only way to change $\rho$ is to go through a phase transition, which violates the adiabatic connectability within a single disorder ensemble. It is then useful to think of each state in the disorder ensembles as a topological order with exact $A$ symmetry (i.e. fluctuating $A$ symmetry defects) and a static $G$ defect network. 

Generalizing the discussions in Sec.~\ref{sec:localization}, in the presence of disorder it is possible to localize Abelian anyons (as long they do not carry any zero modes protected by the exact symmetry, see below). As a result, there is no longer any energetic requirement to have fixed Abelian anyons decorated on $G$ defect junctions. This is in parallel with dropping the 0D charge decoration in the classification of disordered ASPT phases. In fact, one can gauge the exact symmetry $A$ in a disordered ASPT state to get a disordered ASET state. 

We first consider a simpler problem, where the entire symmetry group $\tilde{G}$ becomes average (i.e. $A$ is trivial). In this case, all Abelian anyons can be localized and both the $[O_3]$ obstruction class and the symmetry fractionalization class (i.e. decorations of defect junctions by Abelian anyons) lose their meaning. The same is true for the $\H^4(\tilde{G}, \U)$ anomaly and $\H^3(\tilde{G}, \U)$ torsor since the disordered SET is an ensemble of defects. We thus conclude that with a trivial $A$, disordered ASET phases are completely classified by the maps $\rho$, including those with nontrivial $\H^3$ obstructions. An example of such an intrinsically disordered ASET is presented below in Sec. \ref{D16}.

Next we consider ASETs with a nontrivial $A$, with a given $\rho: \tilde{G}\rightarrow {\rm Aut}(\cal{C})$. The map $\rho$ is associated with an obstruction class $[O_3]\in\H^3_\rho(\tilde{G}, \cal{A})$. The group $\H^3_\rho(\tilde{G}, \cal{A})$ can also be decomposed using the Leray-Serre spectral sequence, whose $E_2$ page consists of
\begin{equation}
    E_2^{p,3-p}=\bigoplus_{p=0}^3 \H^p(G, \H^{3-p}(A, \cal{A})).
\end{equation}
Here we suppress the group action subscripts for clarity. It should be understood that $A$ acts on $\cal{A}$ through $\rho$, and $G$ acts on the coefficient group $\H^{3-p}$ via both $\rho$ and the $G$ action on $A$.

Let us start with the component that only involves $G$, i.e. $E^{3,0}=\H^3_\rho(G, \cal{A})$. Since now we only have an ensemble of $G$ defects, if the additional Abelian anyon from an $F$ move of $G$ defects can be localized then this component of the obstruction class no longer makes sense. However, given the exact $A$ symmetry, localization requires that the Abelian anyons involved in the $F$ move do not transform under $A$ symmetry action, and carry no projective multi-dimensional representations of $A$ (a projective one-dimensional representation, i.e. a fractional charge, is allowed). 

The other components in the decomposition with $p<3$ all involve $A$ defects, and hence even when $G$ becomes an average symmetry they still represent nontrivial obstructions.

Having dealt with the $\H^3$ obstruction, we move to the $\H^2_\rho(\tilde{G}, \cal{A})$ torsor. Again we can decompose it using the LHS spectral sequence:
\begin{equation}
    E_2^{p,2-p}=\H^2_\rho(G, \mathcal{A})\oplus \H^1(G, \H^1(A, \mathcal{A}))\oplus \H^2_\rho(A, \mathcal{A}).
\end{equation}
Here we also suppress the group action subscripts in the middle term.
Via the same reasoning, with localization of Abelian anyons, $\H^2_\rho(G, \mathcal{A})$ may become trivialized, provided that the Abelian anyons involved in this decoration transform trivially (at most as 1D projective reps) under $A$. The next two terms are still meaningful for disordered ASETs. Perhaps the most interesting part is the $\H^1(G, \H^1(A, \cal{A}))$ torsor. When $\rho$ is the identity, in a clean SET this term means that $G$ and $A$ symmetries do not commute when acting on certain anyons. When the $G$ symmetry becomes average, one can interpret this term as the $A$ charge carried by the anyon changes when it passes through certain $G$ defects. Examples of such fractionalizations will be discussed below in Sec. \ref{semion}.

We can similarly discuss what changes need to be made for the $\H^3$ torsor and $\H^4$ anomaly, however, this is identical to the ASPT classification and there is no need to repeat it. 

\subsection{Example: $\Z_2\times\Z_2$ toric code with $\Z_2^{ A}$ symmetry}
We consider an example from a (2+1)$d$ ASPT phase with $A=\Z_2\times\Z_2$ and $G=\Z_2$ while the extension is trivial. The ASPT phase corresponds to the nontrivial element in $\H^1(G, \H^2(A, \U))$: $G$ domain walls are decorated by (1+1)$d$ $A$--SPTs.

\begin{figure}
\begin{tikzpicture}[use Hobby shortcut,xscale=0.66,yscale=0.7]
\tikzstyle{sergio}=[rectangle,draw=none]
\filldraw[fill=blue!20, draw=black] (-1,-1)--(-1,3.5)--(4,3.5)--(4,-1)--cycle;
\coordinate[] (A1) at (-1,0);
\coordinate[] (A2) at (1,1.7);
\coordinate[] (A3) at (2,1.8);
\coordinate[] (A4) at (3,2.5);
\coordinate[] (A5) at (4,2.8);
\filldraw[fill=red!20, line width=.8pt, ultra thick] (A1) .. (A2) .. (A3) .. (A4) .. (A5);
\draw[->,ultra thick] (0.3,2.1) -- (0.3,2.7);
\filldraw[fill=red!20, draw=black] (-1,0)--(-1,-1)--(4,-1)--(4,2.8);
\draw[->,ultra thick] (3,0.6) -- (3,0);
\path (4.4,2.7) node [style=sergio]{\large$\mathcal{D}$};
\coordinate[] (A6) at (1.25,0);
\coordinate[] (A7) at (2,0.5);
\coordinate[] (A8) at (2.8,1.5);
\coordinate[] (A9) at (2.5,2.2);
\coordinate[] (A10) at (3,3);
\draw[color=DarkGreen,line width=.8pt,ultra thick] (A6) .. (A7) .. (A8) .. (A9) .. (A10);
\filldraw[fill=black, draw=black] (1.25,0)circle (3.5pt);
\filldraw[fill=black, draw=black] (3,3)circle (3.5pt);
\path (2.52,2.1) node [style=sergio]{\Large$\bs{+}$};
\path (1.25,-0.5) node [style=sergio]{$m_1$};
\path (2.25,3.1) node [style=sergio]{$e_2m_1$};
\path (3.1,1.98) node [style=sergio]{$e_2$};
\filldraw[fill=blue!20, draw=black] (6,-1)--(6,3.5)--(11,3.5)--(11,-1)--cycle;
\coordinate[] (A11) at (-1+7,0);
\coordinate[] (A12) at (1+7,1.7);
\coordinate[] (A13) at (2+7,1.8);
\coordinate[] (A14) at (3+7,2.5);
\coordinate[] (A15) at (4+7,2.8);
\filldraw[fill=red!20, line width=.8pt, ultra thick] (A11) .. (A12) .. (A13) .. (A14) .. (A15);
\draw[->,ultra thick] (0.3+7,2.1) -- (0.3+7,2.7);
\filldraw[fill=red!20, draw=black] (-1+7,0)--(-1+7,-1)--(4+7,-1)--(4+7,2.8);
\draw[->,ultra thick] (3+7,0.6) -- (3+7,0);
\path (4.4+7,2.7) node [style=sergio]{\large$\mathcal{D}$};
\coordinate[] (A16) at (1.25+7,0);
\coordinate[] (A17) at (2+7,0.5);
\coordinate[] (A18) at (2.8+7,1.5);
\coordinate[] (A19) at (2.5+7,2.2);
\coordinate[] (A20) at (3+7,3);
\draw[color=DarkGreen,line width=.8pt,ultra thick] (A16) .. (A17) .. (A18) .. (A19) .. (A20);
\filldraw[fill=black, draw=black] (1.25+7,0)circle (3.5pt);
\filldraw[fill=black, draw=black] (3+7,3)circle (3.5pt);
\path (2.52+7,2.1) node [style=sergio]{\Large$\bs{+}$};
\path (1.25+7,-0.5) node [style=sergio]{$m_2$};
\path (2.25+7,3.1) node [style=sergio]{$e_1m_2$};
\path (3.1+7,1.98) node [style=sergio]{$e_1$};
\end{tikzpicture}
\caption{Anyon permutation of double toric code model with average symmetry $\Z_2^A$. $\mathcal{D}$ is the symmetry domain wall of $\Z_2^A$ separating the red and blue regimes (with Ising spin-$\uparrow$ and spin-$\downarrow$), and the green curve depicts the string operator of a $\Z_2$ gauge group connecting two anyons $m_1/m_2$ and $e_2m_1/e_1m_2$.}
\label{anyon permutation}
\end{figure}
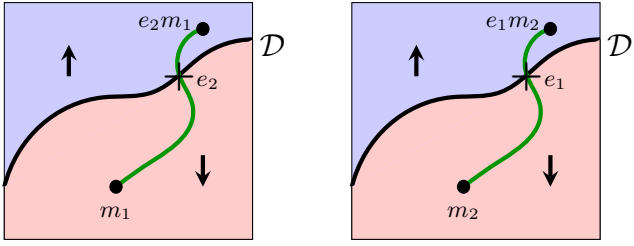

If we gauge the exact symmetry $A$ to obtain a double toric code topological order, with anyons labeled as $\{e_1,m_1,e_2,m_2\}$ and their combinations, then the average symmetry $G$ will permute the anyons according to
\begin{align}
\rho:~
\left\{
\begin{aligned}
&e_1\leftrightarrow e_1\\
&e_2\leftrightarrow e_2
\end{aligned}
\right.~,~~\left\{
\begin{aligned}
&m_1\leftrightarrow m_1e_2\\
&m_2\leftrightarrow m_2e_1
\end{aligned}
\right.
\end{align}
See Fig. \ref{anyon permutation}. This anyon permutation can be seen by firstly considering moving an $m_1$ anyon across a $\Z_2^A$ domain wall. For the (1+1)d SPT state decorated on the domain wall, moving $m_1$ across is equivalent to inserting a $\Z_2$ symmetry defect, which binds a charge of the other $\Z_2$, i.e. $e_2$. Thus the conservation of topological charge requires $m_1\rightarrow m_1e_2$ when passing the $\Z_2^A$ domain wall. Similarly we can show $m_2\rightarrow m_2e_1$.

We can relabel the anyons as $\tilde{e}_1=e_1m_2$, $\tilde{e}_2=e_2m_1$, $\tilde{m}_1=m_1$, and $\tilde{m}_2=m_2$, and the theory is rephrased as two copies of $\Z_2$ gauge theories, with $\Z_2^A$ simply exchanges the two copies.

\subsection{Intrinsically disordered ASET with $\Z_2^A$ symmetry}
\label{D16}
It is instructive to study a lattice model for an intrinsically disordered $\Z_2^A$ ASET phase. The model realizes a $\mathbb{D}_{16}=\Z_8\rtimes\Z_2$ gauge theory, denoted by D$(\mathbb{D}_{16})$. The group $\mathbb{D}_{16}$ has two generators $a$ and $r$ that satisfy $a^8=r^2=1, rar=a^{-1}$.

For a topological gauge theory with a finite gauge group $H$, a large class of anyonic symmetries can be understood as outer automorphisms ${\rm Out}(H)$ of the gauge group. Here ${\rm Out}(H)$ is the quotient ${\rm Aut}(H)/{\rm Inn}(H)$, where ${\rm Aut}(H)$ is the group of automorphisms of $H$, and Inn$(H)$ is the group of inner automorphisms (i.e. conjugation by a group element). Physically, the inner automorphisms correspond to gauge transformations, so they do not lead to faithful symmetry actions. Each element in ${\rm Out}(H)$ corresponds uniquely to an element in ${\rm Aut}({\rm D}(H))$ that does not mix electric and magnetic charges of the gauge theory. 

Adopting the general framework in Sec. \ref{sec: SETGeneral}, different ASETs with an average symmetry group $G$ are classified by homomorphisms $\rho: G\rightarrow {\rm Out}(H)$. Mathematically, $\rho$ is associated with a $\H^3(G, Z(H))$ class where $Z(H)$ is the center of the group $H$. This obstruction arises naturally when classifying the extension of $G$ by $H$: such an extension always gives a homomorphism from $G$ to Out$(H)$, but the converse is not necessarily true. Only when the $\H^3(G, Z(H))$ obstruction vanishes the group extension can be defined.  Physically, gauging the $G$ symmetry whose action is given by $\rho: G\rightarrow {\rm Out}(H)$ should result in a new gauge theory, whose gauge group is an extension of $G$ by $H$.  It turns out that this obstruction of group extension is equivalent to the $\H^3$ obstruction of anyonic symmetry induced by $\rho$~\cite{FidkowskiPRB2017}, with $Z(H)$ identified as the group of Abelian pure flux anyons. 

We now specialize to the $\mathbb{D}_{16}$ gauge theory. Write $G=\Z_2=\{1,g\}$. We assume that the image of $g$ under $\rho$ is the following automorphism:
\begin{equation}
    \rho_g: a\rightarrow a^5, r\rightarrow ra.
    \label{H3rho}
\end{equation}
The permutation action on anyons is order 2 because $\rho_g^2$ is the conjugation by $a^{-3}$.  Ref. [\onlinecite{FidkowskiPRB2017}] showed that this $\rho$ has a nontrivial $\H^3(\Z_2, \Z_2)$ class: $O_3(g,g,g)=[a^4]$. Here $[a^4]$ should be understood as the $a^4$ gauge flux in the $\mathbb{D}_{16}$ gauge theory, which is a $\Z_2$ Abelian boson. Intuitively, inserting a $g$ defect loop introduces an additional $[a^4]$ anyon.

In Ref. [\onlinecite{FidkowskiPRB2017}] a generalization of Kitaev's quantum double model~\cite{Kitaev:1997wr} is introduced for this anomalous SET state. The model is defined on a quasi-2D lattice, where each link has a 16-dimensional Hilbert space, with an orthonormal basis labeled by elements of $\mathbb{D}_{16}$. They can be viewed as lattice gauge fields. Schematically, the Hamiltonian takes the following form:
\begin{equation}
    H_\text{clean}=-\sum_v A_v-\sum_p (\delta_{F_p,1}+\delta_{F_p,a^4}).
\end{equation}
Here $A_v$ implements the Gauss's law at each vertex $v$, and $F_p$ is the gauge flux through a plaquette $p$. For the complete description of the model we refer the readers to \cite{FidkowskiPRB2017}.

Note that the second term imposes the condition that through each plaquette the gauge flux is either $1$ or $a^4$ (note $Z(\mathbb{D}_{16})=\{1,a^4\}$). This is distinct from the standard quantum double construction where the plaquette term enforces $F_p$ to be $1$. $H_\text{clean}$ has an extensive ground state degeneracy since we can have $1$ or $a^4$ flux through each plaquette. However, if the fluxes are fixed, the ground state is indeed a $\mathbb{D}_{16}$ gauge theory (with background fluxes).

The reason for this unusual magnetic term is to incorporate the obstructed $\Z_2$ symmetry. Interested readers can find more details in \cite{FidkowskiPRB2017}, here we just note that a $\Z_2$ symmetry transformation can be defined so that it implements the $\rho$ symmetry in the $\mathbb{D}_{16}$ gauge theory. However, under this $\Z_2$ symmetry transformation, we have $F_p\rightarrow F_p a^4$, which explains the form of the plaquette term. 

Now we lift this extensive number of ground states using a disordered perturbation:
\begin{equation}
\begin{split}
    H[\sigma_p]=H_\text{clean} -\sum_p \varepsilon_p( \delta_{F_p,1}-\delta_{F_p,a^4}).
\end{split}
\end{equation}
Here $\varepsilon_p$ are independent random variables drawn symmetrically from the $[-W, W]$ with $W< 1$. The $\Z_2$ symmetry remains as an average symmetry, under which $\varepsilon_p\rightarrow -\varepsilon_p$. 

In accordance with the general principle, the model supports gapless modes for the flux anyon $[a^4]$ localized at certain plaquettes, in order to accommodate the $\H^3$ obstruction. We can think of the ground state as a localized state of the $[a^4]$ flux anyons.

\subsection{Another example: disorder-enabled quantum spin liquid}
\label{sec:deqsl}

We now discuss another example of intrinsically disordered ASET, with physically realistic symmetries: $2d$ lattice translations $\Z^{(x)}\times\Z^{(y)}$, and time-reversal $\mathcal{T}$. The topological order is a simple $\Z_2$ spin liquid ($\Z_2$ topological order). 

To be concrete, let us consider a square lattice toric code Hamiltonian, but with random coefficient for the vertex terms:
\begin{equation}
\label{eq:detc}
    H=-\sum_p \prod_{l\in p} Z_l-\sum_v\epsilon_v\prod_{l\in v} X_l,
\end{equation}
where the distribution of the random coupling $P[\epsilon_v]$ is identical and independent for each vertex $v$, and satisfies $P[\epsilon_v]=P[-\epsilon_v]$. The model has average lattice translation symmetry $\Z^{(x)}\times\Z^{(y)}$ and an average time-reversal symmetry $\mathcal{T}=K\prod_{x=2n+1,y}Z_{(x,y),\hat{x}}$ ($K$ is complex conjugation, and the particular form is chosen so that $\epsilon_v$ is odd under $\mathcal{T}$)\footnote{Note that our time reversal does not commute with $\hat{x}$-translation $T_x$. We will neglect this aspect, for example by working in the Hilbert space sector with $\mathcal{T}T_x\mathcal{T}^{-1}T^{-1}_x=\prod_{x,y}Z_{(x,y),\hat{x}}=1$}. 

The $\Z_2$ spin liquid described by Eq.~\eqref{eq:detc} has similar physics as the example in Sec.~\ref{D16}: under a time-reversal transform, each unit cell (i.e. defect junction of the $\Z^{(x)}\times\Z^{(y)}$ translation symmetry) flips its $\Z_2$ gauge charge 
 measured by the vertex term. If $\mathcal{T}$ was exact, this symmetry action on the unit cell would forbid the gauge charge to be localized, so Eq.~\eqref{eq:detc} is indeed disorder-enabled as an ASET. As long as the distribution $P[\epsilon_v]$ is bounded, the ground state will be gapless when $\prod_v\epsilon_v<0$. These features are robust against weak perturbations that preserve the symmetries on average.

In contrast, we can have another toric code Hamiltonian that is more conventional from symmetry point of view:
\begin{equation}
\label{eq:trivialtc}
     H_0=-\sum_p \prod_{l\in p} X_l-\sum_v\prod_{l\in v} Z_l,
\end{equation}
which is exactly symmetric under $\Z^{(x)}\times\Z^{(y)}\times\mathcal{T}$. Our result indicates that there is no statistically symmetric smooth path interpolating between Eq.~\eqref{eq:detc} and Eq.~\eqref{eq:trivialtc}.

\subsection{ASET with 't Hooft anomaly: an example with $\Z_2\times\Z_2^A$}
\label{semion}
\begin{figure}
\centering
\begin{tikzpicture}[xscale=0.45,>=latex,yscale=0.5]
\path[draw,name path=border1] (0,0+1) to[out=-10,in=150] (6,-2+1.5);
\path[draw,name path=border2] (12,1+1.5) to[out=150,in=-10] (5.5,3.7+1);
\draw[draw,thick,name path=line1] (6,-2+1.5) -- (12,1+1.5);
\path[draw,name path=line2] (5.5,3.7+1) -- (0,0+1);
\shade[left color=iro10,right color=iro60]
  (0,0+1) to[out=-10,in=150] (6,-2+1.5) --
  (12,1+1.5) to[out=150,in=-10] (5.5,3.7+1) -- cycle;

\path[draw,name path=border3] (-1,-4) to[out=20,in=220] (3,3);
\path[draw,name path=border4] (6,-7) to[out=40,in=210] (9,1);
\path[draw,name path=border5] (-1,-4) to[out=0,in=80] (6,-7);
\path[draw,name path=border6] (3,3) to[out=10,in=140] (9,1);

\node at (10,-4.5) (label) {\large Levin-Gu};
\draw[thick,->] (label) -- +(185:3.35);

\shade[top color=iro10,bottom color=iro90,opacity=.30]
  (-1,-4) to[out=20,in=220] (3,3)  to[out=10,in=140] (9,1)
 to[out=210,in=40] (6,-7) to[out=80,in=0] (-1,-4);

\path[draw,->] (1,0+1.2) to[out=-10,in=150] (5.6,0);
\draw[draw,->] (6.2,0) -- (8,1);
\path[draw,->] (8,1.4) to[out=140,in=10] (3.2,2.7);
\draw[draw,->] (2.8,2.5) -- (1.3,1.5);
\draw[draw,->] (9.2,1.5) -- (11,2.5);
\path[draw,->] (4.1,3.3) to[out=10,in=150] (8.8,1.6);
\path[draw,->] (10.5,2.85) to[out=150,in=-10] (5.6,4.4);
\draw[draw,->] (5.2,4.3) -- (4,3.5);
\node at (5,1.5) {\large CSL$_1$};
\node at (8.5,2.8) {\large CSL$_2$};
\end{tikzpicture}
\caption{Surface topological order of (3+1)$d$ ASPT with $\Z_2\times\Z_2^A$ symmetry. The indigo surface depicts the $\Z_2^A$ domain wall decorated by a Levin-Gu state, and the violet surface depicts the surface chiral spin liquid enriched by $\Z_2\times\Z_2^A$ symmetry.}
\label{Z2*Z2A}
\end{figure}
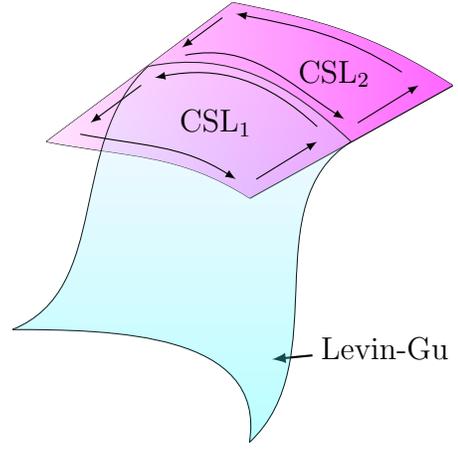

We consider the surface of a (3+1)$d$ ASPT with $\Z_2\times\Z_2^A$ symmetry. The generator of the $\Z_2$ ($\Z_2^A$) will be denoted by $g$ ($g_A$). In the bulk ASPT state, on each (2+1)$d$ domain wall of the average $\Z_2^A$ symmetry we decorate a Levin-Gu state \cite{LevinGu} of the exact $\Z_2$ (see Fig. \ref{Z2*Z2A}), with the following topological action on a 4-manifold $X_4$:
\begin{align}
S=\int_{X_4}a^3\cup b
\end{align}
where $a$ and $b$ are background gauge fields of $\Z_2$ and $\Z_2^A$, respectively. Notice that when both symmetries are exact, there are two other nontrivial topological terms. One of them is given by
\begin{equation}
    S'=\int_{X_4}a\cup b^3.
\end{equation}
It corresponds to decoration of exact $\Z_2$ charges on junctions of the $\Z_2^A$ symmetry. As explained in Sec. \ref{sec:localization}, this action $S'$ is trivialized when $\Z_2^A$ becomes average in a disordered ensemble. The other action is $S+S'$, which is now identified with $S$.

On the (2+1)$d$ surface of this state, we have a random network of $\Z_2^A$ domain walls, each decorated with a Levin-Gu edge theory described by a Luttinger liquid [$\varphi^T=(\varphi^1,\varphi^2)$],
\begin{align}
\mathcal{L}=\frac{K_{IJ}}{4\pi}\left(\partial_x\varphi^I\right)(\partial_t\varphi^J)+\frac{V_{IJ}}{8\pi}\left(\partial_x\varphi^I\right)(\partial_x\varphi^J)
\label{Luttinger formulation}
\end{align}
where the $K$-matrix $K=\sigma^x$. The nontrivial $\Z_2$ symmetry action is defined as 
\begin{align}
\varphi^{1,2}\mapsto\varphi^{1,2}+\pi.
\label{Z2 symmetry}
\end{align}

We now want to gap out these domain wall modes in a $\Z_2$ symmetric way, which can be achieved by placing a semion topological order/chiral spin liquid (CSL) on the surface. Then on each domain wall, we have not only the Levin-Gu edge state but also two counter-propagating chiral Luttinger liquids as the edge modes of the semion topological orders on both sides of the domain wall. The total (1+1)$d$ domain wall theory is
\begin{align}
\mathcal{L}=\frac{K_{IJ}'}{4\pi}\left(\partial_x\varphi^I\right)(\partial_t\varphi^J)+\frac{V_{IJ}'}{8\pi}\left(\partial_x\varphi^I\right)(\partial_x\varphi^J)
\end{align}
where $\varphi^T=(\varphi^1,\varphi^2,\varphi^3,\varphi^4)$, and the $K$-matrix $K'=\sigma^x\oplus2\sigma^z$. A semion is created by the operator $e^{i\varphi^3}$ on one side of the domain wall, or $e^{i\varphi^4}$ on the other side. Originally, without Levin-Gu edge modes, the domain wall can be gapped by adding a Higgs term $\cos(2\varphi^3-2\varphi^4)$, which induces coherent tunneling of the semions across the domain walls. Alternatively, with the Levin-Gu edge modes, we can gap out the theory in a $\Z_2$ symmetric way by the following Higgs terms,
\begin{align}
\cos(\varphi^1+\varphi^2-2\varphi^4)+\cos(\varphi^1-\varphi^2-2\varphi^3).
\label{Higgs}
\end{align}
In order for this term to preserve the $\Z_2$ symmetry, $e^{2i\varphi^4}$ and $e^{2i\varphi^3}$ should be invariant under $\Z_2$.

We now analyze the symmetry fractionalization between $\Z_2$ and $\Z_2^A$ in the semion surface theory.  It is instructive to start with the clean case, when both symmetries are exact (we will continue denote the symmetry group as $\Z_2\times\Z_2^A$). Denote the local $h$ symmetry action on a semion by $U_h$ for every $h\in \Z_2\times\Z_2^A$.  The group relations in $\Z_2\times\Z_2^A$ lead to the following invariants:
\begin{equation}
\begin{split}
     \lambda_h = U_h^2, h\in \Z_2\times\Z_2^A\\
     \eta = U_g U_{g_A}U_{g}^{-1}U_{g_A}^{-1}.
\end{split}
\end{equation}
All of them take $\pm 1$ value. They are related by the algebraic identity: $\lambda_g\lambda_{g_A}\lambda_{gg_A}=\eta$.
So there are three independent $\Z_2$ invariants for the projective symmetry action, consistent with the $\H^2(\Z_2\times\Z_2,\Z_2)=\Z_2^3$ classification. The four $\eta=1$ classes can all be realized by one-dimensional representations, while the $\eta=-1$ classes must be realized by at least two-dimensional representations. In particular, the one with $\lambda_g=\lambda_{g_A}=\lambda_{gg_A}=-1$ is realized in the semion chiral spin liquid, where $\Z_2\times\Z_2^A$ is a subgroup of the SO(3) symmetry. The other three classes with $\eta=-1$ are all anomalous with respect to the (exact) $\Z_2\times \Z_2^A$ symmetry. Their 't Hooft anomalies were computed in \cite{XieChen2015} and we recall the results in Table \ref{tab:semion_anomaly}.

\begin{table}
\renewcommand\arraystretch{1.2}
\centering
\begin{tabular}{|c|c|c|c|}
\hline
$\eta$ & $\lambda_{g}$ & $\lambda_{g_A}$ & $S_{\rm anomaly}$ \\
\hline
$1$ & $\pm 1$ & $\pm 1$ & 0\\
\hline
~$-1$~  & $1$  & $1$ & ~$a\cup b^3+a^3\cup b$~\\
\hline
$-1$ & $1$ & ~$-1$~ & $a^3\cup b$\\
\hline
$-1$  & ~$-1$~ & $1$ & $a\cup b^3$\\
\hline
$-1$ & $-1$ & $-1$ & 0\\
\hline
\end{tabular}
\caption{'t Hooft anomalies for the projective $\Z_2\times\Z_2^A$ symmetry actions in clean semion topological order.}
\label{tab:semion_anomaly}
\end{table}

When $\Z_2^A$ becomes average, as discussed in Sec. \ref{sec: SETGeneral}, the invariants $\lambda_{g_A}$ and $\lambda_{gg_A}$ are ambiguous due to the localization of semions. However, $\lambda_g$ and $\eta$ should remain well-defined. The former should be fairly clear since $\Z_2$ is an exact symmetry, so we now explain how to define $\eta$. Since $U_g U_{g_A}=\eta U_{g_A}U_g$, under the $g_A$ action the $\Z_2$ charge, measured by the eigenvalue of $U_g$, changes by $\eta$. Therefore, $\eta$ can be measured as the change of the $\Z_2$ charge when when adiabatically moving a semion across a $g_A$ defect (which implements the $g_A$ symmetry on the semion).

First we determine $\lambda_g$. In the Luttinger liquid formulation (\ref{Luttinger formulation}), $\lambda_g=-1$ means that under the $\Z_2$ symmetry, $\varphi^3$ and $\varphi^4$ must shift by an odd multiples of $\pi/2$, and therefore $e^{2i\varphi^{3/4}}$ should be odd under $\Z_2$. However, we have seen that for the Higgs term to preserve symmetry $e^{2i\varphi^{3/4}}$ have to be $\Z_2$ even. Thus we must have $\lambda_g=1$.

From Eq. (\ref{Z2 symmetry}) and \eqref{Higgs}, we find that at the domain wall $\varphi^4$ is identified with $\varphi^3+\varphi^2$,  therefore under the $\Z_2$ symmetry, $e^{i\varphi^4}$ has opposite $\Z_2$ charge to that of $e^{i\varphi^3}$. In other words, adiabatically moving a semion across a domain wall changes its global $\Z_2$ charge by $-1$.  This is the manifestation of the non-commutativity between $\Z_2$ and $\Z_2^A$ acting on the semion in the ensemble.

Let us now discuss how the anomaly inflow works in the average SET. We have shown that $\lambda_g$ and $\eta$ remain good invariants, and $\lambda_{g_A}$ and $\lambda_{gg_A}$ are not well-defined individually, but their ratio is fixed by $\lambda_g\eta$. Interestingly, with $\eta=-1$ fixed, the anomalies in Table \ref{tab:semion_anomaly} separate into two groups corresponding to $\lambda_g=\pm 1$. The two classes inside each group differs just by the $a\cup b^3$ term,    which describes decorations of $\Z_2$ charges on $\Z_2^A$ defect junctions, and becomes trivial due to the localization. This ties nicely with the observation that $\lambda_{g_A}$ is now ambiguous. The new bulk-boundary correspondence for the semion theory with average $\Z_2\times \Z_2^A$ symmetry is now summarized in Table \ref{tab:semion_anomaly2}.
\begin{table}
\renewcommand\arraystretch{1.2}
\centering
\begin{tabular}{|c|c|c|}
\hline
$\eta$ & $\lambda_{g}$  & ~$S_{\rm anomaly}$~ \\
\hline
$1$ & $\pm 1$ & 0\\
\hline
~$-1$~ & $1$  & $a\cup b^3$\\
\hline
$-1$  & ~$-1$~  & $0$\\
\hline
\end{tabular}
\caption{Average 't Hooft anomalies  in disordered semion topological order. }
\label{tab:semion_anomaly2}
\end{table}

This example illustrates that for a mixed group of exact and average symmetry, ``symmetry fractionalization'' can be well-defined on anyons. We have also derived the complete average 't Hooft anomaly matching for $\Z_2\times\Z_2^A$ symmetry in the semion topological order.

\subsection{An example with Lieb-Schultz-Mattis anomaly}

We consider a (2+1)$d$ lattice system with average $\mathbb{Z}\times\Z$ translation symmetry, exact spin SO(3) rotation symmetry and a spin-$1/2$ moment per lattice unit cell. This system has a 't Hooft anomaly from Lieb-Schultz-Mattis (LSM) constraints~\cite{ChengPRX2016, Metlitski:2017fmd}
\begin{equation}
    S=\int_{X_4}x\cup y\cup w_2^{\mathrm{SO}(3)},
\end{equation}
where $x,y$ are background gauge fields of the $x,y$ translation symmetries, and $w_2^{\mathrm{SO}(3)}$ is the second Stiefel-Whitney class of the background SO(3) gauge field. This anomaly remains nontrivial as the $\Z\times\Z$ translation symmetry becomes average \cite{MaWangASPT}.

In the clean limit, one of the most well-known topological orders that match with the LSM anomaly is a $\Z_2$ topological order. In this $\Z_2$ topological order, the $e$ particle carries a projective representation of SO(3), corresponding to the nontrivial symmetry fractionalization class in $\H^2(\mathrm{SO}(3),\Z_2)$ (in common term, it carries spin-$1/2$); the $m$ particle transforms projectively under $\Z\times\Z$ (the nontrivial class in $\H^2(\Z\times\Z,\Z_2)$), in the sense that $T_xT_y=-T_yT_x$ when acting on $m$, where $T_{x,y}$ are the generators of the $x,y$ translations. These symmetry fractionalizations are required for the $\Z_2$ topological order to match the anomaly. While the spin-$1/2$ moment on the $e$ particle is not affected by disorders, we do need to explain the meaning of ``$T_xT_y=-T_yT_x$" on the $m$ particle. As discussed in Sec.~\ref{sec: SETGeneral}, the expression $T_xT_y=-T_yT_x$ on $m$ can be interpreted as having an $e$ particle localized at each unit cell (the intersection of a $T_x$ domain wall and a $T_y$ domain wall). If $e$ particle does not carry any degeneracy, then we can deform the state by randomly distributing localized $e$ particles among all unit cells, and states with different $e$ particle distributions will be smoothly connected. However, since $e$ particle carries spin-$1/2$ moment, the additional localized $e$ particles must find their way to form an SO(3) invariant (singlet) state. We do not expect the singlet state to be achievable without developing further long-range entanglement. Therefore the notion of ``one $e$ particle per unit cell'' is robust. Equivalently, $T_xT_y=-T_yT_x$ on $m$ particle and the $\Z_2$ topological order indeed matches the LSM anomaly.

\section{Summary and outlook}
\label{sec:summary}

In this work we have developed a systematic framework, based on the physical picture of defect decoration and the mathematical tool of spectral sequence, to classify and characterize average symmetry-protected topological (ASPT) phases, in both decohered and disordered systems. We have also studied average symmetry-enriched topological (ASET) phases in disordered systems. Our main results are:
\begin{itemize}
\item We emphasized the subtle differences between ASPT phases in decohered and disordered systems, which leads to different classifications of ASPT in the two scenarios. Nevertheless, they can both be classified and characterized under the same framework of spectral sequence (decorated defects).
\item We discovered a plethora of ASPT phases that are intrinsically mixed, in the sense that they can only appear in mixed state systems (decohered or disordered) with part of the symmetry being average. In other words, these states cannot be viewed as clean SPT states deformed by decoherence or disorder. Some of these states can, however, be viewed as intrinsically gapless pure SPT deformed by decoherence or disorder. We discussed many examples, in both bosonic and fermionic systems.
\item We developed a systematic theory of ASET phases in disordered $(2+1)d$ bosonic systems. Compared to clean SET phases, disordered ASET with average (and exact) symmetries have some distinct features:
\begin{enumerate}[1.]
\item While an average symmetry can still permute different anyons, its fractional representation on the anyons cannot be robustly defined, unless the fractionalization pattern involves some 't Hooft anomaly.
\item An average symmetry and an exact symmetry can jointly have fractional representation on the anyons.
\item The $\H^3$ obstructions of symmetry-enrichment patterns are lifted when the relevant symmetries become average. This leads to intrinsically disordered ASET phases without clean limits. The ground states of such ASET phases contain localized anyons, which leads to gapless (yet still localized) excitation spectra.
\end{enumerate}
\end{itemize}

We end with some open questions:

\begin{itemize}

\item {\color{black}\textit{Stability of disordered ASPT/ASET}: in our study we demanded the ground state wavefunctions of disordered ASPT/ASET states to be adiabatically connected to some gapped clean system. It appears physically reasonable to expect that this condition could provide some level of stability against weak perturbations on the Hamiltonian. However, since the actual disordered Hamiltonians are in general gapless, a rigorous proof of stability would be difficult. The issue may also be related to the stability of (ground-state) localization. It is an interesting future direction to demonstrate the stability of such topological phases with disorder-induced gaplessness. }

\item \textit{Boundary physics}: In this work, we mostly focused on the bulk classification and characterization of ASPT phases. In \cite{MaWangASPT}, it was shown that the boundary of an ASPT state has an average 't Hooft anomaly. We have discussed a few examples of bulk-boundary correspondence, including (1+1)$d$ $\Z_2\times\Z_2^{\rm ave}$ ASPT phase and the surface ASET of a (3+1)$d$ ASPT phase. It will be useful to understand more systematically how average 't Hooft anomalies constrain the behavior of the boundary states, especially for the intrinsically mixed/disordered ASPTs with no clean limit.

\item \textit{Fermionic ASET}: In this work we studied bosonic ASETs in (2+1)$d$ disordered systems. A natural question is to extend the theory to fermionic systems. In the clean limit, classifications of fermionic SET phases have been developed recently in \cite{Bulmash2022a, Bulmash2022b, Aasen:2021vva}. While certain basic elements of the theory are parallel to the bosonic case, there are important new ingredients and subtleties, especially when 't Hooft anomalies are concerned.

\item \emph{Topological order in mixed states}: We have limited ourselves to topological order in disorder ensembles, where the notion of SRE ensembles can be naturally extended to LRE ensembles. Defining the notion of topological order for a general mixed state is an important question. In fact, a large family of examples can be obtained by ``classically" gauging average symmetry and ``quantum-mechanically" gauging exact symmetry in ASPT phases.   

\item \textit{Phase transitions}: quantum phase transitions of intrinsically mixed ASPT or ASET will necessarily involve decoherence or disorders in important manners. This makes the study of such (necessarily non-unitary) quantum phase transitions both challenging and exciting.

\item \textit{Physical realizations}: an important task is to realize some ASPT or ASET phases, especially the intrinsically mixed ones, in experimental platforms such as NISQ simulators or disordered solid-state systems. On this front, simple preparation protocols such as those outlined in Sec.~\ref{sec:BerryFree} are particularly promising.

\end{itemize}

\begin{acknowledgements}

We thank Vladimir Calvera, Tyler Ellison, Dominic Else, Zheng-Cheng Gu, David Long, Sagar Vijay, Chenjie Wang, Cenke Xu, Yichen Xu, Carolyn Zhang, and Yijian Zou for the helpful discussions. { We are especially grateful to Yijian Zou for pointing out a mistake in the previous version of the manuscript.} R.M. acknowledges support from the Natural Sciences and Engineering Research Council of Canada (NSERC) through Discovery Grants. Research at Perimeter Institute (R.M. and C.W.) is supported in part by the Government of Canada through the Department of Innovation, Science and Industry Canada and by the Province of Ontario through the Ministry of Colleges and Universities. J.H.Z. and Z.B. are supported by a startup fund at Penn State University. M.C. acknowledges support from NSF under award number DMR-1846109. J.H.Z., Z.B., and C.W. thank the hospitality of the Kavli Institute for Theoretical Physics, which is supported in part by the National Science Foundation under Grant No. NSF PHY-1748958.

\end{acknowledgements}

\appendix

\section{Classification of decohered ASPT with $A\times G$ symmetry}
\label{App:ASPTclassification}

We can understand the nontriviality of the states in Eq.~\eqref{eq:decoheredclassificationdp}, $\H^{D+1}(A\times G,\U)/\H^{D+1}(G,\U)$, by focusing on pure states. Consider a pure state $|\Psi\rangle$, symmetric under $A\times G$. We can understand Eq.~\eqref{eq:decoheredclassificationdp} in two steps:
\begin{enumerate}[1.]
    \item If $|\Psi\rangle$ is an invertible state protected solely by symmetry $G$, then it is two-way connected to a pure product state $|00...\rangle$ via FD local channels that are weakly symmetric in $G$.  The channel $\E_1: |\Psi\rangle\to|00...\rangle$ can be constructed by starting with an ancillary Hilbert space that is isomorphic to the physical one $\H'\simeq\H$ with identical $G$-action (allowed since $G$ is only a weak symmetry). Then we can have 
    \begin{equation}
        |\Psi\rangle\otimes |00...\rangle \xrightarrow{U_{\rm SWAP}}|00...\rangle\otimes |\Psi\rangle \xrightarrow{{\rm tr}_{\H'}}|00...\rangle,
    \end{equation}
    where $U_{\rm SWAP}$ is the on-site gate swapping the physical and ancillary system. The reverse channel $\E_2:|00...\rangle\to |\Psi\rangle$ can be constructed also using an isomorphic ancilla $\H'\simeq\H$:
    \begin{equation}
|00...\rangle\otimes|00...\rangle\xrightarrow{U}|\Psi\rangle\otimes|\tilde{\Psi}\rangle\xrightarrow{{\rm tr}_{\H'}}|\Psi\rangle,
    \end{equation}
    where $U$ creates $|\Psi\rangle$ and its inverse $|\tilde{\Psi}\rangle$. 
    
    \item If $|\Psi\rangle$ is protected jointly by $A\times G$ (or by $A$ alone), then we can show that neither $\E_1$ nor $\E_2$ can exist if $\E_{1,2}$ are strongly (weakly) symmetric under $A$ ($G$). To see this, note that if $\E_1: |\Psi\rangle\to |00\dots\rangle$, then $\E_1$ can be purified to a symmetric FD unitary
    \begin{equation}     
    \label{eq:pureform}
    U|\Psi\rangle\otimes |00...\rangle=|00...\rangle\otimes |\Psi'\rangle,
    \end{equation}
    where the tensor product form of the right hand side is required by the fact that tracing out the ancilla gives a pure state $|00...\rangle$. Eq.~\eqref{eq:pureform} then requires $|\Psi'\rangle\in\H'$ to belong to the same SPT phase as $|\Psi\rangle$ -- but this is impossible since, by definition, $A$ acts trivially on $\H'$, so $\H'$ cannot support any SPT that requires $A$ symmetry. The same logic can be applied to show that we cannot have $\E_2:|00...\rangle\to |\Psi\rangle$.
\end{enumerate}

To summarize, states in $\H^{D+1}(G,\U)$ are trivial, while nontrivial states in 
\begin{equation}
    \H^{D+1}(A\times G,\U)/\H^{D+1}(G,\U)
\end{equation}
remain nontrivial in the decohered context.

\section{On the definition of disordered ASPT}
\label{App:DefASPT}

The definition of disordered average SPT originally given in Ref.~\cite{MaWangASPT} was slightly more complicated than that given in Sec.~\ref{sec:disorderedReview}. In particular, in Ref.~\cite{MaWangASPT} a short-range entangled ensemble requires any two ground states to be adiabatically connected, namely for any disorder realization $I, I'$, it is required that $|\Psi_I\rangle=U_{II'}|\Psi_I\rangle$ for some finite-depth local unitary $U_{II'}$. Instead, in Sec.~\ref{sec:disorderedReview} we only demand that any ground state in the ensemble $|\Psi_I\rangle$ to be short-range entangled with correlation length $\xi_I<\xi_{m}$ for some finite maximal correlation length in the ensemble $\xi_{m}$. We now discuss to what extend our new definition is equivalent to the old one.

Consider a ground state in the ensemble $|\Psi_I\rangle$ in a large but finite system without boundary. As a short-range entangled state, it belongs to an invertible phase labeled by $\omega_I^{D+1}\in h^{D+1}(A)$, where $h^{D+1}(A)$ classifies invertible phases with exact symmetry $A$ in $D$ space dimensions. On a finite system, we could effectively compactify the space by viewing certain dimensions as points -- for example, the entire system can be viewed as a point if we zoom out far enough. When the system is compactified this way to a $(D-p)$ dimensional space, it should be labeled as a $(D-p)$ invertible state labeled by $\omega_I^{D-p+1}\in h^{D-p+1}(A)$ (it should also depend on the compactification cycle but we will omit this information in the notation). In general $\omega_I^{D-p+1}$ is not directly decided by the bulk state $\omega_I^{D+1}$. In particular, when $p=D$ the system reduces to a point and the only nontrivial information left is the charge under $A$ symmetry.

Now the condition in Ref.~\cite{MaWangASPT} is the statement that $\omega_I^{D-p+1}$ is identical for any disorder realization $I$, for all $0\leq p\leq D$. We now show that the new condition in Sec.~\ref{sec:disorderedReview} automatically implies the old condition for $0\leq p\leq D-1$: if for some $0\leq p\leq D-1$, $\omega_I\neq\omega_I'$ for two disorder realizations $H_I$ and $H_{I'}$, then we can choose a third disorder realization in the ensemble $I''$, which has $H_{I''}=H_I$ in some region $R$, and $H_{I''}=H_{I'}$ in the complement $\bar{R}$. Notice that the spatial independence of the disorder potential is crucial for this construction. The difference in $\omega$ in $R$ and $\bar{R}$, for suitably chosen $R$, implies a long-range entangled boundary state on $\partial R$, which leads to a correlation length $\sim|\partial R|$. This leads to unbounded correlation length, and therefore violates the new condition in Sec.~\ref{sec:disorderedReview}.

The above argument does not apply for $p=D$, since two regions with different $A$ charges do not need to have nontrivial boundary state. However, our definition of exact symmetry in Sec.~\ref{sec:disorderedReview} requires that the exact charge to be identical for each disorder realization, so the $p=D$ condition is automatically satisfied. There is still one subtle difference between the old and new definition. In Ref.~\cite{MaWangASPT} we consider the ensemble of ground states $\{|\Psi^{(0)}_I\rangle\}$ of the Hamiltonians $\{H_I\}$, and if the states have different total $A$ charges, the ensemble is not considered symmetric SRE. In this work, however, we consider ground states $\{|\Psi_I\rangle\}$ of the Hamiltonians $\{H_I\}$, subject to the condition that the $A_I$ charge is some $I$-independent fixed value $Q$. This is analogous to canonical ensemble (instead of grand canonical ensemble) in statistical mechanics. In such ``canonical ensemble'', an individual state $|\Psi_I\rangle$ may not be the absolute ground state of $H_I$. In this case $|\Psi_I\rangle$ has a finite excitation energy, namely the excitation energy density should vanish. If the excitation is localized, in the sense that $|\Psi_I\rangle$ can be obtained from the absolute ground state $|\Psi_I^{(0)}\rangle$ using a finite-depth local unitary circuit $|\Psi_I\rangle=U_{FD}|\Psi_I^{(0)}\rangle$, then $|\Psi_I\rangle$ is still short-range entangled if the absolute ground state is short-range entangled. As long as all $|\Psi_I\rangle$'s are short-range entangled and adiabatically connected to each other, we can still consider the ensemble as symmetric SRE, and the discussions on topological phases can be proceeded with no difficulty. A consequence of using the ``canonical ensemble'' is that states constructed by decorating zero-dimensional average symmetry defects are no longer considered nontrivial, as discussed in Sec.~\ref{sec:disorderedReview}.

\section{A brief review of spectral sequence}
\label{App: spectral sequence}
This section provides a brief overview of the LHS spectral sequence of group cohomology, which is relevant for classifying bosonic ASPT and ASET phases. We also extend our discussion to the more general Atiyah-Hirzebruch (AH) spectral sequence.

A spectral sequence consists of an assembly of Abelian groups $E_r^{p,q}$ ($0\leq r,p,q\in\mathbb{Z}$). For a fixed $r$, the collection of all $E_r^{p,q}$ are called $E_r$ \textit{page}. The differentials are defined as the group endomorphism of the $E_r$ page as:
\begin{align}
\mathrm{d}_r:E_{r}^{p,q}\rightarrow E_{r}^{p+r,q-r+1},
\label{r differential}
\end{align}
which should satisfy $\mathrm{d}_r^2=0$. Therefore, the $E_r$ pages and the differentials $\mathrm{d}_r$ form a cochain complex, with the following isomorphism:
\begin{align}
E_{r+1}^{p,q}\simeq\frac{\mathrm{Ker}(\mathrm{d}_r^{p,q})}{\mathrm{Im}(\mathrm{d}_r^{p-r,q+r-1})}=H_*(E_r,\mathrm{d}_r),
\label{r+1 page}
\end{align}
where $H_*(E_r,\mathrm{d}_r)$ is the homology group of the cochain complex $\{E_r,\mathrm{d}_r\}$. 

For the LHS spectral sequence with the symmetry group given by Eq. (\ref{group extension}), we denote the set of $n$-cochains, $n$-cocycles, and $n$-coboundaries of a group $G_0$ with coefficients in $M$ by $\mathcal{C}^n[G_0, M]$, $\mathcal{Z}^n[G_0, M]$, and $\mathcal{B}^n[G_0, M]$, respectively. 

\begin{figure*}
\begin{tikzpicture}[scale=0.85]
\tikzstyle{sergio}=[rectangle,draw=none]
\draw[draw=black] (-7.5,-1.5)--(-7.5,3.5)--(-2.5,3.5)--(-2.5,-1.5)--cycle;
\draw[densely dashed] (-2.5,-0.5)--(-7.5,-0.5);
\draw[densely dashed] (-2.5,2.5)--(-7.5,2.5);
\draw[densely dashed] (-2.5,1.5)--(-7.5,1.5);
\draw[densely dashed] (-2.5,0.5)--(-7.5,0.5);
\draw[densely dashed] (-6.5,-1.5)--(-6.5,3.5);
\draw[densely dashed] (-3.5,-1.5)--(-3.5,3.5);
\draw[densely dashed] (-4.5,-1.5)--(-4.5,3.5);
\draw[densely dashed] (-5.5,-1.5)--(-5.5,3.5);
\path (-8,3.75) node [style=sergio]{\large$p$};
\path (-2.25,-2) node [style=sergio]{\large$q$};
\path (-8,3) node [style=sergio]{\large$4$};
\path (-8,-1) node [style=sergio]{\large$0$};
\path (-8,2) node [style=sergio]{\large$3$};
\path (-8,1) node [style=sergio]{\large$2$};
\path (-8,0) node [style=sergio]{\large$1$};
\path (-7,-2) node [style=sergio]{\large$0$};
\path (-6,-2) node [style=sergio]{\large$1$};
\path (-5,-2) node [style=sergio]{\large$2$};
\path (-4,-2) node [style=sergio]{\large$3$};
\path (-3,-2) node [style=sergio]{\large$4$};
\path (-6,0) node [style=sergio]{$E_0^{1,1}$};
\path (-6,1) node [style=sergio]{$E_0^{1,2}$};
\draw[->,color=red,thick] (-6,0.25)--(-6,0.75);
\path (-5.7,0.5) node [style=sergio,color=red]{$\mathrm{d}_0$};
\path (-4,1) node [style=sergio]{$E_0^{3,2}$};
\path (-4,2) node [style=sergio]{$E_0^{3,3}$};
\draw[->,color=red,thick] (-4,1.25)--(-4,1.75);
\path (-3.7,1.5) node [style=sergio,color=red]{$\mathrm{d}_0$};
\draw[draw=black] (-0.5,-1.5)--(-0.5,3.5)--(4.5,3.5)--(4.5,-1.5)--cycle;
\draw[densely dashed] (4.5,-0.5)--(-0.5,-0.5);
\draw[densely dashed] (4.5,2.5)--(-0.5,2.5);
\draw[densely dashed] (4.5,1.5)--(-0.5,1.5);
\draw[densely dashed] (4.5,0.5)--(-0.5,0.5);
\draw[densely dashed] (0.5,-1.5)--(0.5,3.5);
\draw[densely dashed] (3.5,-1.5)--(3.5,3.5);
\draw[densely dashed] (2.5,-1.5)--(2.5,3.5);
\draw[densely dashed] (1.5,-1.5)--(1.5,3.5);
\path (-1,3.75) node [style=sergio]{\large$p$};
\path (4.75,-2) node [style=sergio]{\large$q$};
\path (-1,3) node [style=sergio]{\large$4$};
\path (-1,-1) node [style=sergio]{\large$0$};
\path (-1,2) node [style=sergio]{\large$3$};
\path (-1,1) node [style=sergio]{\large$2$};
\path (-1,0) node [style=sergio]{\large$1$};
\path (0,-2) node [style=sergio]{\large$0$};
\path (1,-2) node [style=sergio]{\large$1$};
\path (2,-2) node [style=sergio]{\large$2$};
\path (3,-2) node [style=sergio]{\large$3$};
\path (4,-2) node [style=sergio]{\large$4$};
\path (1,0) node [style=sergio]{$E_1^{1,1}$};
\path (2,0) node [style=sergio]{$E_1^{2,1}$};
\draw[->,color=red,thick] (1.25,0)--(1.7,0);
\path (1.5,-0.25) node [style=sergio,color=red]{$\mathrm{d}_1$};
\path (2,2) node [style=sergio]{$E_1^{2,3}$};
\path (3,2) node [style=sergio]{$E_1^{3,3}$};
\draw[->,color=red,thick] (2.25,2)--(2.7,2);
\path (2.5,1.75) node [style=sergio,color=red]{$\mathrm{d}_1$};
\draw[draw=black] (6.5,-1.5)--(6.5,3.5)--(11.5,3.5)--(11.5,-1.5)--cycle;
\draw[densely dashed] (11.5,-0.5)--(6.5,-0.5);
\draw[densely dashed] (11.5,2.5)--(6.5,2.5);
\draw[densely dashed] (11.5,1.5)--(6.5,1.5);
\draw[densely dashed] (11.5,0.5)--(6.5,0.5);
\draw[densely dashed] (7.5,-1.5)--(7.5,3.5);
\draw[densely dashed] (10.5,-1.5)--(10.5,3.5);
\draw[densely dashed] (9.5,-1.5)--(9.5,3.5);
\draw[densely dashed] (8.5,-1.5)--(8.5,3.5);
\path (6,3.75) node [style=sergio]{\large$p$};
\path (11.75,-2) node [style=sergio]{\large$q$};
\path (6,3) node [style=sergio]{\large$4$};
\path (6,-1) node [style=sergio]{\large$0$};
\path (6,2) node [style=sergio]{\large$3$};
\path (6,1) node [style=sergio]{\large$2$};
\path (6,0) node [style=sergio]{\large$1$};
\path (7,-2) node [style=sergio]{\large$0$};
\path (8,-2) node [style=sergio]{\large$1$};
\path (9,-2) node [style=sergio]{\large$2$};
\path (10,-2) node [style=sergio]{\large$3$};
\path (11,-2) node [style=sergio]{\large$4$};
\path (8,1) node [style=sergio]{$E_2^{2,1}$};
\path (10,0) node [style=sergio]{$E_2^{3,1}$};
\draw[->,color=red,thick] (8.1705,0.9213)--(9.644,0.1807);
\path (8.75,0.25) node [style=sergio,color=red]{$\mathrm{d}_2$};
\path (9,3) node [style=sergio]{$E_1^{2,3}$};
\path (11,2) node [style=sergio]{$E_1^{3,3}$};
\draw[->,color=red,thick] (9.1706,2.8951)--(10.6249,2.1817);
\path (9.75,2.25) node [style=sergio,color=red]{$\mathrm{d}_2$};
\path (-5,-2.75) node [style=sergio]{\large(a). $E_0$ page};
\path (2,-2.75) node [style=sergio]{\large(b). $E_1$ page};
\path (9,-2.75) node [style=sergio]{\large(c). $E_2$ page};
\end{tikzpicture}
\caption{LHS spectral sequence and differentials in $E_0$, $E_1$, and $E_2$ pages.}
\label{LHS spectral sequence}
\end{figure*}
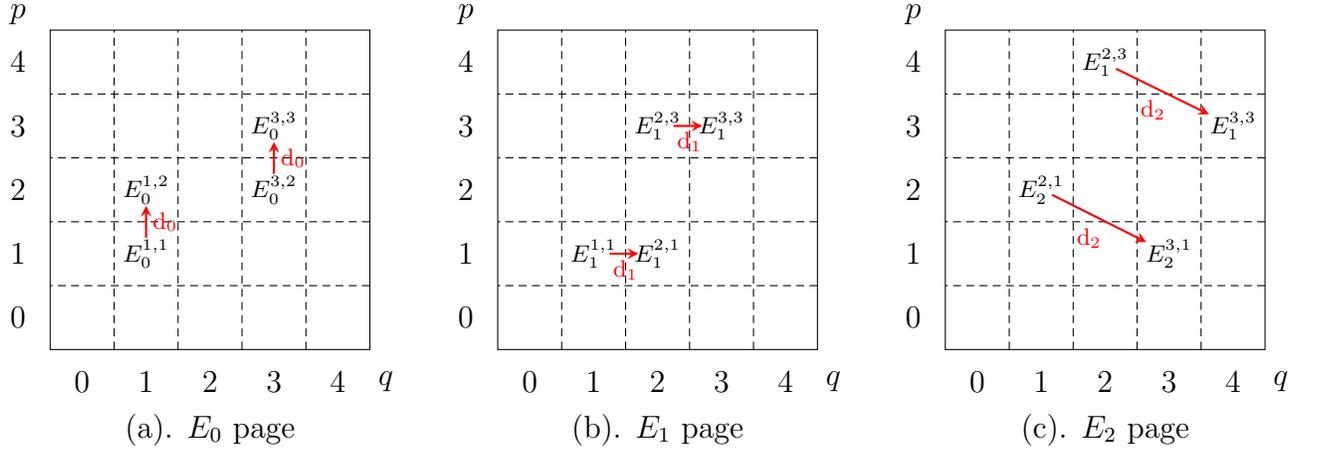

The $E_0$ page of the LHS spectral sequence is defined as a group of cochains $E_0^{p,q}=\mathcal{C}^p\left(G, \mathcal{C}^q[A, M]\right)$, and the $\mathrm{d}_0$ differential maps a cochain in $E_0^{p,q}$ to a cochain $E_0^{p,q+1}$ (see Fig. \ref{LHS spectral sequence}(a) for illustration). The kernel of $\mathrm{d}_0^{p,q}$ is $\mathcal{C}^p\left(G, \mathcal{Z}^q[A, M]\right)$, while the image of $\mathrm{d}_0^{p,q-1}$ is just $\mathcal{C}^p\left(G, \mathcal{B}^{q}[A, M]\right)$. Hence the $E_1$ page is given according to Eq. (\ref{r+1 page}) by
\begin{align}
E_1^{p, q}&=\frac{\mathrm{Ker}(\mathrm{d}_0^{p,q})}{\mathrm{Im}(\mathrm{d}_0^{p,q-1})}=\frac{\mathcal{C}^p\left(G, \mathcal{Z}^q[A, M]\right)}{\mathcal{C}^p\left(G, \mathcal{B}^{q}[A, M]\right)}\nonumber\\
&=\mathcal{C}^p\left(G, \mathcal{H}^q[A, M]\right).
\end{align}
We recognize that $E_1$ page is a subgroup of $E_0$ page: $E_1\subset E_0$. More precisely, if we label the elements of $E_0^{p,q}$ by $w_0^{p,q}$, the elements of $E_1^{p,q}$ are the equivalence classes of elements in $E_0^{p,q}$ that satisfy the condition $\mathrm{d}_0^{p,q}w_0^{p,q}=0$. 

Subsequently, for the $E_1$ page $E_1=\mathcal{C}^p\left(G, \mathcal{H}^q[A, M]\right)$, the differential $\mathrm{d}_1$ maps a cochain in $\mathcal{C}^p\left(G, \mathcal{H}^q[A, M]\right)$ to a cochain in $\mathcal{C}^{p+1}\left(G, \mathcal{H}^q[A, M]\right)$ [see Fig. \ref{LHS spectral sequence}(b)]. The kernel of $\mathrm{d}_1^{p,q}$ is cocycles in $\mathcal{Z}^p\left(G, \mathcal{H}^q[A, M]\right)$, and the image of $\mathrm{d}_1^{p-1,q}$ is coboundries in $\mathcal{B}^p\left(G, \mathcal{H}^q[A, M]\right)$. Then according to Eq. (\ref{r+1 page}), the $E_2$ page is given by
\begin{align}
E_2^{p,q}&=\frac{\mathrm{Ker}(\mathrm{d}_1^{p,q})}{\mathrm{Im}(\mathrm{d}_1^{p-1,q})}=\frac{\mathcal{Z}^p\left(G, \mathcal{H}^q[A, M]\right)}{\mathcal{B}^p\left(G, \mathcal{H}^q[A, M]\right)}\nonumber\\
&=\mathcal{H}^p\left(G, \mathcal{H}^q[A, M]\right).
\end{align}
And $E_2$ page is a subset of $E_1$ page,
\begin{align}
E_2\subset E_1\subset E_0.
\end{align}
More precisely, elements of $E_2$ page are those elements in $E_0$ page that satisfy the conditions
\begin{align}
\mathrm{d}_0w_0=0,~~\mathrm{d}_1w_0=0.
\end{align}
Following the above paradigm, we can further define the arbitrary $E_r$ page, satisfying the following condition:
\begin{align}
E_r\subset E_{r-1}\subset \cdot\cdot\cdot\subset E_2\subset E_1\subset E_0.
\end{align}
The elements in $E_r$ page should satisfy the following $r$ conditions,
\begin{align}
\mathrm{d}_qw_0=0,~~0\leq q\leq r-1.
\end{align}
In particular, if there is a large enough integer $r$ such that the condition $\mathrm{d}_{r}w_0=0$ is satisfied over the entire $E_r$ page, then the $E_{r+1}$ page is essentially identical to the $E_r$ page: $E_{r+1}=E_r$, and all higher pages are the same. It is then said that the spectral sequence stabilizes at the $E_r$ page. For the LHS spectral sequence, the $E_\infty$ page is isomorphic to the group cohomology $\mathcal{H}^{p+q}[\tilde{G}, M]$ as a set. In this paper, we set $M=U(1)$ for the classification of ASPT phases.

The generalization to the AH spectral sequence is straightforward: We can construct the $E_r$ pages of the AH spectral sequence from LHS spectral sequence by substituting the term $\H^q[A, U(1)]$ characterizing the classification of $A$-symmetric bosonic group-cohomology SPT phases in $q$-dimensional spacetime by the generalized cohomology group $h^q(A)$ characterizing the classification of $A$-symmetric invertible topological phases. For example, the $E_2$ page is defined as
\begin{align}
E_2^{p,q}=\mathcal{H}^p[G,h^q(A)],~p+q=d+1.
\end{align}
Likewise, we can also define the differentials as Eq. (\ref{r differential}), and the AH spectral sequence will converge to the generalized cohomology group $h^{d+1}(\tilde{G})$, i.e., $E_\infty$ is isomorphic to $h^{d+1}(\tilde{G})$.

 Below we will see how the AH spectral sequence emerges from a construction of fixed-point wavefunctions of (2+1)$d$ fermionic SPT phases in Appendix~\ref{App: construction}.

\section{Constructions of (2+1)$d$ fermionic SPTs and ASPTs}
\label{App: construction}
 We review the construction of fixed-point wavefunctions for fermionic SPT (FSPT) phases in (2+1)$d$ interacting fermionic systems following \cite{general2}, and describe how the construction can be modified for fermionic ASPT phases. This is also a good example to illustrate the AH spectral sequence for decorated domain wall construction. Below $G_b$ denotes the physical symmetry group, assumed to be finite for simplicity.  

The fixed-point wavefunction is constructed by proliferating decorated domain wall: the quantum superposition of all possible $G_b$ symmetry-breaking pattern $|\{g_i\}\rangle$ with fermionic decoration, as
\begin{align}
|\Psi_{\mathrm{FSPT}}\rangle=\sum\limits_{\{g_i\}}\Psi(\{g_i\})|\{g_i\}\rangle
\label{FSPT wavefunction}
\end{align}
The basis state $|\{g_i\}\rangle$ is a state decorated by Majorana chains (the nontrivial element in $h^2(\Z_2^f)=\Z_2$) on the $G_b$ symmetry domain walls, and complex fermions (the nontrivial element in $h^1(\Z_2^f)=\Z_2$) on the junctions of the $G_b$ symmetry domain walls. Therefore, on the triangle lattice, there are three layers of degrees of freedom:
\begin{enumerate}[1.]
\item $|G_b|$ level bosonic state $|g_i\rangle$ ($g_i\in G_b$) on each vertex;
\item $|G_b|$ species of complex fermions $c_{ijk}^\sigma$ ($\sigma\in G_b$) at the center of each triangle $\langle ijk\rangle$;
\item $|G_b|$ species of complex fermions (split to pairs of Majorana fermions) $a_{ijk}^\sigma=(\gamma_{ij, A}^\sigma+i\gamma_{ij, B}^\sigma)/2$ on two sides of the link $\langle ij\rangle$.
\end{enumerate}
All degrees of freedom on a triangle are summarized as follows
\begin{align}
\begin{tikzpicture}[scale=1.2,decoration={
    markings,
    mark=at position 0.5 with {\arrow{>}}}]
\tikzstyle{sergio}=[rectangle,draw=none]
\path (-7.75,-1.75) node [style=sergio]{$g_0$};
\path (-6,1.5) node [style=sergio]{$g_2$};
\path (-4.25,-1.75) node [style=sergio]{$g_1$};
\draw[postaction={decorate},thick] (-7.5,-1.5)--(-6,1.25);
\draw[postaction={decorate},thick] (-7.5,-1.5)--(-4.5,-1.5);
\draw[postaction={decorate},thick] (-4.5,-1.5)--(-6,1.25);
\shade[ball color=green] (-6,-0.6) circle (0.12);
\path (-6,0) node [style=sergio,color=DarkGreen]{\small$c_{012}^\sigma$};
\filldraw[fill=red, draw=red] (-6,-1.2)circle (1.5pt);
\filldraw[fill=red, draw=red] (-6.5,-0.25)circle (1.5pt);
\filldraw[fill=red, draw=red] (-5.5,-0.25)circle (1.5pt);
\draw[postaction={decorate},densely dashed,color=red] (-6.5,-0.25)--(-6,-1.2);
\draw[postaction={decorate},densely dashed,color=red] (-5.5,-0.25)--(-6,-1.2);
\draw[postaction={decorate},densely dashed,color=red] (-6.5,-0.25)--(-5.5,-0.25);
\filldraw[fill=red, draw=red] (-7,0)circle (1.5pt);
\filldraw[fill=red, draw=red] (-5,0)circle (1.5pt);
\filldraw[fill=red, draw=red] (-6,-1.8)circle (1.5pt);
\path (-5.5373,-1.1992) node [style=sergio,color=red]{\footnotesize$\gamma_{01,A}^\sigma$};
\path (-5.5504,-1.8127) node [style=sergio,color=red]{\footnotesize$\gamma_{01,B}^\sigma$};
\path (-6.6705,-0.6049) node [style=sergio,color=red]{\footnotesize$\gamma_{02,B}^\sigma$};
\path (-5.2705,-0.6049) node [style=sergio,color=red]{\footnotesize$\gamma_{12,A}^\sigma$};
\path (-7,0.3) node [style=sergio,color=red]{\footnotesize$\gamma_{02,A}^\sigma$};
\path (-5,0.3) node [style=sergio,color=red]{\footnotesize$\gamma_{12,B}^\sigma$};
\draw[postaction={decorate},densely dashed,color=red] (-7,0)--(-6.5,-0.25);
\draw[postaction={decorate},densely dashed,color=red] (-5.5,-0.25)--(-5,0);
\draw[postaction={decorate},densely dashed,color=red] (-6,-1.2)--(-6,-1.8);
\end{tikzpicture}
\label{triangulation}
\end{align}

The $G_b$ symmetry transformations of the degrees of freedom are ($g,g_i,\sigma\in G_b$)
\begin{align}
\begin{gathered}
U(g)|g_i\rangle=|gg_i\rangle\\
U(g)c_{ijk}^\sigma U^\dag(g)=(-1)^{\omega_2(g,\sigma)}c_{ijk}^{g\sigma}\\
U(g)\gamma_{ij,A}^\sigma U^\dag(g)=(-1)^{\omega_2(g,\sigma)}\gamma_{ij,A}^{g\sigma}\\
U(g)\gamma_{ij,B}^\sigma U^\dag(g)=(-1)^{\omega_2(g,\sigma)+s_1(g)}\gamma_{ij,B}^{g\sigma}
\end{gathered},
\label{symmetry fermion}
\end{align}
where $\omega_2(g,\sigma)$ is the factor system of the group extension (\ref{group extension}) characterizing the group structure of $\tilde{G}$, and $s_1(g)$ is defined as
\begin{align}
s_1(g)=
\begin{cases}
0 & ~\mathrm{if}~g~\text{is unitary}\\
1 & \mathrm{if}~g~\text{is antiunitary}
\end{cases}.
\end{align}

We first consider the Majorana chain decoration on the domain walls of $G_b$. 
The configuration of the decorated Majorana chain is specified by $n_1(g_i,g_j)\in\Z_2$ on the link $\langle ij\rangle$ . If the Majorana chain does not go across the link $\langle i,j\rangle$, we say $n_1(g_i,g_j)=0$, and the Majorana fermions $\gamma_{ij,A}^\sigma$ and $\gamma_{ij,B}^\sigma$ on two sides of the link are paired up as $-i\gamma_{ij,A}^\sigma\gamma_{ij,B}^\sigma=1$ for $\forall\sigma\in G_b$. If the Majorana chain goes across the link $\langle i,j\rangle$, Majorana fermions $\gamma_{ij, A}^{g_i}$ (and $\gamma_{ij, B}^{g_i}$) will form Majorana entanglement pairs with other Majoranas inside the triangle, and all other Majorana fermions $\gamma_{ij, A}^\sigma$ and $\gamma_{ij, B}^\sigma$ with $\sigma\ne g_i$ are still paired up as $-i\gamma_{ij, A}^\sigma\gamma_{ij, B}^\sigma=1$. The quantity $n_1(g_0,g_1)n_1(g_1,g_2)=0,1$ indicates if $\gamma_{12,A}^{g_1}$ and $\gamma_{01,A}^{g_0}$ are paired up or not [see Eq. (\ref{triangulation})]. The arrows in Eq.(\ref{triangulation}) specify the direction of the Majorana pair.

The total number of decorated Majorana chains going through the edges of a specific triangle $\langle012\rangle$ is
\begin{align}
\mathrm{d}n_1(g_0,g_1,g_2)=n_1(g_0,g_1)+n_1(g_0,g_2)+n_1(g_1,g_2).
\end{align}
Towards a gapped SPT wavefunction, the above quantity should be even to avoid dangling Majorana fermion, i.e.,
\begin{align}
\mathrm{d}n_1=0~(\mathrm{mod}~2).
\end{align}
Therefore $n_1$ is an element of $\mathcal{Z}^1[G, h^2(\Z_2^f)]$, the $E_2^{1,2}$ term on the $E_2$ page of the AH spectral sequence.

We see that the symmetry action may change the directions of the Majorana entanglement pairs, and thereby the fermion parity of a specific triangle.  One can find that the fermion parity inside a specific triangle under the symmetry action $U(g_0)$ changes by
\begin{align}
\Delta P_f^\gamma=(-1)^{(\omega_2\cup n_1+s_1\cup n_1\cup n_1)(g_0,g_0^{-1}g_1,g_1^{-1}g_2)}.
\label{n1 obstruction}
\end{align}

In order to get a fermionic SPT wavefunction, the wavefunction (\ref{FSPT wavefunction}) should have a definite fermion parity, i.e., states $\ket{\{g\}}$ with different bosonic configurations $\{g\}$ should have the same fermion parity. We take a specific triangulated state $\ket{\{g_0\}}$ and consider a global symmetry transformation $U(g)$, say $U(g)\ket{\{g_0\}}=\ket{\{gg_0\}}$, the fermion parity difference between $\ket{\{g_0\}}$ and $\ket{\{gg_0\}}$ would be the product of Eq. (\ref{n1 obstruction}) over the whole triangulation, say
\begin{align}
\Delta P_f=\prod\limits_\Delta(-1)^{(\omega_2\cup n_1+s_1\cup n_1\cup n_1)(g_1,g_2,g_3)},
\end{align}
where $g_1, g_2, g_3\in \{g\}$. Towards a well-defined fixed-point wavefunction of FSPT, we should enforce $\Delta P_f=1$, and obtain that the following expression should be a 3-coboundary with $\Z_2$ coefficient,
\begin{align}
\omega_2\cup n_1+s_1\cup n_1\cup n_1.
\end{align}
It can be shown that this 3-coboundary is actually $\mathrm{d}n_2$ [cf. Eq. (\ref{fsptconsistency})], where $n_2$ is the topological index of complex fermion decoration \cite{general2}.


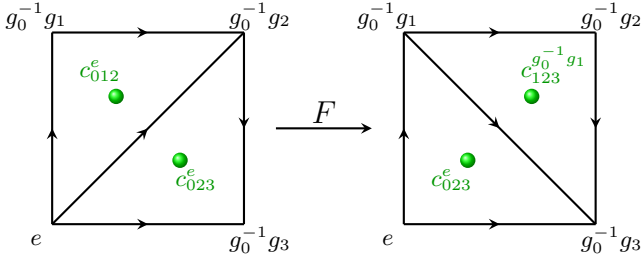
\begin{figure}
\begin{tikzpicture}[scale=0.85,decoration={
    markings,
    mark=at position 0.5 with {\arrow{>}}}]
\tikzstyle{sergio}=[rectangle,draw=none]
\path (-7.75,-1.75) node [style=sergio]{$e$};
\path (-7.75,1.75) node [style=sergio]{$g_0^{-1}g_1$};
\path (-4.25,-1.75) node [style=sergio]{$g_0^{-1}g_3$};
\path (-4.25,1.75) node [style=sergio]{$g_0^{-1}g_2$};
\draw[postaction={decorate},thick] (-7.5,-1.5)--(-7.5,1.5);
\draw[postaction={decorate},thick] (-7.5,-1.5)--(-4.5,-1.5);
\draw[postaction={decorate},thick] (-4.5,1.5)--(-4.5,-1.5);
\draw[postaction={decorate},thick] (-7.5,1.5)--(-4.5,1.5);
\draw[postaction={decorate},thick] (-7.5,-1.5)--(-4.5,1.5);
\shade[ball color=green] (-6.5,0.5) circle (0.12);
\shade[ball color=green] (-5.5,-0.5) circle (0.12);
\path (-6.75,0.9) node [style=sergio,color=DarkGreen]{$c_{012}^e$};
\path (-5.25,-0.8) node [style=sergio,color=DarkGreen]{$c_{023}^e$};
\draw[->,thick] (-4,0) -- (-2.5,0);
\path (-3.25,0.25) node [style=sergio]{\large$F$};
\draw[postaction={decorate},thick] (-2,-1.5)--(1,-1.5);
\draw[postaction={decorate},thick] (-2,-1.5)--(-2,1.5);
\draw[postaction={decorate},thick] (-2,1.5)--(1,1.5);
\draw[postaction={decorate},thick] (1,1.5)--(1,-1.5);
\draw[postaction={decorate},thick] (-2,1.5)--(1,-1.5);
\shade[ball color=green] (0,0.5) circle (0.12);
\shade[ball color=green] (-1,-0.5) circle (0.12);
\path (-2.25,-1.75) node [style=sergio]{$e$};
\path (-2.25,1.75) node [style=sergio]{$g_0^{-1}g_1$};
\path (1.25,-1.75) node [style=sergio]{$g_0^{-1}g_3$};
\path (1.25,1.75) node [style=sergio]{$g_0^{-1}g_2$};
\path (-1.25,-0.8) node [style=sergio,color=DarkGreen]{$c_{023}^e$};
\path (0.35,1) node [style=sergio,color=DarkGreen]{$c_{123}^{g_0^{-1}g_1}$};
\end{tikzpicture}
\caption{$F$-move of complex fermion decoration. The Majorana chain decorations are omitted.}
\label{F-move}
\end{figure}

Lastly, we need to check the pentagon identity is satisfied for the $F$-moves (see Fig. \ref{F-move}), to make sure that there is no nontrivial Berry phase in the space of decorated domain wall states. As shown in Ref. \cite{general2}, this condition requires a certain U(1) 4-cocycle $O_4[n_2]$ to be cohomologically trivial. The expression for $O_4[n_2]$ can be found in \cite{general2}.  Note that $n_1$ does not directly contribute to the obstruction function. 

To summarize, the input data to the construction is a triple $(n_1, n_2, \omega_3)$, where $n_1\in {\mathcal{C}}^1(G_b, \Z_2), n_2\in \mathcal{C}^2(G_b, \Z_2)$ and $\omega_3\in {\cal C}^3(G_b, \U)$, satisfying the following consistency conditions:
\begin{align}
    \mathrm{d}n_1 &=0,\nonumber \\
    \mathrm{d}n_2 &=\omega_2\cup n_1+s_1\cup n_1\cup n_1, \nonumber \\
    \mathrm{d}\omega_3 &=O_4[n_2]
    \label{fsptconsistency}
\end{align}
This is precisely the structure expected from the AH spectral sequence, where $n_1,n_2$ and $\omega_3$ correspond to the $E_2$ page, and we can identify the differentials. For example:
\begin{align}
    d_2n_1&\equiv \omega_2\cup n_1+ s_1\cup n_1\cup n_1,\\
    d_2n_2&\equiv n_2\cup n_2.
\end{align}

We now modify the construction to get disordered ASPT states. First, the complex fermions $c_{ijk}^\sigma$ is decoupled from the domain wall configurations, and forms an Anderson insulator. Concretely, we can use the following Hamiltonian for the $c_{ijk}$ fermions:
\begin{equation}
    H_{\rm AI}[\{\varepsilon\}]=-\sum_{f}\sum_\sigma \varepsilon_{f\sigma} (c_f^\sigma)^\dag c_f^\sigma.
\end{equation}
Here $f$ enumerates all the triangles. $\varepsilon_{f\sigma}$ is a random variable drawn symmetrically from $[-W,W]$ where $W>0$ (the probability distribution should be independent of $\sigma$).

The $G_b$ spins $\{g\}$ are now treated as quenched disorders, assumed to be short-range correlated. The Majorana decorations on domain walls proceed as before, so we include Hamiltonian terms that enforce the decoration pattern, denoted by $H_D[\{g\}]$. The full Hamiltonian reads
\begin{equation}
    H[\{g\},\{\varepsilon\}]=H_D[\{g\}]+H_{\rm AI}[\{\varepsilon\}].
\end{equation}
The ensemble consists of ground state of $H[\{g\},\{\varepsilon\}]$ in the total fermion parity even sector. 

For decohered SPTs, it is still necessary to impose the fermion parity conservation Eq. \eqref{n1 obstruction}. Then the ``fixed-point" density matrix is given by
\begin{equation}
    \rho=\sum_{\{g\}}p(\{g\})\ket{\{g\}}\bra{\{g\}}.
\end{equation}
Here $\ket{\{g\}}$ denote the decorated domain wall states, and $p(\{g\})$ is the probability distribution of domain walls.

\bibliography{apssamp}

\end{document}